\documentclass[aos]{imsart}

\usepackage{amsmath, amsthm, amssymb}
\usepackage{graphicx}
\usepackage{verbatim}
\usepackage{natbib}
\usepackage{caption}
\usepackage{subcaption}
\usepackage{fancyvrb}
\usepackage{enumerate}
\usepackage{relsize}
\usepackage{algorithm}
\usepackage[noend]{algpseudocode}
\usepackage{ragged2e}

\usepackage{tikz}
\usetikzlibrary{trees}
\usepackage{minibox}

\usepackage[colorinlistoftodos]{todonotes}
\newcommand\todoitem[1]{}%{\textcolor{red}{#1}}

\usepackage{hyperref}
\usepackage[margin=1.5in]{geometry}
\hypersetup{colorlinks,citecolor=blue,urlcolor=blue,linkcolor=blue}

\usepackage{stefan_tex}
\graphicspath{{./figures/}}

\makeatletter
\def\BState{\State\hskip-\ALG@thistlm}
\makeatother

\tikzstyle{level 1}=[level distance=3.5cm, sibling distance=3.5cm]
\tikzstyle{level 2}=[level distance=3.5cm, sibling distance=2cm]

\tikzstyle{bag} = [text width=4em, text centered]
\tikzstyle{end} = [circle, minimum width=3pt,fill, inner sep=0pt]

\theoremstyle{plain}
\newtheorem{prop}{Proposition}

\newtheorem{lemm}[prop]{Lemma}
\newtheorem{theo}[prop]{Theorem}

\theoremstyle{definition}
\newtheorem{exam}{Example}

\newtheorem{assu}{Assumption}
\newtheorem{spec}{Specification}

\newtheorem*{rema*}{Remark}

\theoremstyle{remark}

\newcommand{\FIGW}{0.45}

\arxiv{1610.01271}

\begin{document}

\begin{frontmatter}

\title{Generalized Random Forests}
\runtitle{Generalized Random Forests}

\author{\fnms{Susan} \snm{Athey}\ead[label=e1]{athey@stanford.edu}}
\address{Stanford Graduate School of Business \\ 655 Knight Way \\ Stanford, CA-94305, USA \\\printead{e1}}
\author{\fnms{Julie} \snm{Tibshirani}\ead[label=e2]{julietibs@gmail.com}}
\address{Elasticsearch BV \\ 800 West El Camino Real, Suite 350  \\ Mountain View, CA-94040 \\\printead{e2}}
\and
 \author{\fnms{Stefan} \snm{Wager}\corref{}\ead[label=e3]{swager@stanford.edu}}
 \address{Stanford Graduate School of Business \\ 655 Knight Way \\ Stanford, CA-94305, USA \\ \printead{e3}}
 \affiliation{Stanford University and Elasticsearch BV}

\runauthor{Athey, Tibshirani and Wager}

\begin{abstract}
\sloppy{
We propose generalized random forests, a method for non-parametric statistical estimation based
on random forests (Breiman, 2001) that can be used to fit any quantity of interest
identified as the solution to a set of local moment equations.
Following the literature on local maximum likelihood estimation, our method considers a weighted set of
nearby training examples; however, instead of using classical kernel weighting functions
that are prone to a strong curse of dimensionality, we use an adaptive
weighting function derived from a forest designed to express heterogeneity in
the specified quantity of interest.
We propose a flexible, computationally efficient algorithm for growing generalized random forests,
develop a large sample theory for our method showing that our estimates are consistent and asymptotically Gaussian,
and provide an estimator for their asymptotic variance that enables valid confidence intervals.
We use our approach to develop new methods for three statistical tasks:
non-parametric quantile regression,
conditional average partial effect estimation,
and heterogeneous treatment effect estimation via instrumental variables.
A software implementation, \texttt{grf} for \texttt{R} and \texttt{C++}, is
available from \texttt{CRAN}.}
\end{abstract}

\end{frontmatter}

\section{Introduction}

Random forests, introduced by \citet{breiman2001random}, are a widely used algorithm for statistical learning.
Statisticians usually study random forests as a practical method for non-parametric conditional mean estimation:
Given a data-generating distribution for $(X_i, \, Y_i) \in \xx \times \RR$, forests are used to estimate
%\begin{equation}
%\label{eq:conditional_mean}
$\mu(x) = \EE{Y_i \cond X_i = x}$.
%\end{equation}
Several theoretical results are available on the asymptotic behavior of such forest-based
estimates \smash{$\hmu(x)$}, including consistency
\citep{arlot2014analysis,biau2008consistency,biau2012analysis,denil2014narrowing,lin2006random,
scornet2015consistency,wager2015uniform},
second-order asymptotics \citep{mentch2016quantifying},
and confidence intervals \citep{wager2015estimation}.

This paper extends Breiman's random forests into a flexible
method for estimating any quantity $\theta(x)$ identified via local moment conditions. Specifically,
given data $(X_i, \, O_i) \in \xx \times \oo$, we seek forest-based estimates of $\theta(x)$ defined by a local
estimating equation of the form
\begin{equation}
\label{eq:conditional_theta}
\EE{\psi_{\theta(x), \, \nu(x)}\p{O_i} \cond X_i = x} = 0 \ \text{ for all } \ x \in \xx, 
\end{equation}
\sloppy{where $\psi(\cdot)$ is some scoring function and $\nu(x)$ is an optional nuisance parameter.
This setup encompasses several key statistical problems. For example, if we model
the distribution of $O_i$ conditionally on $X_i$ as having a density $f_{\theta(x), \, \nu(x)}(\cdot)$
then, under standard regularity conditions, the moment condition \eqref{eq:conditional_theta} with
\smash{$\psi_{\theta(x), \, \nu(x)}(O) = \nabla  \log\p{f_{\theta(x), \, \nu(x)}(O_i)}$}
identifies the local maximum likelihood parameters $(\theta(x), \, \nu(x))$. More generally, we can use moment
conditions of the form \eqref{eq:conditional_theta} to identify conditional means, quantiles, average partial effects, etc., and to develop robust regression procedures via Huberization.
Our main substantive application of generalized random forests
involves heterogeneous treatment effect estimation with instrumental variables.}

Our aim is to build a family of non-parametric
estimators that inherit the desirable empirical properties of regression forests---such as
stability, ease of use, and flexible adaptation to different functional forms as in, e.g.,
\citet{biau2016random} or \citet{varian2014big}---but can be
used in the wide range of statistical settings characterized by \eqref{eq:conditional_theta}
in addition to standard conditional mean estimation.
This paper addresses the resulting conceptual and methodological challenges
and establishes formal asymptotic results.

Regression forests are typically understood as ensemble methods, i.e., forest
predictions \smash{$\hmu(x)$} are written as the average of $B$ noisy tree-based predictors \smash{$\hmu_b(x)$},
%\begin{equation}
%\label{eq:simple_avg}
\smash{$\hmu(x) = B^{-1} \sum_{b = 1}^B \hmu_b(x)$};
%\end{equation}
and, because individual trees \smash{$\hmu_b(x)$} have low bias but high variance,
such averaging meaningfully stabilizes predictions \citep{buhlmann2002analyzing,scornet2015consistency}.
However, noisy solutions to moment
equations as in \eqref{eq:conditional_theta} are generally biased, and averaging
would do nothing to alleviate the bias.

To avoid this issue, we cast forests as a type of adaptive locally weighted estimators that
first use a forest to calculate a weighted set of neighbors for each test point $x$, and then
solve a plug-in version of the estimating equation \eqref{eq:conditional_theta} using these neighbors. Section \ref{sec:ANN} gives a detailed treatment of this perspective.
This locally weighting view of random forests was previously advocated by
\citet{hothorn2004bagging} in the context of survival analysis and
by \citet{meinshausen2006quantile} for quantile regression, and also
underlies theoretical analyses of regression forests \citep[e.g.,][]{lin2006random}.
For conditional mean estimation, the averaging and weighting views of forests are equivalent;
however, once we move to more general settings, the weighting-based perspective proves substantially more effective,
and also brings forests closer to the literature on local maximum likelihood estimation
\citep{fan1996local,loader1999local,newey1994kernel,stone1977consistent,tibshirani1987local}.

A second challenge in generalizing forest-based methods is that their success hinges on whether the adaptive
neighborhood function obtained via partitioning adequately captures the
heterogeneity in the underlying function $\theta(x)$ we want to estimate.
Even within the same class of statistical tasks, different types of questions can require different neighborhood functions.
For example, suppose that two scientists are studying the effects
of a new medical treatment: One is looking at how the treatment affects long-term
survival, and the other at its effect on the length of hospital stays.
It is plausible that the treatment heterogeneity in each setting would be based on
disparate covariates, e.g., a patient's smoking habits for long-term survival,
and the location and size of the hospital for the length of stay.

Thus, each time we apply random forests to a new scientific task, it is important to use rules for recursive partitioning
that are able to detect and highlight heterogeneity in the signal the researcher is interested in.
In prior work, such problem-specific rules have largely been designed on a case by case basis.
Although the CART rules of \citet{breiman1984classification}
have long been popular for classification and regression tasks, there has been a steady
stream of papers proposing new splitting rules for other problems, including \citet{athey2016recursive}
and \citet{su2009subgroup} for treatment effect estimation, \citet{beygelzimer2009offset}
and \citet{kallus2016learning} for personalized policy allocation, and
\citet{gordon1985tree}, \citet{leblanc1992relative}, \citet{molinaro2004tree} as well as several others
for survival analysis. \citet{zeileis2008model} propose a method
for constructing a single tree for general maximum likelihood problems, where splitting is based on
hypothesis tests for improvements in goodness of fit.

In contrast, we seek a unified, general framework for computationally efficient problem-specific splitting rules,
optimized for the primary objective of capturing heterogeneity in a key parameter of interest.
In the spirit of gradient boosting \citep{friedman2001greedy}, our recursive partitioning
method begins by computing a linear, gradient-based approximation to the
non-linear estimating equation we are trying to solve, and then uses
this approximation to specify the tree-split point. Algorithmically, our procedure
reduces to iteratively applying a labeling step where we generate pseudo-outcomes
by computing gradients using parameters estimated in the parent node, and a regression step where we pass this labeled data
to a standard CART regression routine.  Thus, we can make use of pre-existing,
optimized tree software to execute the regression step, and obtain high quality
neighborhood functions while only using computational
resources comparable to those required by standard CART algorithms.  
In line with this approach, our generalized random forest software package
builds on the carefully optimized \texttt{ranger}
implementation of regression forest splitting rules \citep{wright2015ranger}.

Moment conditions of the form \eqref{eq:conditional_theta} typically arise in
scientific applications where rigorous statistical inference is required.
The bulk of this paper is devoted to a theoretical analysis of generalized random
forests, and to establishing asymptotic consistency and Gaussianity of the resulting
estimates \smash{$\htheta(x)$}. We also develop methodology for asymptotic confidence
intervals. Our analysis is motivated by classical results for local estimating equations,
in particular \citet{newey1994kernel}, paired with machinery from
\citet{wager2015estimation} to address the adaptivity of the random forest weighting function.

The resulting framework presents a flexible method for non-parametric
statistical estimation and inference with formal asymptotic guarantees.
In this paper, we develop applications to quantile regression, conditional average partial
effect estimation and heterogeneous treatment effect estimation with instrumental variables; however,
there are many other popular statistical models that fit directly into our framework,
including panel regression, Huberized robust regression, models of consumer choice, etc.
In order to fit any of these models with generalized random forests,
the analyst simply needs to provide the problem-specific routines to calculate gradients of the
moment conditions evaluated at different observations in the dataset for the ``label'' step of our algorithm.
Moreover, we emphasize that our method is in fact
a proper generalization of regression forests: If we apply our framework to build a forest-based method for
local least-squares regression, 
we exactly recover a regression forest.
A high-performance software implementation of generalized random forests,
\texttt{grf} for \texttt{R} and \texttt{C++}, is available from \texttt{CRAN}.

\subsection{Related Work}

The idea of local maximum likelihood (and local generalized method of moments) estimation has a long history,
including \citet{fan1998local}, \citet{newey1994kernel}, \citet{staniswalis1989kernel}, \citet{stone1977consistent},
\citet{tibshirani1987local} and \citet{lewbel2007local}. In economics, popular applications of these techniques include
multinomial choice modeling in a panel data setting \citep[e.g.,][]{honore2000panel} and instrumental variables
regression \citep[e.g.,][]{su2013local}. The core idea is that when estimating
parameters at a particular value of covariates, a kernel weighting function is used to place more weight on nearby observations
in the covariate space. A challenge facing this approach is that if the covariate space has more than two or three dimensions, performance can suffer due to the ``curse of dimensionality'' \citep[e.g.,][]{robins1997toward}.

Our paper replaces the kernel weighting with forest-based weights, that is,
weights derived from the fraction of trees in which an observation
appears in the same leaf as the target value of the covariate vector.  The original random
forest algorithm for non-parametric classification and regression
was proposed by \citet{breiman2001random}, building on insights from the ensemble learning
literature \citep{amit1997shape,breiman1996bagging,dietterich2000experimental,ho1998random}.
The perspective we take on random forests as a form of
adaptive nearest neighbor estimation, however, most closely builds on the proposals of
\citet{hothorn2004bagging} and \citet{meinshausen2006quantile} for forest-based survival
analysis and quantile regression. This adaptive
nearest neighbors perspective also underlies several statistical analyses of random forests,
including \citet{arlot2014analysis}, \citet{biau2010layered}, and \citet{lin2006random}.

Our gradient-based splitting scheme draws heavily from a long tradition in the statistics and
econometrics literatures of using gradient-based test statistics to detect change points in likelihood models
\citep{andrews1993tests,hansen1992testing,hjort2002tests,nyblom1989testing,
ploberger1992cusum,zeileis2005unified,zeileis2007generalized}.
In particular, \citet{zeileis2008model} consider the use of such methods for
model-based recursive partitioning. Our problem setting differs from the above
in that we are not focused on running a hypothesis test, but rather seek an adaptive
nearest neighbor weighting that is as sensitive as possible to heterogeneity in our
parameter of interest; we then rely on the random forest resampling mechanism
to achieve statistical stability \citep{mentch2016quantifying,scornet2015consistency,wager2015estimation}.
In this sense, our approach is related to gradient boosting
\citep{friedman2001greedy}, which uses gradient-based approximations to guide
a greedy, non-parametric regression procedure.

Our asymptotic theory relates to an extensive recent literature on the statistics of random forests, most
of which focuses on the regression case
\citep{arlot2014analysis,biau2012analysis,biau2008consistency,biau2016random,
buhlmann2002analyzing,denil2014narrowing,geurts2006extremely,
ishwaran2010consistency,lin2006random,meinshausen2006quantile,mentch2016quantifying,
scornet2015consistency,sexton2009standard,wager2015estimation,wager2015uniform,zhu2015reinforcement}.
Our present paper complements this body of work by showing how methods developed
to study regression forests can also be used understand estimated solutions to local
moment equations obtained via generalized random forests.

%Finally we note that the problem we study, namely estimating how a function $\theta(x)$ varies with covariates,
%is distinct from the problem of estimating a single, low-dimensional
%parameter---such as an average treatment effect---while controlling 
%for a non-parametric or high-dimensional set of covariates. Recent contributions to the latter
%include \citet{athey2016efficient}, \citet{chernozhukov2016double},
%\citet{robins2017minimax}, and \citet{van2006targeted}.
%In particular, \citet{chernozhukov2016double} and \citet{van2006targeted} discuss how
%traditional machine learning methods like regression forests can be used as sub-components
%in efficient inference about such low-dimensional parameters.

\section{Generalized Random Forests}

%\subsection{Review of Breiman's Forests}

%As with any algorithm widely used by practitioners, there are many perspectives
%on what random forests are and how they work. \citet{breiman2001random} followed
%the perspective of \citet{amit1997shape}, \citet{breiman1996bagging}, \citet{dietterich2000experimental}, \citet{ho1998random}, etc., where the forest prediction is viewed as a consensus decision made by an ensemble of trees. For example, when discussing classification forests, \citet{breiman2001random} lets each tree ``vote'' on how to classify a given example; the forest then goes with the plurality of trees.

In standard classification or regression forests as proposed by \citet{breiman2001random}, the prediction for a
particular test point $x$ is determined by averaging predictions across an ensemble of different trees
\citep{amit1997shape,breiman1996bagging,dietterich2000experimental,ho1998random}.
Individual trees are grown by greedy recursive partitioning, i.e.,
we recursively add axis-aligned splits to the tree, where each split it chosen to maximize the
improvement to model fit \citep{breiman1984classification};
see Figure \ref{fig:tree} in the Appendix for an example of a tree. The trees are randomized using
bootstrap (or subsample) aggregation, whereby each tree is grown on a different random subset of the training data,
and random split selection that restricts the variables available at each step of the algorithm. For an introductory overview of random forests, we recommend the chapter of \citet{hastie2009elements} dedicated to the method.  As discussed below, in generalizing
random forests, we 
preserve several core elements of Breiman's forests---including recursive partitioning,
subsampling, and random split selection---but we abandon the idea that our final estimate
is obtained by averaging estimates from each member of an ensemble.
Treating forests as a type of
adaptive nearest neighbor estimator is much more amenable
to statistical extensions.

\subsection{Forest-Based Local Estimation}
\label{sec:ANN}

Suppose that we have $n$ independent and identically distributed samples, indexed
$ i = 1, \, ..., \, n$. For each sample, we have
access to an observable quantity $O_i$ that encodes information
relevant to estimating $\theta(\cdot)$, along with a set of auxiliary covariates $X_i$. In the case of non-parametric
regression, this observable just consists of an outcome $O_i = \cb{Y_i}$ with $Y_i \in \RR$; in
general, it may contain richer information. In the case of treatment effect
estimation with exogenous treatment assignment, $O_i = \cb{Y_i, \, W_i}$ also includes the treatment
assignment $W_i$.
Given this type of data, our goal is to estimate solutions to local estimation equations
of the form
%\begin{equation}
\smash{$\mathbb{E}[\psi_{\theta(x), \, \nu(x)}\p{O_i} \cond X_i = x] = 0$} for all $\in \xx$, 
%\end{equation}
where $\theta(x)$ is the parameter we care about and $\nu(x)$ is an optional
nuisance parameter.

One approach to estimating such functions $\theta(x)$ is to first define
some kind of similarity weights $\alpha_i(x)$ that measure the relevance of the $i$-th training example to
fitting $\theta(\cdot)$ at $x$, and then fit the target of interest via an empirical version of
the estimating equation
\citep{fan1998local,newey1994kernel,staniswalis1989kernel,stone1977consistent,tibshirani1987local}:
\begin{equation}
\label{eq:gee_estimate}
\p{\htheta(x), \, \hnu(x)} \in \argmin_{\theta, \, \nu} \cb{\Norm{\sum_{i = 1}^n \alpha_i(x) \, \psi_{\theta, \, \nu}\p{O_i}}_2}.
\end{equation}
When the above expression has a unique root, we can simply say that \smash{$(\htheta(x), \, \hnu(x))$}
solves \smash{$\sum_{i = 1}^n \alpha_i(x) \, \psi_{\htheta(x), \, \hnu(x)}\p{O_i} = 0$}.
The weights $\alpha_i(x)$ used to specify the above solution to the heterogeneous estimating
equation are traditionally obtained via a deterministic kernel function, perhaps with an adaptively chosen
bandwidth parameter \citep{hastie2009elements}. Although methods of the above kind often work well in low dimensions, they
are sensitive to the curse of dimensionality.

\begin{figure}
\begin{center}
\begin{tabular}{cccc}
\includegraphics[width = 0.25\textwidth, trim=15mm 40mm 20mm 40mm, clip = TRUE]{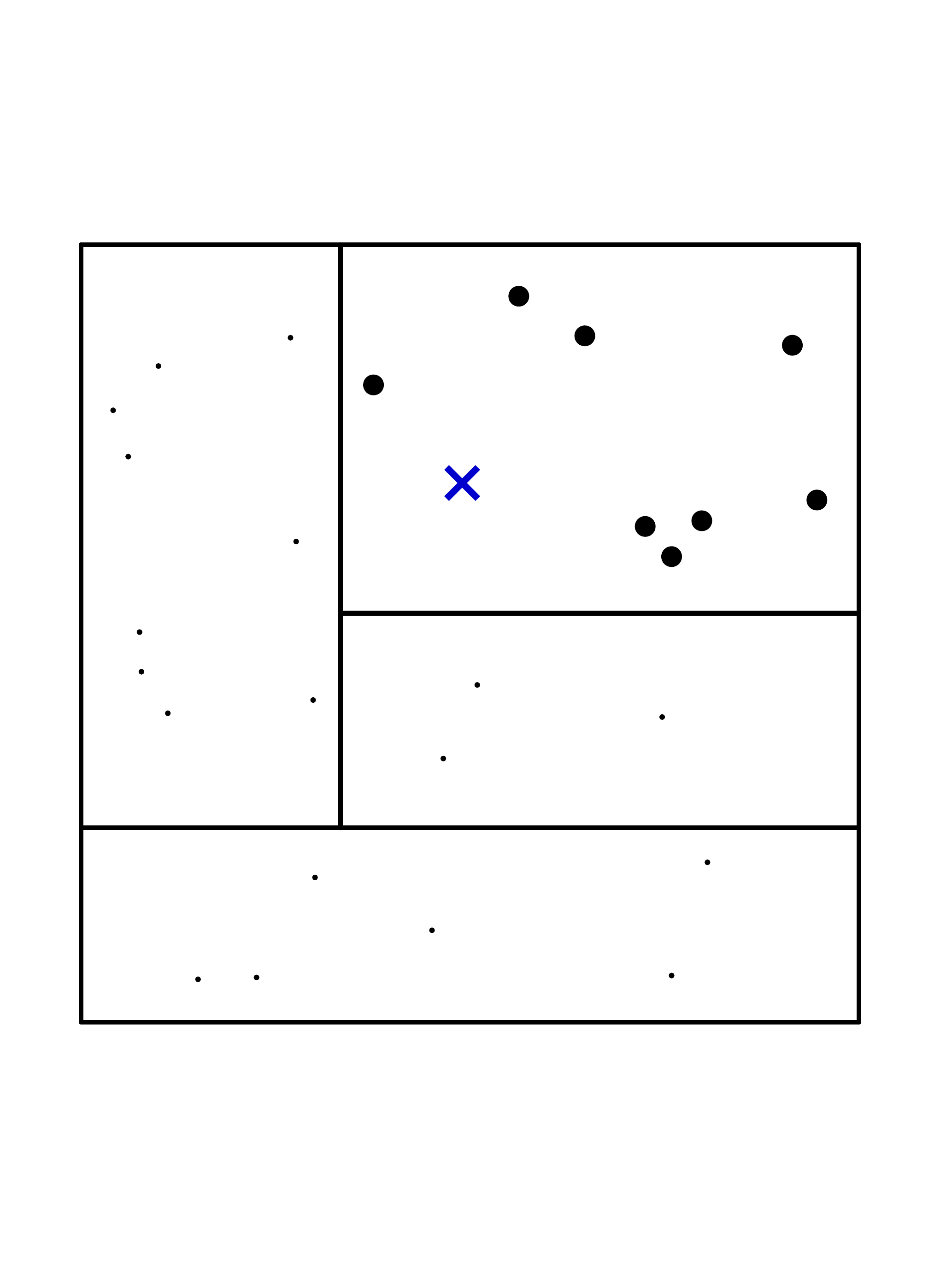} &
\includegraphics[width = 0.25\textwidth, trim=15mm 40mm 20mm 40mm, clip = TRUE]{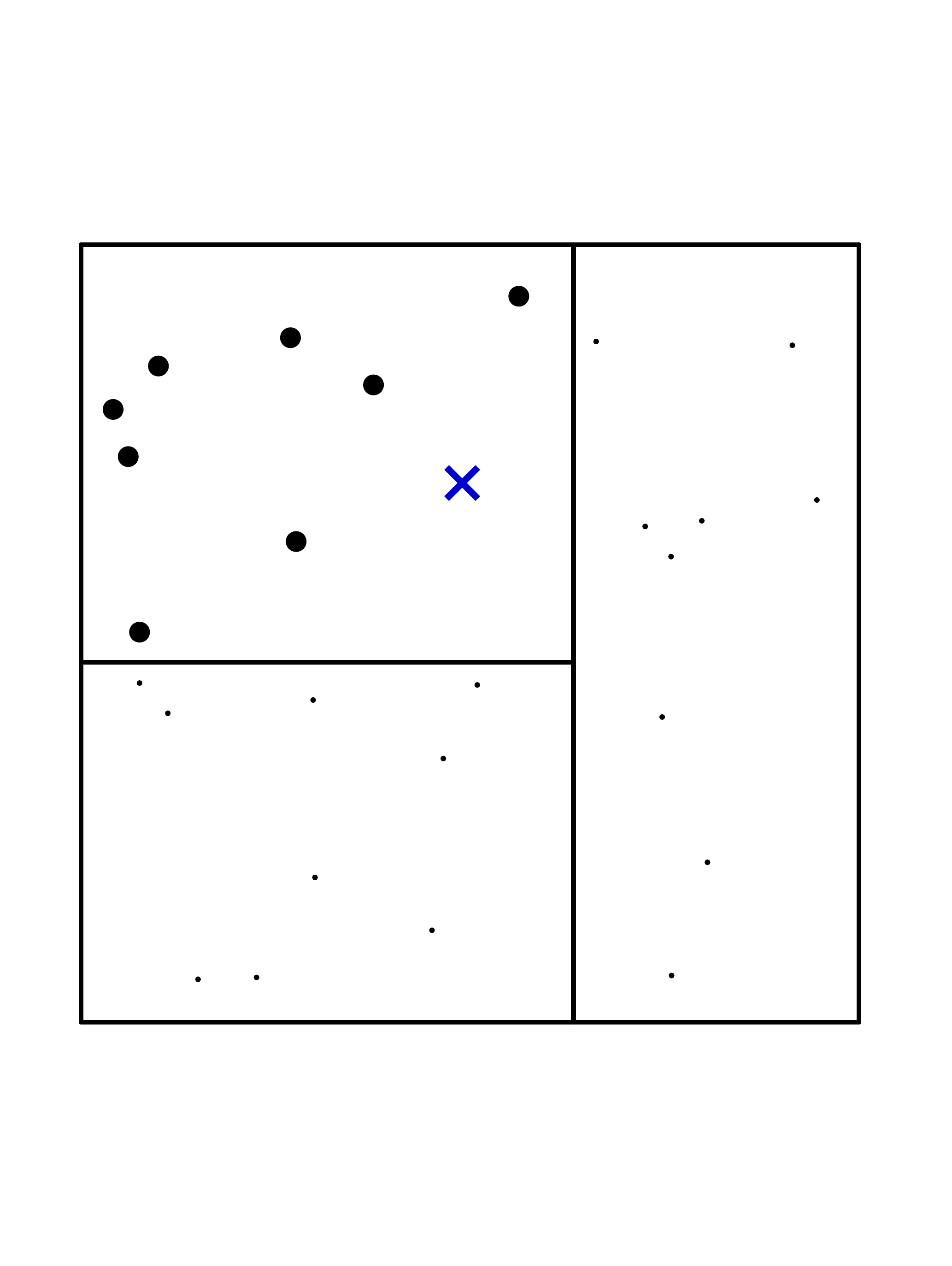} &
\includegraphics[width = 0.25\textwidth, trim=15mm 40mm 20mm 40mm, clip = TRUE]{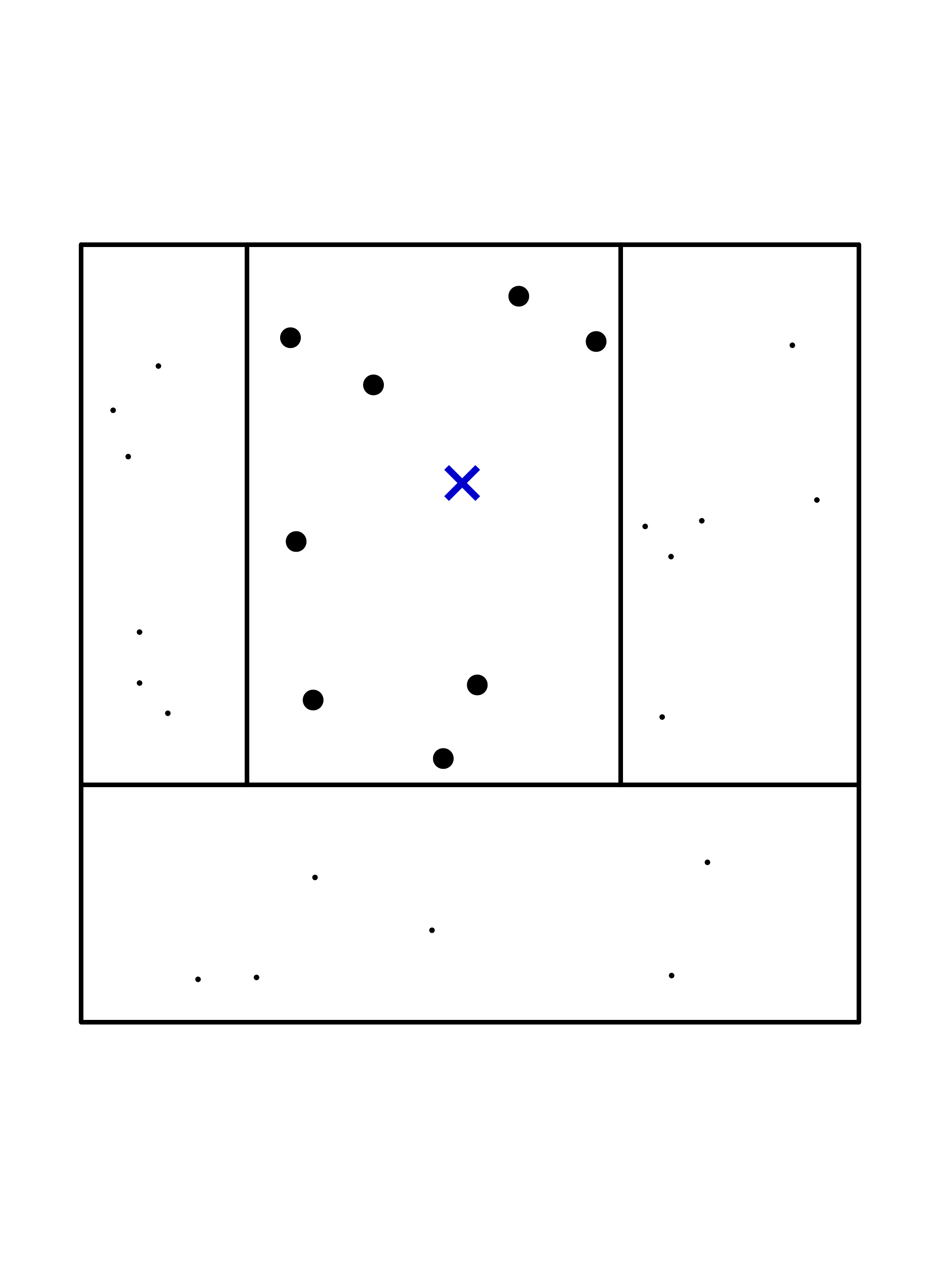} &
\parbox[b][0.25\textwidth][c]{0.1\textwidth}{$$\cdots$$}
\end{tabular}

\vspace{-\baselineskip}

\begin{tabular}{cc}
\parbox[b][0.25\textwidth][c]{0.12\textwidth}{$$\mathlarger{\mathlarger{\Longrightarrow}}$$} &
\includegraphics[width = 0.25\textwidth, trim=15mm 40mm 20mm 40mm, clip = TRUE]{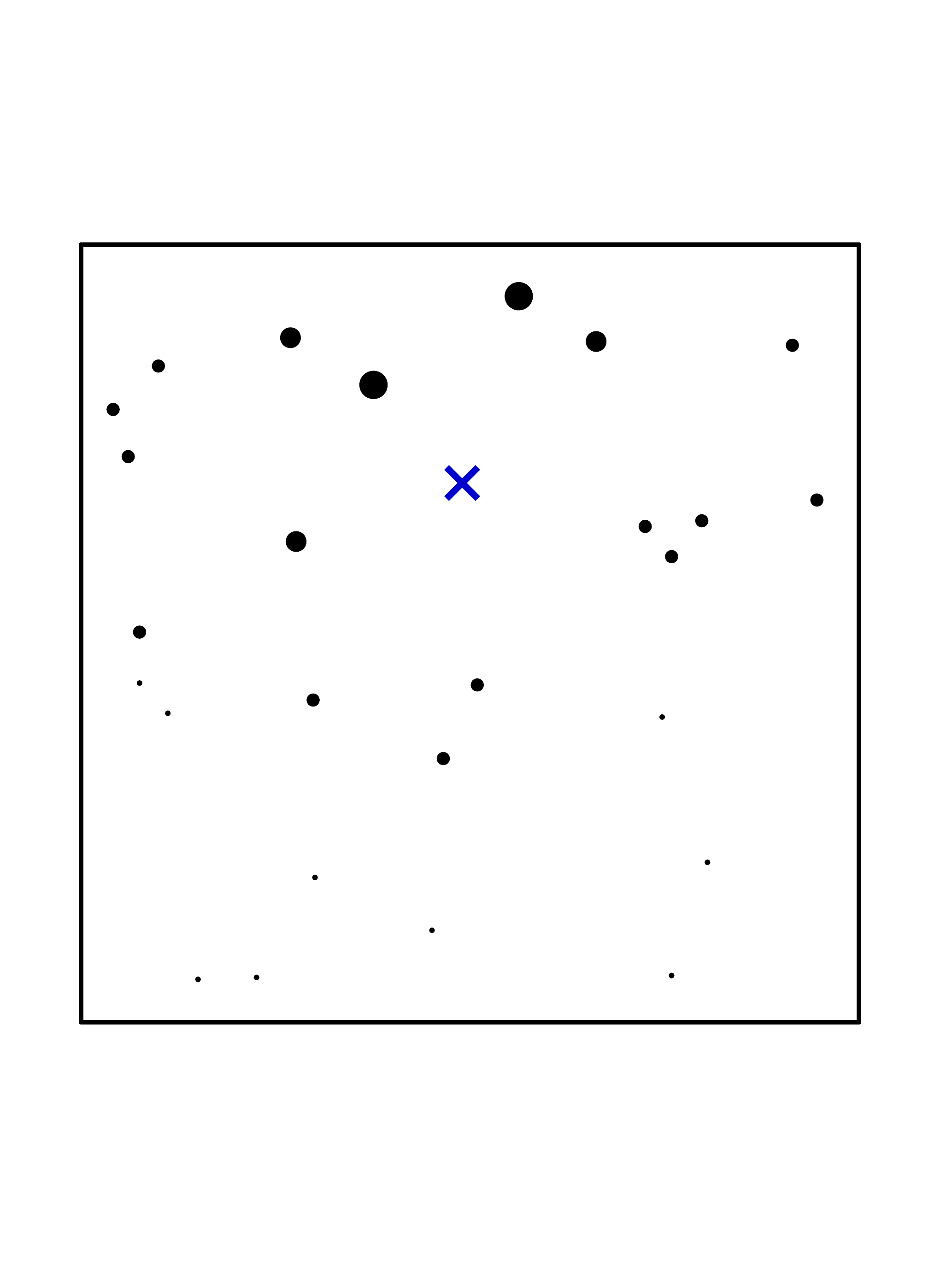}
\end{tabular}
\end{center}
\caption{Illustration of the random forest weighting function. The rectangles depticted above correspond
to terminal nodes in the dendogram representation of Figure \ref{fig:tree}. Each tree starts by giving equal
(positive) weight to the training examples in the same leaf as our test point $x$ of interest, and zero
weight to all the other training examples. Then, the forest averages all these tree-based weightings,
and effectively measures how often each training example falls into the same leaf as $x$.}
\label{fig:weighting}
\vspace{-1.5\baselineskip}
\end{figure}

Here, we seek to use forest-based algorithms to adaptively learn better, problem-specific,
weights $\alpha_i(x)$ that can be used in conjunction with \eqref{eq:gee_estimate}.
As in \citet{hothorn2004bagging} and \citet{meinshausen2006quantile}, our generalized random forests obtain such weights
by averaging neighborhoods implicitly produced by different trees. First, we grow a set of
$B$ trees indexed by $b = 1, \, ..., \, B$ and, for each such tree, define $L_b(x)$
as the set of training examples falling in the same ``leaf'' as $x$. The weights $\alpha_i(x)$ then
capture the frequency with which the $i$-th training example falls into the same leaf as $x$:
\begin{equation}
\label{eq:forest_weights}
\alpha_{bi}(x) = \frac{\1\p{\cb{X_i \in L_b(x)}}}{\abs{L_b(x)}}, \ \ \alpha_i(x) = \frac{1}{B} \sum_{b = 1}^B \alpha_{bi}(x).
\end{equation}
These weights sum to 1, and define the forest-based adaptive neighborhood of $x$; see Figure \ref{fig:weighting} for an illustration of this weighting function.

There are some subtleties in how the sets $L_b(x)$ are defined---in particular,
as discussed in Section \ref{sec:implementation}, our construction will rely on both
subsampling and a specific form of sample-splitting to achieve consistency---but at a high level
the estimates \smash{$\htheta(x)$} produced by a generalized random forests are simply
obtained by solving \eqref{eq:gee_estimate} with weights \eqref{eq:forest_weights}.

Finally, for the special case of regression trees, our weighting-based definition of a random forest is equivalent
to the standard ``average of trees'' perspective taken in \citet{breiman2001random}:
If we estimate the conditional mean function $\mu(x) = \EE{Y_i \cond X_i = x}$, as
identified in \eqref{eq:conditional_theta} using
$\psi_{\mu(x)}(Y_i) = Y_i - \mu(x)$, then we see that
%\begin{equation}
\smash{$\sum_{i = 1}^n \frac{1}{B} \sum_{b = 1}^B \alpha_{bi}(x) \p{Y_i - \hmu(x)} = 0$}
%\ \ \iff \ \
if and only if
\smash{$\hmu(x) = \frac{1}{B} \sum_{b = 1}^B \hmu_b(x)$},
%\end{equation}
where \smash{$\hmu_b(x) = \sum_{\cb{i : X_i \in L_b(x)}} Y_i \, \big/ \, \abs{L_b(x)}$} is the
prediction made by a single CART regression tree.

\subsection{Splitting to Maximize Heterogeneity}

We seek trees that, when combined into a forest, induce weights
$\alpha_i(x)$ that lead to good estimates of $\theta(x)$. The main difference between
random forests relative to other non-parametric regression techniques
is their use of recursive partitioning on subsamples to generate these weights $\alpha_i(x)$.
Motivated by the empirical success of regression forests across several application areas, our approach
mimics the algorithm of \citet{breiman2001random} as closely as possible, while tailoring
our splitting scheme to focus on heterogeneity in the target functional $\theta(x)$.

Just like in Breiman's forests, our search for good splits proceeds greedily, i.e., we seek splits that
immediately improve the quality of the tree fit as much as possible.
Every split starts with a parent node $P \subseteq \xx$; given a sample of data $\mathcal{J}$, we define
\smash{$(\htheta_P, \, \hnu_P)(\mathcal{J})$} as the solution to the estimating equation, as follows
(we suppress dependence on $\mathcal{J}$ when unambiguous):
\begin{equation}
\label{eq:leaf_solve}
\p{\htheta_P, \, \hnu_P}(\mathcal{J}) \in \argmin_{\theta, \, \nu} \cb{\Norm{\sum_{\cb{i\in\mathcal{J} : X_i \in P}} \, \psi_{\theta, \, \nu}\p{O_i}}_2}.
\end{equation}
We would like to divide $P$ into two children $C_1, \, C_2 \subseteq \xx$ using an axis-aligned
cut such as to improve the accuracy of our $\theta$-estimates as much as possible;
formally, we seek to minimize \smash{$\err\p{C_1, \, C_2}$} defined as
%\begin{equation}
%\label{eq:err_criterion}
 \smash{$\err\p{C_1, \, C_2} = \sum_{j = 1, \, 2}
 \mathbb{P}[X \in C_j \cond X \in P] \, \mathbb{E}[(\htheta_{C_j}(\mathcal{J}) - \theta(X))^2 \cond X \in C_j]$},
%\end{equation}
where \smash{$\htheta_{C_j}(\mathcal{J})$} are fit over children $C_j$ in analogy to \eqref{eq:leaf_solve},
and expectations are taken over both the randomness in \smash{$\htheta_{C_j}(\mathcal{J})$} and a
new test point $X$.

Many standard regression tree implementations, such as CART \citep{breiman1984classification},
choose their splits by simply minimizing the in-sample prediction error of the node, which corresponds to 
\smash{$\err\p{C_1, \, C_2}$} with plug-in estimators from the training sample. 
In the case of estimating the effect of a binary treatment,
\citet{athey2016recursive} study sample-splitting trees, and
propose an unbiased, model-free estimate of  \smash{$\err\p{C_1, \, C_2}$} using
an overfitting penalty in the spirit of \citet{mallows1973some}.
In our setting, however, this kind of direct loss minimization is not an option: If $\theta(x)$ is only
identified through a moment condition, then we do not in general have access to unbiased,
model-free estimates of the criterion  \smash{$\err\p{C_1, \, C_2}$}. To address this issue, we
rely on the following more abstract characterization of our target criterion. 

\begin{prop}
\label{prop:motivation}
Suppose that basic assumptions detailed in Section \ref{sec:theory} hold,
and that the parent node $P$ has a radius smaller than
$r$ for some value $r > 0$. We write $n_P = \abs{\cb{i\in\mathcal{J} : X_i \in P}}$ for the number of
observations in the parent and $n_{C_j}$ for the number of observations in each child, and define
\begin{equation}
\label{eq:delta_criterion}
\Delta(C_1, \, C_2) := {n_{C_1} n_{C_2}}\,/\,{n_P^2}  \p{\htheta_{C_1}(\mathcal{J}) - \htheta_{C_2}(\mathcal{J})}^2,
\end{equation}
where \smash{$\htheta_{C_1}$} and \smash{$\htheta_{C_2}$} are solutions to the
estimating equation computed in the children, following \eqref{eq:leaf_solve}.
Then, treating the child nodes $C_1$ and $C_2$ as well as the corresponding counts $n_{C_1}$ and $n_{C_2}$
as fixed, and assuming that $n_{C_1}, \, n_{C_2} \gg r^{-2}$, we have
%\begin{align}
%\label{eq:motivation}
\smash{$\err\p{C_1, \, C_2} = K(P) - \EE{\Delta(C_1, \, C_2)} + o\p{r^2}$}
%\end{align}
where $K(P)$ is a deterministic term that measures the purity of the parent node
that does not depend on how the parent is split, and the $o$-term incorporates terms that depend on
sampling variance.
\end{prop}

Motivated by this observation, we consider splits that make the above $\Delta$-criterion
\eqref{eq:delta_criterion} large. A special case of the above idea also underlies
the splitting rule for treatment effect estimation proposed by \citet{athey2016recursive}.
At a high level, we can think of this $\Delta$-criterion as favoring splits that
increase the heterogeneity of the in-sample $\theta$-estimates as fast as possible.
The dominant bias term in \smash{$\err\p{C_1, \, C_2}$} is due to the
sampling variance of regression trees, and is the same term that appears in the analysis
of \citet{athey2016recursive}. Including this error term in the splitting criterion may
stabilize the construction of the tree, and further it can prevent the splitting criterion
from favoring splits that make the model difficult to estimate.

\subsection{The Gradient Tree Algorithm}

The above discussion provides conceptual guidance on how to pick
good splits. But actually optimizing the
criterion $\Delta(C_1, \, C_2)$ over all possible axis-aligned splits
while explicitly solving for $\htheta_{C_1}$ and $\htheta_{C_2}$ in each candidate child
using an analogue to \eqref{eq:leaf_solve} may be quite expensive computationally.
To avoid this issue, we instead optimize an approximate criterion \smash{$\tDelta(C_1, \, C_2)$}
built using gradient-based approximations for $\htheta_{C_1}$ and $\htheta_{C_2}$.
For each child $C$, we use $\ttheta_C \approx \htheta_C$ as follows:
We first compute $A_P$ as any consistent estimate for the gradient of the expectation of the $\psi$-function, i.e.,
$\nabla \mathbb{E}[\psi_{\htheta_P, \, \hnu_P}(O_i) \cond X_i \in P]$, and then set
\begin{align}
\label{eq:ttheta}
&\ttheta_C = \htheta_P - \frac{1}{\abs{\cb{i : X_i \in C}}} \sum_{\cb{i : X_i \in C}} \xi^\top A^{-1}_P \ \psi_{\htheta_P, \, \hnu_P}\p{O_i},
\end{align}
where $\htheta_P$ and $\hnu_P$ are obtained by solving \eqref{eq:leaf_solve} once in the parent node,
and $\xi$ is a vector that picks out the $\theta$-coordinate from the $(\theta, \, \nu)$ vector.
When the $\psi$-function itself is continuously differentiable, we use
\begin{equation}
\label{eq:AP}
A_P= \frac{1}{\abs{\cb{i : X_i \in P}}} \sum_{\cb{i : X_i \in P}} \nabla \psi_{\htheta_P, \, \hnu_P}\p{O_i},
\end{equation}
and the quantity \smash{$\xi^\top A^{-1}_P \ \psi_{\htheta_P, \, \hnu_p}\p{O_i}$} corresponds
to the influence function of the $i$-th observation for computing \smash{$\htheta_P$} in the parent.
Cases where $\psi$ is non-differentiable, e.g., with quantile regression, require more care.

Algorithmically, our recursive partitioning scheme now reduces to alternatively
applying the following two steps. First, in a {\bf labeling step}, we compute
$\htheta_P$, $\hnu_P$, and the derivative matrix $A_P^{-1}$ on the parent
data as in \eqref{eq:leaf_solve}, and use them to get pseudo-outcomes
\begin{equation}
\label{eq:relabel}
\rho_i = -\xi^\top A_P^{-1} \ \psi_{\htheta_P, \, \hnu_P}\p{O_i} \in \RR.
\end{equation}
Next, in a {\bf regression step}, we run a standard CART regression split on the pseudo-outcomes
$\rho_i$. Specifically, we split $P$ into two axis-aligned children $C_1$ and $C_2$
such as to maximize the criterion
\begin{equation}
\label{eq:gradient_criterion}
\tDelta(C_1, \, C_2) =  \sum_{j = 1}^2\frac{1}{\abs{\cb{i : X_i \in C_j}}}\p{ \sum_{\cb{i : X_i \in C_j}} \rho_i}^2.
\end{equation}
Once we have executed the regression step, we relabel observations in each child by solving the
estimating equation, and continue on recursively.

For intuition, it is helpful to examine
the simplest case of least-squares regression, i.e., with $\psi_{\theta(x)}(Y) = Y - \theta(x)$.
Here, the labeling step \eqref{eq:relabel} doesn't change anything---we get
$\rho_i = Y_i - \bY_p$, where $\bY_p$ is the mean outcome in the parent---while
the second step maximizing \eqref{eq:gradient_criterion} corresponds to the usual way
of making splits as in \citet{breiman2001random}.
Thus, the special structure of the type of problem we are trying to solve is
encoded in \eqref{eq:relabel}, while the second scanning step is a universal step shared
across all different types of forests.

We expect this approach to provide more consistent computational
performance than optimizing \eqref{eq:delta_criterion} at each split directly.
When growing a tree, the computation is typically dominated by the split-selection step,
and so it is critical for this step to be implemented as efficiently as possible (conversely, the 
labeling step \eqref{eq:relabel} is only solved once per node, and so is less performance sensitive).
From this perspective, using a regression splitting criterion as in \eqref{eq:gradient_criterion} is very
desirable, as it is possible to evaluate all possible split points along a given feature with only
a single pass over the data in the parent node (by representing the criterion in terms of cumulative sums).
In contrast, directly optimizing the original criterion \eqref{eq:delta_criterion} may require
solving intricate optimization problems for each possible candidate split.

This type of gradient-based approximation also underlies other popular  statistical algorithms,
including gradient boosting \citep{friedman2001greedy} and the model-based recursive
partitioning algorithm of \citet{zeileis2008model}.
Conceptually, tree splitting bears some connection to change-point detection if we imagine
tree splits as occurring at detected change-points in $\theta(x)$; and, from this perspective,
our approach is closely related to standard techniques for moment-based change-point detection
\citep{andrews1993tests,hansen1992testing,hjort2002tests,nyblom1989testing,
ploberger1992cusum,zeileis2005unified,zeileis2007generalized}.

In our context, we can verify that the error from using the approximate criterion \eqref{eq:gradient_criterion} instead of
the exact $\Delta$-criterion \eqref{eq:delta_criterion} is within the tolerance used
to motivate the $\Delta$-criterion in Proposition \ref{prop:motivation}, thus suggesting
that our use of \eqref{eq:ttheta} to guide splitting may not result in too much inefficiency.
Note that consistent estimates of $A_P$ can in general be derived directly via, e.g., \eqref{eq:AP},
without relying on Proposition \ref{prop:approximation}.

\begin{prop}
\label{prop:approximation}
Under the conditions of Proposition \ref{prop:motivation}, if 
$|A_P - \nabla \mathbb{E}[\psi_{\htheta_P, \, \hnu_P}(O_i) \cond X_i \in P]| \rightarrow_P 0$,
i.e., $A_P$ is consistent,
then \smash{$\Delta(C_1, \, C_2)$} and \smash{$\tDelta(C_1, \, C_2)$} are approximately equivalent, in that
\smash{$\tDelta(C_1, \, C_2) = \Delta(C_1, \, C_2) + o_P(\max\cb{r^2, \, {1}\,/\,{n_{C_1}}, \, {1}\,/\,{n_{C_2}}})$}.
\end{prop}

\newcommand{\pluseq}{\mathrel{+}=}

\begin{algorithm}[t]
\caption{Generalized random forest with honesty and subsampling}\label{alg:forest}
\vspace{-0.4\baselineskip}
\justify
All tuning parameters are pre-specified, including the number of trees $B$ and the sub-sampling $s$ rate used in
\textsc{Subsample}. This function is implemented in the package \texttt{grf}
for \texttt{R} and \texttt{C++}.
\begin{algorithmic}[1]
\Procedure{GeneralizedRandomForest}{set of examples $\set$, test point $x$}
\State weight vector $\alpha \gets$ \Call{Zeros}{$\abs{\set}$}
\For{$b = 1$ to total number of trees $B$}
\State set of examples $\ii \gets$ \Call{Subsample}{$\set$, $s$}
\State sets of examples $\jj_1, \, \jj_2 \gets$ \Call{SplitSample}{$\ii$}
\State tree $\tset \gets$ \Call{GradientTree}{$\jj_1$, $\xx$}
\Comment{\parbox[t]{.44\linewidth}{See Algorithm \ref{alg:tree}.}}
\State $\nn \gets $\Call{Neighbors}{$x, \, \tset, \, \jj_2$} 
\Comment{\parbox[t]{.44\linewidth}{Returns those elements of $\jj_2$ that fall into the same leaf as $x$ in the tree $\tset$.}}
\ForAll{example $e \in \nn$}
\State $\alpha[{e}] \pluseq 1/\abs{\nn}$
\EndFor
\EndFor
\State {\bf output} $\htheta(x)$, the solution to \eqref{eq:gee_estimate} with weights $\alpha/B$
\EndProcedure
\end{algorithmic}
\justify
The function \textsc{Zeros} creates a vector of zeros of length $\abs{\set}$;
\textsc{Subsample} draws a subsample of size $s$ from $\set$ without replacement; and
\textsc{SplitSample} randomly divides a set into two evenly-sized, non-overlapping halves.
The step \eqref{eq:gee_estimate} can be solved using any numerical estimator.
Our implementation \texttt{grf} provides an explicit plug-in point where a user can write
a solver for \eqref{eq:gee_estimate} appropriate for their $\psi$-function.
$\xx$ is the domain of the $X_i$.
In our analysis, we consider a restricted class of generalized random forests satisfying
Specification \ref{spec:forest}.
\end{algorithm}

\newcommand{\qset}{\mathcal{Q}}

\begin{algorithm}[t]
\caption{Gradient tree}\label{alg:tree}
\vspace{-0.4\baselineskip}
\justify
Gradient trees are grown as subroutines of a generalized random forest.
\begin{algorithmic}[1]
\Procedure{GradientTree}{set of examples $\jj$, domain $\xx$}
\State node $P_0 \gets$ \Call{CreateNode}{$\jj$, $\xx$}
\State queue $\qset \gets$ \Call{InitializeQueue}{$P_0$}
\While{\Call{NotNull}{node $P \gets$ \textsc{Pop}($\qset$)}}
%\State set of samples $\jj_P \gets$ \Call{GetExamplesInNode}{$\jj$, $P$}
\State $(\htheta_P, \, \hnu_P, \, A_P) \gets$ \Call{SolveEstimatingEquation}{$P$}
\Comment{\parbox[t]{.28\linewidth}{Computes \eqref{eq:leaf_solve} and \eqref{eq:AP}.}}
\State vector $R_P \gets$ \Call{GetPseudoOutcomes}{$\htheta_P, \, \hnu_P, \, A_P$}
\Comment{\parbox[t]{.28\linewidth}{Applies \eqref{eq:relabel} over $P$.}}
\State split $\Sigma \gets$ \Call{MakeCartSplit}{$P$, $R_P$}
\Comment{\parbox[t]{.28\linewidth}{Optimizes \eqref{eq:gradient_criterion}.}}
\If{\Call{SplitSucceeded}{$\Sigma$}}
\State \Call{SetChildren}{$P$, \textsc{GetLeftChild}($\Sigma$), \textsc{GetRightChild}($\Sigma$)}
\State \Call{AddToQueue}{$\qset$, \textsc{GetLeftChild}($\Sigma$)}
\State \Call{AddToQueue}{$\qset$, \textsc{GetRightChild}($\Sigma$)}
\EndIf
\EndWhile
\State {\bf output} tree with root node $P_0$
\EndProcedure
\end{algorithmic}
\justify
The function call \textsc{InitializeQueue} initializes a queue with a single element;
\textsc{Pop} returns and removes the oldest element of a queue $\qset$,
unless $\qset$ is empty in which case it returns null. \textsc{MakeCartSplit} runs a CART split
on the pseudo-outcomes, and either returns two child nodes or a failure message that no legal
split is possible.
\end{algorithm}

\subsection{Building a Forest with Theoretical Guarantees}
\label{sec:implementation}

Now, given a practical splitting scheme for growing individual trees, we want to
grow a forest that allows for consistent estimation of $\theta(x)$ using \eqref{eq:gee_estimate}
paired with the forest weights \eqref{eq:forest_weights}. We expect each
tree to provide small, relevant neighborhoods for $x$ that give us noisy estimates of $\theta(x)$;
then, we may hope that
forest-based aggregation will provide a single larger but still relevant neighborhood for $x$
that yields stable estimates $\htheta(x)$.

To ensure good statistical behavior,
we rely on two conceptual ideas that have proven to be successful in the literature
on forest-based least-squares regression: Training trees on subsamples of the training data
\citep{mentch2016quantifying,scornet2015consistency,wager2015estimation}, and a sub-sample
splitting technique that we call honesty \citep{biau2012analysis,denil2014narrowing,wager2015estimation}.
Our final algorithm for forest-based solutions to heterogeneous estimating equations is given as
Algorithm \ref{alg:forest}; we refer to Section 2.4 of \citet{wager2015estimation} for a more in-depth discussion
of honesty in the context of forests.
As shown in Section \ref{sec:theory}, 
assuming regularity conditions, the estimates $\htheta(x)$ obtained using
a generalized random forest as described in Algorithm \ref{alg:forest} are consistent for $\theta(x)$.
Moreover, given appropriate subsampling rates, we establish asymptotic normality of the resulting forest estimates $\htheta(x)$.

\section{Asymptotic Analysis}
\label{sec:theory}

%\subsection{Theoretical Setup}

We now turn to a formal characterization of generalized random forests, with the aim of
establishing asymptotic Gaussianity of the \smash{$\htheta(x)$}, and of providing
tools for statistical inference about $\theta(x)$.
We first list assumptions underlying our theoretical results.
Throughout, the covariate space and the parameter space are both subsets of Euclidean space;
specifically, $\xx = [0, \, 1]^p$ and $(\theta, \, \nu) \in \bb \subset \RR^k$
for some $p, \, k > 0$, where $\bb$ is a compact subset of $\RR^k$.
Moreover, we assume that the features $X$ have a density that is bounded away from 0 and $\infty$; as
argued in, e.g., \citet{wager2015uniform}, this is equivalent to imposing a weak dependence condition on the
individual features $(X_i)_j$ because trees and forests are invariant to monotone rescaling of the features.
All proofs are in the appendix.

Some practically interesting cases, such as quantile regression, involve discontinuous score functions $\psi$,
which makes the analysis more intricate. Here, we follow standard practice, and
assume that the \emph{expected} score function,
\begin{equation}
\label{eq:M}
M_{\theta, \, \nu}(x) := \EE{\psi_{\theta, \, \nu}(O) \cond X = x},
\end{equation}
varies smoothly in the parameters, even though $\psi$ itself may be
discontinuous. For example, with quantile regression
\smash{$\psi_\theta(Y) = 1\p{\cb{Y > \theta}} - (1 - q)$} is discontinuous in $q$,
but \smash{$M_\theta(x) = \PP{Y > \theta \cond X = x} - (1 - q)$} will be smooth
whenever $Y\cond X = x$ has a smooth density.

\begin{assu}[Lipschitz $x$-signal]
\label{assu:lip}
For fixed values of $(\theta, \, \nu)$, we assume that $M_{\theta, \, \nu}(x)$ 
as defined in \eqref{eq:M} is Lipschitz continuous in $x$.
\end{assu}

\begin{assu}[Smooth identification]
\label{assu:identification}
When $x$ is fixed, we assume that the $M$-function is twice continuously differentiable
in $(\theta, \, \nu)$ with a uniformly bounded second derivative, and
that $V(x): = V_{\theta(x), \, \, \nu(x)}(x)$ is invertible for all $x \in \xx$, with
%\begin{equation}
%\label{eq:V}
\smash{$V_{\theta,\,\nu}(x) := {\partial}/{\partial (\theta, \, \nu)} \, M_{\theta, \, \nu}(x) \cond_{\theta(x), \ \nu(x)}$}.
%\end{equation}
\end{assu}

Our next two assumptions control regularity properties of the $\psi$-function itself.
Assumption \ref{assu:covariance} holds trivially when $\psi$ itself is
Lipschitz in $(\theta, \, \nu)$ (in fact, having $\psi$ be 0.5-H{\"o}lder would be enough),
while Assumption \ref{assu:donsker} is used to show that a certain empirical process is
Donsker. Examples are given at the end of this section.

\begin{assu}[Lipschitz $(\theta, \, \nu)$-variogram]
\label{assu:covariance}
The score functions $\psi_{\theta, \, \nu}(O_i)$ 
have a continuous covariance structure.
Writing $\gamma$ for the worst-case variogram
and $\Norm{\cdot}_F$ for the Frobenius norm,
then for some $L > 0$,
\begin{equation}
\label{eq:variogram}
\begin{split}
&\gamma\p{ \begin{pmatrix} \theta \\ \nu \end{pmatrix}, \,  \begin{pmatrix} \theta' \\ \nu' \end{pmatrix}} \leq
L \Norm{\begin{pmatrix} \theta \\ \nu \end{pmatrix} -   \begin{pmatrix} \theta' \\ \nu' \end{pmatrix}}_2
\text{ for all } (\theta,\, \nu), \ (\theta',\,\nu'), \\
&\gamma\p{ \begin{pmatrix} \theta \\ \nu \end{pmatrix}, \,  \begin{pmatrix} \theta' \\ \nu' \end{pmatrix}}
:= \sup_{x \in \xx} \cb{\Norm{\Var{\psi_{\theta, \, \nu}\p{O_i} - \psi_{\theta', \, \nu'}\p{O_i} \cond X_i = x}}_F}.
\end{split}
\end{equation}
\end{assu}

\begin{assu}[Regularity of $\psi$]
\label{assu:donsker}
\sloppy{
The $\psi$-functions can be written as
\smash{$ \psi_{\theta, \, \nu}(O) = \lambda\p{\theta, \, \nu; \, O_i} + \zeta_{\theta, \, \nu}\p{g(O_i)} $},
such that $\lambda$ is Lipschitz-continuous in $(\theta, \, \nu)$,
$g : \cb{O_i} \rightarrow \RR$ is a univariate summary of $O_i$,
and $\zeta_{\theta, \, \nu} : \RR \rightarrow \RR$ is any family of
monotone and bounded functions.
}
\end{assu}

\begin{assu}[Existence of solutions]
\label{assu:existence}
We assume that, for any weights $\alpha_i$ with $\sum \alpha_i = 1$,
the estimating equation \eqref{eq:gee_estimate} returns a
minimizer \smash{$(\htheta, \, \hnu)$} that at least approximately
solves the estimating equation:
%\begin{equation}
%\label{eq:approx_minimizer}
\smash{$\lVert\sum_{i = 1}^n \alpha_i \, \psi_{\htheta, \, \hnu}\p{O_i}\rVert_2 \leq C\max\cb{\alpha_i}$},
for some constant $C \geq 0$.
%\end{equation}
\end{assu}

All the previous assumptions only deal with local properties of the estimating
equation, and can be used to control the behavior of \smash{$(\htheta(x), \, \hnu(x))$}
in a small neighborhood of the population parameter value $(\theta(x), \, \nu(x))$.
Now, to make any use of these assumptions, we first need to verify that 
\smash{$(\htheta(x), \, \hnu(x))$} be consistent. Here, we use the following assumption
to guarantee consistency; this setup is general enough to cover both instrumental
variables regression and quantile regression.

\begin{assu}[Convexity]
\label{assu:convexity}
The score function $\psi_{\theta, \, \nu}(O_i)$ is a negative sub-gradient of a convex function,
and the expected score $M_{\theta, \, \nu}(X_i)$ is the negative gradient of a strongly convex function.
\end{assu}

Finally, our consistency and Gaussianty results require using some specific settings
for the trees from Algorithm \ref{alg:forest}.
In particular, we require that all trees be honest and regular in the sense of \citet{wager2015estimation}, as follows.
In order to satisfy the minimum split probability condition below,
our implementation relies on the device of \citet{denil2014narrowing}, whereby the number splitting variables
considered at each step of the algorithm is random; specifically, we try
$\min\cb{\max\cb{\text{Poisson$(m)$}, \, 1}, \, p}$ variables at each step,
where $m > 0$ is a tuning parameter.

\begin{spec}
\label{spec:forest}
All trees are symmetric, in that their output is invariant to permuting
the indices of training examples;
make balanced splits, in the sense that every split puts at least a fraction
$\omega$ of the observations in the parent node into each child, for some $\omega > 0$; and
are randomized in such a way that, at every split, the probability that
the tree splits on the $j$-th feature is bounded from below by some $\pi > 0$.
The forest is honest and built via subsampling
with subsample size $s$ satisfying $s/n \rightarrow 0$ and $s \rightarrow \infty$,
as described in Section \ref{sec:implementation}.
\end{spec}

For generality, we set up Assumptions \ref{assu:lip}--\ref{assu:convexity}
in an abstract way. We end this section by showing that, in the context of our
main problems of interest requiring Assumptions \ref{assu:lip}--\ref{assu:convexity} is
not particularly stringent. Further examples that satisfy the above assumptions will
be discussed in Sections \ref{sec:cape} and \ref{sec:iv}.

\begin{exam}[Least squares regression]
\label{exam:regression}
In the case of least-squares regression, i.e., $\psi_{\theta}(Y_i) = Y_i - \theta$,
Assumptions \ref{assu:identification}--\ref{assu:convexity} hold immediately from the
definition of $\psi$. In particular, $V = 1$ in Assumption \ref{assu:identification},
$\gamma(\theta, \, \theta') = 0$ in Assumption \ref{assu:covariance},
$\psi$ itself is Lipschitz for Assumption \ref{assu:donsker}, and 
\smash{$\psi_{\theta}(y) = - \frac{d}{d\theta} (y - \theta)^2/2$} for Assumption \ref{assu:convexity}.
Meanwhile, Assumption \ref{assu:lip} simply means that
the conditional mean function $\EE{Y_i \cond X_i = x}$ must be Lipschitz in $x$;
this is a standard assumption in the literature on regression forests.
\end{exam}

\begin{exam}[Quantile regression]
For quantile regression, we have
\smash{$\psi_{\theta}(Y_i) = q - \1\p{\cb{Y_i \leq \theta}}$}
and \smash{$M_{\theta}(x) = q - F_{x}(\theta)$},
where $F_x(\cdot)$ denotes the cumulative distribution function of $Y_i$ given $X_i = x$.
Assumption \ref{assu:lip} is equivalent to assuming that the
conditional exceedance probabilities $\PP{Y_i > y \cond X_i = x}$
be Lipschitz-continuous in $x$ for all $y \in \RR$, while Assumption \ref{assu:identification}
holds if the conditional density $f_x(y)$ has a continuous uniformly bounded first derivative,
and is bounded away from 0 at the quantile of interest $y = F^{-1}_x(q)$.
Assumption \ref{assu:covariance} holds if $f_x(y)$ is uniformly bounded from above
(specifically, $\gamma(\theta, \, \theta') \leq \max_x \cb{f_x(y)} \abs{\theta - \theta'}$),
Assumption \ref{assu:donsker} holds because $\psi$ is monotone and $O_i = Y_i$ is univariate,
Assumption \ref{assu:existence} is immediate, and Assumption \ref{assu:convexity} holds
because $ - \frac{d}{d\theta} M_\theta(x) = f_x(\theta) > 0$ and \smash{$\psi_{\theta}(Y_i)$}
is the negative sub-gradient of a V-shaped function with elbow at $Y_i$.
\end{exam}

\subsection{A Central Limit Theorem for Generalized Random Forests}
\label{sec:gauss}

Given these assumptions, we are now ready to provide an asymptotic characterization
of generalzed random forests. In doing so, we note that existing asymptotic analyses of regression forests,
including \citet{mentch2016quantifying}, \citet{scornet2015consistency} and \citet{wager2015estimation},
were built around the fact that regression forests are averages of regression
trees grown over sub-samples, and can thus be analyzed as $U$-statistics \citep{hoeffding1948class}.
Unlike regression forest predictions, however, the parameter estimates \smash{$\htheta(x)$}
from generalized random forests are not averages of estimates made by different trees;
instead, we obtain \smash{$\htheta(x)$} by solving a single weighted moment equation
as in \eqref{eq:gee_estimate}. Thus, existing proof strategies do not apply in our setting.

We tackle this problem using the method of influence functions as described by \citet{hampel1974influence};
in particular, we are motivated by the analysis of \citet{newey1994kernel}. The core idea of these methods
is to first derive a sharp, linearized approximation to the local estimator \smash{$\htheta(x)$}, and
then to analyze the linear approximation instead.
In our setup, the influence function heuristic motivates a natural approximation \smash{$\ttheta^*(x)$}
to \smash{$\htheta(x)$} as follows. Let $\rho_i^*(x)$ denote the influence function of the
$i$-th observation with respect to the true parameter value $\theta(x)$,
%\begin{equation}
%\label{eq:opt_pseudo}
\smash{$\rho_i^*(x) :=  -\xi^\top V(x)^{-1} \psi_{\theta(x), \, \nu(x)}(O_i)$}.
%\end{equation}
These quantities are closely related to the pseudo-outcomes \eqref{eq:relabel} used in our gradient tree
splitting rule; the main difference is that, here, the $\rho_i^*(x)$ depend on the unknown true
parameter values at $x$ and are thus inaccessible in practice. We use the $*$-superscript to remind ourselves
of this fact.

Then, given any set of forest weights $\alpha_i(x)$ used to define the generalized random forest
estimate \smash{$\htheta(x)$} by solving \eqref{eq:gee_estimate}, we can also define a
pseudo-forest
\begin{equation}
\label{eq:pseudo_forest}
\ttheta^*(x) := \theta(x) + \sum_{i = 1}^n \alpha_i(x) \rho_i^*(x),
\end{equation}
which we will use as an approximation for  \smash{$\htheta(x)$}. We note that,
formally, this pseudo-forest estimate \smash{$\ttheta^*(x)$} is equivalent to the output of an (infeasible)
regression forest with weights $\alpha_i(x)$ and outcomes $\theta(x) + \rho_i^*(x)$.

The upshot of this approximation is that, unlike \smash{$\htheta(x)$}, the pseudo-forest
\smash{$\ttheta^*(x)$} is a $U$-statistic.
Because $\ttheta^*(x)$ is a linear function of the pseudo-outcomes $\rho_i^*(x)$,
we can write it as an average of pseudo-tree predictions
%\begin{equation}
\smash{$\ttheta^*(x) = \frac{1}{B} \sum_{b= 1}^B \ttheta^*_b(x)$} with
\smash{$\ttheta^*_b(x) = \sum_{i = 1}^n \alpha_{ib}(x) \p{\theta(x) + \rho_i^*(x)}$}.
%\end{equation}
Then, because each individual pseudo-tree prediction \smash{$\ttheta_b^*(x)$} is trained on a size-$s$ subsample
of the training data drawn without replacement (see Section \ref{sec:implementation}), 
\smash{$\ttheta^*(x)$} is an infinite-order $U$-statistic whose order corresponds to the subsample size,
and so the arguments of \citet{mentch2016quantifying} or \citet{wager2015estimation} can be used to study the averaged estimator
\smash{$\ttheta^*(x)$} using results about $U$-statistics \citep{hoeffding1948class,efron1981jackknife}.

Following this proof strategy, the key difficulty is in showing that our influence-based statistic
\smash{$\ttheta^*(x)$} is in fact a good approximation for \smash{$\htheta(x)$}.
To do so, we start by establishing consistency of \smash{$\htheta(x)$} for $\theta(x)$ given
our assumptions; we note that this is the only point in the paper where we use the fact that
$\psi$ is the negative gradient of a convex loss as in Assumption \ref{assu:convexity}.

\begin{theo}
\label{theo:consistency}
Given Assumptions \ref{assu:lip}--\ref{assu:convexity},
estimates $(\htheta(x), \, \hnu(x))$ from a 
forest satisfying Specification \ref{spec:forest}
converge in probability to $(\theta(x), \, \nu(x))$.
\end{theo}

Building on this consistency result, we obtain a coupling of the desired type in Lemma \ref{lemm:coupling},
the main technical contribution of this paper.
We note that separating the analysis of moment estimators
into a local approximation argument that hinges on consistency and a separate result that establishes
consistency is standard; see, e.g., Chapter 5.3 of \citet{van2000asymptotic}.
The remainder of our analysis assumes that trees are grown on subsamples of size $s$
scaling as $s = n^\beta$ for some $\beta_{\min} < \beta < 1$, with
\begin{equation}
\label{eq:scaling}
\beta_{\min} := 1 - \p{1 + \pi^{-1} \p{\log\p{\omega^{-1}}} \Big/\p{\log\p{\p{1 - \omega}^{-1}}}}^{-1} < \beta < 1,
\end{equation}
where $\pi$ and $\omega$ are as in Specification \ref{spec:forest}.
This scaling guarantees that the errors of forests are variance-dominated.

\begin{lemm}
\label{lemm:coupling}
Given Assumptions \ref{assu:lip}--\ref{assu:existence},
and a forest trained according to Specification \ref{spec:forest} with \eqref{eq:scaling},
suppose that the generalized random forest estimator $\htheta(x)$ is
consistent for $\theta(x)$. Then $\htheta(x)$ and $\ttheta^*(x)$ are coupled
at the following rate, where $s$, $\pi$ and $\omega$ are as in Specification \ref{spec:forest}:
\begin{equation}
\label{eq:coupling}
\sqrt{\frac{n}{s}} \p{\ttheta^*(x) - \htheta(x)} = \oo_P\p{\max\cb{s^{-\frac{\pi}{2}\frac{\log\p{\p{1 - \omega}^{-1}}}{\log\p{\omega^{-1}}}}, \, \p{\frac{s}{n}}^{\frac{1}{6}}}}.
\end{equation}
\end{lemm}

Given this coupling result, it now remains to study the asymptotics of \smash{$\ttheta^*(x)$}.
In doing so, we re-iterate that \smash{$\ttheta^*(x)$} is \emph{exactly} the output of an infeasible
regression forest trained on outcomes $\theta(x) + \rho_i^*(x)$. Thus, the results of
\citet{wager2015estimation} apply directly to this object, and can be used to establish
its Gaussianity. That we cannot actually compute \smash{$\ttheta^*(x)$}
does not hinder an application of their results.
Pursuing this approach, we find that given \eqref{eq:scaling}, \smash{$\ttheta^*(x)$} and
\smash{$\htheta(x)$} are both asymptotically normal. By extending the same
argument, we could also show that the nuisance parameter estimates \smash{$\hnu(x)$} are
consistent and asymptotically normal; however, we caution that the tree splits are not necessarily
targeted to expressing heterogeneity in \smash{$\nu(x)$}, and so the resulting \smash{$\hnu(x)$}
may not be particularly accurate in finite samples.

\begin{theo}
\label{theo:gauss}
Suppose Assumptions \ref{assu:lip}--\ref{assu:convexity} and
a forest trained according to Specification \ref{spec:forest} with
trees are grown on subsamples of size $s = n^\beta $ satisfying \eqref{eq:scaling}.
Finally, suppose that $\operatorname{Var}[\rho_i^*(x) \cond X_i = x] > 0$. Then, there is a sequence
$\sigma_n(x)$ for which
%\begin{equation}
%\label{eq:sigman}
\smash{$(\htheta_n(x) - \theta(x)) \, /\, \sigma_n(x) \Rightarrow \nn\p{0, \, 1}$} and
\smash{$\sigma^2_n(x) = \operatorname{polylog}(n/s)^{-1} \, s/n$},
%\end{equation}
where $\operatorname{polylog}(n/s)$ is a function that is bounded away from 0 and
increases at most polynomially with the log-inverse sampling ratio $\log\p{n/s}$.
\end{theo}

\section{Confidence Intervals via the Delta Method}
\label{sec:delta_method}

Theorem \ref{theo:gauss} can also be used for statistical inference about $\theta(x)$. Given any
consistent estimator $\hsigma_n(x)/\sigma_n(x) \rightarrow_p 1$ of the noise scale
of \smash{$\htheta_n(x)$}, Theorem \ref{theo:gauss} can be paired with Slutsky's lemma
to verify that
%\begin{equation}
%\label{eq:confidence}
\smash{$\limn \, \mathbb{E}[\theta(x) \in (\htheta_n(x) \pm \Phi^{-1}(1 -\alpha/2) \hsigma_n(x))] = \alpha$}.
%\end{equation}
Thus, in order to build asymptotically valid confidence intervals for $\theta(x)$ centered on $\htheta(x)$,
it suffices to derive an estimator for \smash{$\sigma_n(x)$}.

In order to do so, we again leverage coupling with our approximating pseudo-forest \smash{$\ttheta^*(x)$}.
In particular, the proof of Theorem \ref{theo:gauss} implies that 
$\text{Var}[\ttheta^*(x)]/\sigma^2_n(x) \rightarrow_p 1$, and so it again suffices to study
 \smash{$\ttheta^*(x)$}. Moreover, from the definition of \smash{$\ttheta^*(x)$}, we directly see that
\begin{equation}
\label{eq:H}
\begin{split}
&\Var{\ttheta^*(x)} = \xi^\top V(x)^{-1} H_n(x; \, \theta(x), \, \nu(x)) (V(x)^{-1})^\top \xi,
\end{split}
\end{equation}
where \smash{$H_{n}(x; \, \theta, \, \nu) = \Var{\sum_{i = 1}^n \alpha_i(x) \psi_{\theta, \, \nu}(O_i)}$}.
Thus, we propose building Gaussian confidence intervals using
\begin{equation}
\label{eq:sigma_hat}
\hsigma_n^2(x) := \xi^\top \hV_n(x)^{-1} \hH_n(x) (\hV_n(x)^{-1})^\top \xi,
\end{equation}
where \smash{$\hV_n(x)$} and \smash{$\hH_n(x)$} are consistent estimators
for the quantities in \eqref{eq:H}.

The first quanitity $V(x)$ is a problem specific curvature parameter, and is
not directly linked to forest-based methods. It is the same quantity that is needed
to estimate variance of classical local maximum likelihood methods following \citet{newey1994kernel};
e.g., for the instrumental variables problem described in Section \ref{sec:iv},
\begin{equation}
\label{eq:iv_V}
V(x) = \begin{pmatrix}
\EE{Z_iW_i\cond X_i = x} & \EE{Z_i\cond X_i = x} \\
\EE{W_i\cond X_i = x} & 1
\end{pmatrix},
\end{equation}
while for quantile regression, $V(x) = f_x(\theta(x))$.  In both cases,
several different strategies are available for estimating this term. In the
case of instrumental variables forests, we suggest estimating the
entries of \eqref{eq:iv_V} using (honest and regular) regression forests.

The more interesting term is the inner variance term $H_n(x; \, \theta(x), \, \nu(x))$. To
study this quantity, we note that the forest score
\smash{$\Psi(\theta(x), \, \nu(x)) = \sum_{i = 1}^n \alpha_i(x) \psi_{\theta(x), \, \nu(x)}(O_i)$}
is again formally equivalent to the output of a regression forest with weights
$\alpha_i(x)$, this time with effective outcomes $\psi_{\theta(x), \, \nu(x)}(O_i)$.
A number of proposals have emerged for estimating the variance of a regression
forest, including work by \citet{sexton2009standard}, \citet{mentch2016quantifying}
and \citet{wager2014confidence}; and, in principle, any of these methods could be
adapted to estimate the variance of $\Psi$. The only difficulty is that $\Psi$ depends
on the true parameter values $(\theta(x), \, \nu(x))$, and so cannot directly be
accessed in practice.
Here, we present results based on a variant of the bootstrap of little bags algorithm (or noisy bootstrap) proposed
by \citet{sexton2009standard}. As a side benefit, we also obtain the first consistency guarantees for
this method for any type of forest, including regression forests.

%One of the main advantages of the infinitesimal jackknife was that, as emphasized by \citet{wager2014confidence}, it
%operates only on outputs from an already trained forest, and so can be added ``on top'' of an optimized forest implementation
%without needing to modify the internal workings of the implementation. But now we already needed to develop
%our generalized random forest software, \texttt{grf}, from the ground up, so adding support for
%the method of \citet{sexton2009standard} deep within the forest was less of a concern.

\subsection{Consistency of the Bootstrap of Little Bags}
\label{sec:blb}

To motivate the bootstrap of little bags, we first note that
building confidence intervals via half-sampling---whereby we evaluate an estimator on random
halves of the training data to estimate its sampling error---is closely related to the
bootstrap \citep{efron1982jackknife} (throughout this section, we assume that $s \leq \lfloor n/2 \rfloor$).
In our context, the ideal half-sampling estimator would be \smash{$\hH_n^{HS}(x)$} defined as
\begin{equation}
\label{eq:hs_opt}
\binom{n}{\lfloor n/2 \rfloor}^{-1}
\!\!\!\!\!
\sum_{\cb{\hh \, : \, \abs{\hh} = \left\lfloor \frac{n}{2} \right\rfloor}}
\p{\Psi_{\hh}\p{\htheta(x), \, \hnu(x)} - \Psi\p{\htheta(x), \, \hnu(x)}}^2,
\end{equation}
where \smash{$\Psi_{\hh}$} denotes a version of $\Psi$ computed only using all the possible
trees that only rely on data from the half sample $\hh \subset \cb{1, \, ..., \, n}$ (specifically, in terms of Algorithm
\ref{alg:forest}, we only use trees whose full $\ii$-subsample is contained in $\hh$).
If we could evaluate \smash{$\hH_n^{HS}(x)$}, results from \citet{efron1982jackknife}
suggest that it would be a good variance estimator for $\Psi$, but doing so is
effectively impossible computationally as it would require growing very many forests.

Following \citet{sexton2009standard}, however, we can efficiently approximate \smash{$\hH_n^{HS}(x)$}
at almost no computational cost if we are willing to slightly modify our subsampling scheme.
To do so, let $\ell \geq 2$ denote a little bag size and assume, for simplicity, that $B$ is an integer
multiple of it.
Then, we grow our forest as follows: First draw $g = 1, \, ..., \, B/\ell$ random half-samples
$\hh_g \subset \cb{1, \, ..., \, n}$ of size $\lfloor n/2 \rfloor$, and then generate the subsamples
$\ii_b$ used to build the forest in Algorithm \ref{alg:forest} such that $\ii_b \subseteq \hh_{\lceil b/\ell \rceil}$
for each $b = 1, \, ..., \, B$. In other words, we now generate our forest using little bags of
$\ell$ trees, where all the trees in a given bag only use data from the same half-sample.
\citet{sexton2009standard} discuss optimal choices of $\ell$ for minimizing Monte Carlo error,
and show that they depend on the ratio of the sampling variance of a single tree to that of the full forest.

The upshot of this construction is that we can now identify \smash{$\hH_n^{HS}(x)$}
using a simple variance decomposition. Writing $\Psi_b$ for a version of $\Psi$
computed only using the $b$-th tree, we can verify that \smash{$\hH_n^{HS}(x)$} can
be expressed in terms of the ``between groups'' and ``within group'' variance terms,
\begin{equation*}
%\label{eq:anova}
\EE[ss]{\p{\frac{1}{\ell} \sum_{b = 1}^\ell \Psi_b - \Psi}^2} = \hH_n^{HS}(x) + \frac{1}{\ell - 1} \EE[ss]{\frac{1}{\ell} \sum_{b = 1}^\ell\p{\Psi_b -\frac{1}{\ell} \sum_{b = 1}^\ell \Psi_b}^2},
\end{equation*}
where $\mathbb{E}_{ss}$ denotes expectations over the subsampling mechanism while
holding the data fixed. We define our feasible boostrap of little bags variance estimator
\smash{$\hH_n^{BLB}(x)$} via a version of the above ANOVA decomposition that uses empirical moments
and note that, given a large enough number of trees $B$, this converges to the ideal
half-sampling estimator.

The result below verifies that, under the conditions of Theorem \ref{theo:gauss},
the optimal half-sampling estimator \smash{$\hH_n^{HS}(x)$}
with plug-in values for \smash{$(\htheta(x), \, \hnu(x))$} as in \eqref{eq:hs_opt}
consistently estimates the sampling variance of $\Psi(\theta(x), \, \nu(x))$.
We have already seen above that the computationally feasible estimator
\smash{$\hH_n^{BLB}(x)$} will match \smash{$\hH_n^{HS}(x)$} whenever $B$ is large enough and so,
given any consistent estimator \smash{$\hV_n(x)$} for $V(x)$, we find that the confidence
intervals built using \eqref{eq:sigma_hat} will be asymptotically valid.

\begin{theo}
\label{theo:blb}
Given the conditions of Therorem \ref{theo:gauss}, \smash{$\hH_n^{HS}(x)$} is consistent,
%\begin{equation}
%\label{eq:blb_consistency}
\smash{$\lVert\hH_n^{HS}(x) - H_n(x; \, \theta(x), \, \nu(x))\rVert_F \, \big/ \,
\lVert H_n(x; \, \theta(x), \, \nu(x))\rVert_F \rightarrow_p 0$}.
%\end{equation}
Moreover, given any consistent \smash{$\hV_n(x)$} estimator for $V(x)$ such that
\smash{$\lVert\hV(x) - V(x)\rVert_F \rightarrow_p 0$}, Gaussian confidence intervals
built using \eqref{eq:sigma_hat} will asymptotically have nominal coverage.
\end{theo}

One challenge with the empirical moment estimator
based on the above is that, if $B$ is small, the variance estimates
\smash{$\hH_n^{BLB}(x)$} may be negative. In our software, we avoid this problem
by using a Bayesian analysis of variance following, e.g., \citet{gelman2014bayesian},
with an improper uniform prior for \smash{$\hH_n^{HS}(x)$} over $[0, \, \infty)$.
When $B$ is large enough, this distinction washes out.

\section{Application: Quantile Regression Forests}
\label{sec:quantile}

Our first application of generalized random forests is to the classical problem of non-parametric quantile regression.
This problem has also been considered in detail by
\citet{meinshausen2006quantile}, who proposed a consistent forest-based quantile regression
algorithm; his method also fits into the paradigm of solving estimating equations \eqref{eq:gee_estimate}
using random forest weights \eqref{eq:forest_weights}. However, unlike us, \citet{meinshausen2006quantile}
does not propose a splitting rule that is tailored to the quantile regression context, and instead
builds his forests using plain CART regression splits. Thus, a comparison of our method with that of
\citet{meinshausen2006quantile} provides a perfect opportunity for evaluating the value of our
proposed method for constructing forest-based weights $\alpha_i(x)$ that are specifically designed
to express heterogeneity in conditional quantiles.

Recall that, in the language of estimating equations, the $q$-th quantile $\theta_q(x)$
of the distribution of $Y$ conditionally on $X = x$ is identified via \eqref{eq:conditional_theta}, 
using the moment function
%\begin{equation}
\smash{$\psi_{\theta}(Y_i) = q \1\p{\cb{Y_i > \theta}} - (1 - q) \1\p{\cb{Y_i \leq \theta}}$}.
%\end{equation}
Plugging this moment function into our splitting scheme \eqref{eq:relabel} gives us pseudo-outcomes
%\begin{equation}
%\label{eq:quantile_relabel}
\smash{$\rho_i = \1(\{Y_i > \htheta_{q, P(X_i)}\})$}, where \smash{$\htheta_{q, P(X_i)}$} is the $q$-th quantile of the parent node $P(X_i)$
%\end{equation}
containing $X_i$,
up to a scaling and re-centering that do not affect the subsequent regression split on these
pseudo-outcomes.
In other words, gradient-based quantile regression trees try to separate observations that
fall above the $q$-th quantile of the parent from those below it.

\begin{figure}
\centering
\begin{tabular}{ccc}
\includegraphics[width=\FIGW\textwidth]{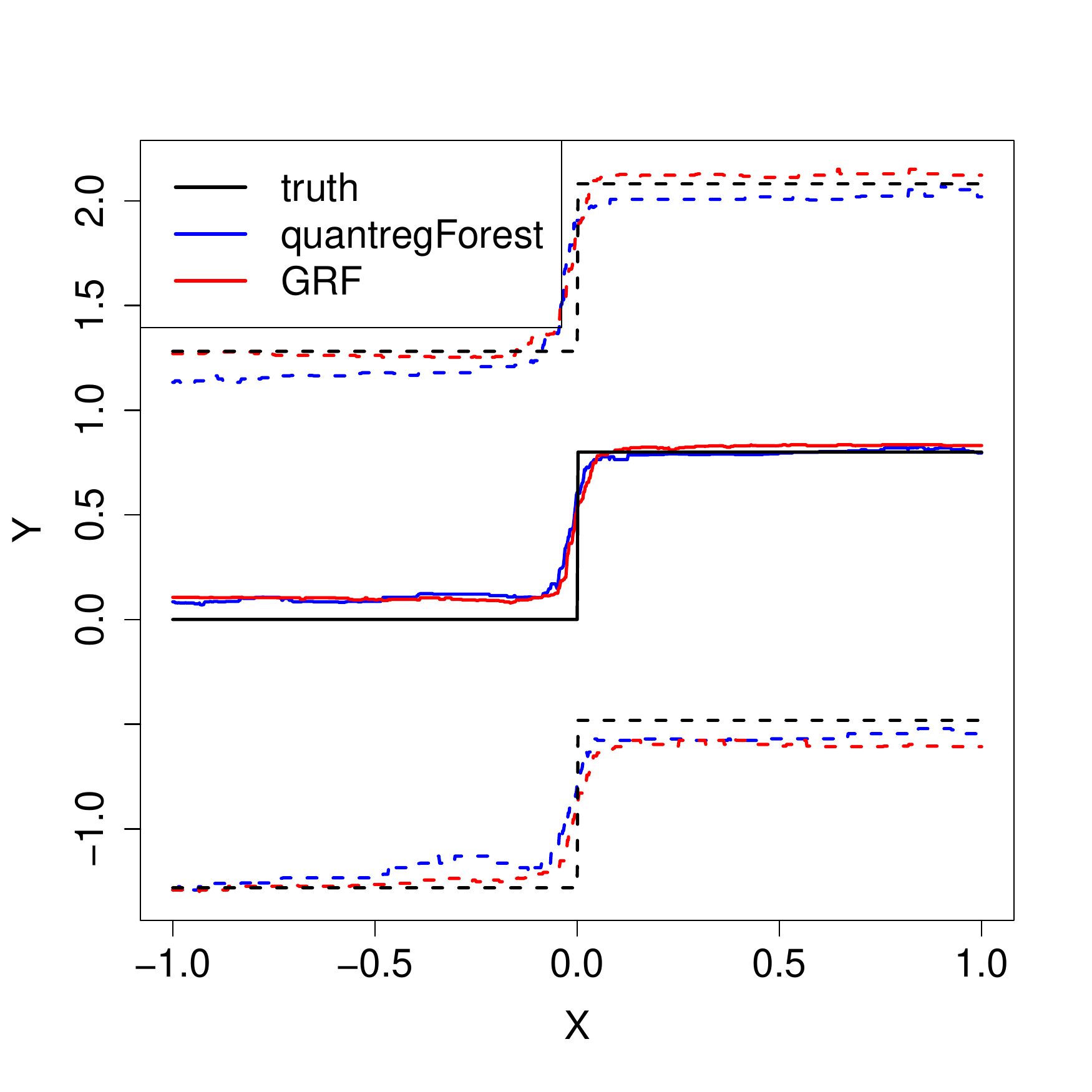} & &
\includegraphics[width=\FIGW\textwidth]{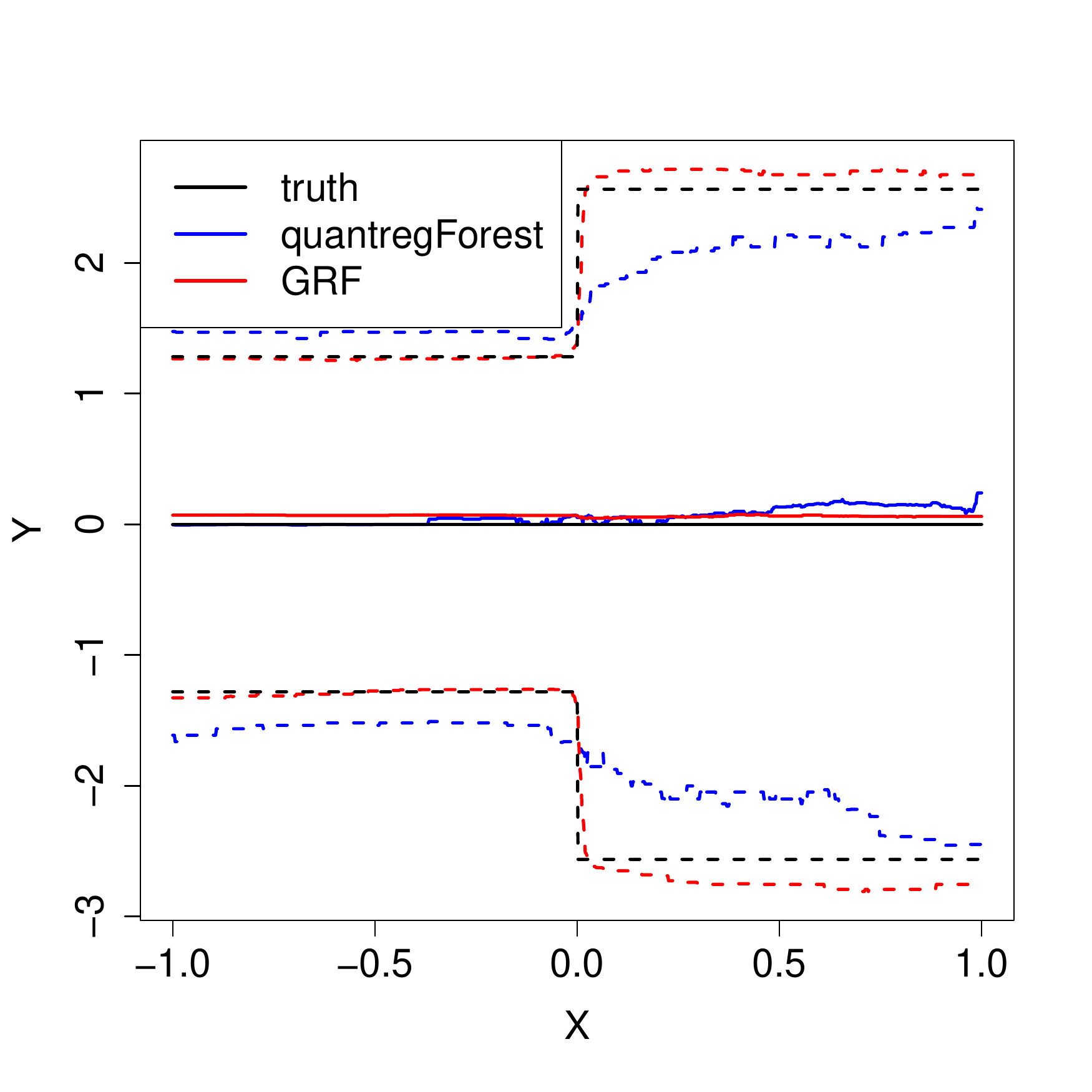} \\
mean shift & & scale shift
\end{tabular}
\caption{Comparison of quantile regression using generalized random forests and
the \texttt{quantregForest} package of \citet{meinshausen2006quantile}. In both
cases, we have $n = 2,000$ independent and identically distributed examples
where $X_i$ is uniformly distributed over $[-1, \, 1]^p$ with $p = 40$, and $Y_i$
is Gaussian conditionally on $(X_i)_1$:
In the left panel, $Y_i \cond X_i \sim \nn\p{0.8 \cdot \mathbf{1}\p{\cb{(X_i)_1 > 0}}, \, 1}$,
while in the right panel $Y_i \cond X_i \sim \nn(0, \, (1 + \mathbf{1}\p{\cb{(X_i)_1 > 0}})^2)$.
The other 39 covariates are noise. 
We estimate the quantiles at $q = 0.1, \, 0.5, \, 0.9$.}
\label{fig:quantile_simu}
\vspace{-1.5\baselineskip}
\end{figure}

We compare our method to that of \citet{meinshausen2006quantile} in Figure \ref{fig:quantile_simu}.
In the left panel, we have a mean shift in the distribution of $Y_i$ conditional on $X_i$ at
$(X_i)_1 = 0$, and both methods are able to pick it up as expected. However, in the right panel,
the mean of $Y$ given $X$ is constant, but there is a scale shift at $(X_i)_1 = 0$. Here, our method still
performs well, as our splitting rule targets changes in the quantiles of the $Y$-distribution. However,
the method of \citet{meinshausen2006quantile} breaks down completely, as it relies on CART regression
splits that are only sensitive to changes in the conditional mean of $Y$ given $X$.
We also note that generalized random forests produce somewhat smoother sample paths than the method of
\citet{meinshausen2006quantile}; this is due to our use of honesty as described in Section
\ref{sec:implementation}. If we run generalized random forests without honesty, then our method still
correctly identifies the jumps at $x = 0$, but has sample paths that oscillate locally
just as much as the baseline method.
The purpose of this example is not to claim that our variant of quantile regression forests
built using gradient trees is always superior to the method of \citet{meinshausen2006quantile} that
uses regression-based splitting to obtain the weights $\alpha_i(x)$; rather, we found that, 
our splitting rule is specifically sensitive to quantile shifts in a way that regression splits are
not---and, moreover, deriving our splitting rule was fully automatic
given the generalized random forest formalism.

In several applications, we want to estimate multiple quantiles
at the same time. For example, in Figure \ref{fig:quantile_simu}, we estimate at $q = 0.1, \, 0.5, \, 0.9$.
Estimating different forests for each quantile separately would be undesirable
for many reasons: It would be computationally expensive, and there is a risk that quantile
estimates might cross in finite samples due to statistical noise. Thus, we need to build a forest using
a splitting scheme that is sensitive to changes at any of our quantiles of interests.
Here, we use a simple heuristic inspired by our relabeling transformation.
Given a set of quantiles of interest $q_1 < ... < q_k$, we first evaluate all these quantiles
\smash{$\htheta_{q_1, P(X_i)} \leq ... \leq \htheta_{q_k, P(X_i)}$}  in the parent node, and label $i$-th point by the
interval \smash{$[\htheta_{q_{j - 1}, P(X_i)}, \, \htheta_{q_{j}, P(X_i)})$} it falls into. Then, we choose the
split point using a multiclass classification rule that classifies each observation into one of the intervals.

%Finally, the purpose of this example is not to claim that our variant of quantile regression forests
%built using gradient trees is always superior to the method of \citet{meinshausen2006quantile} that
%uses regression-based splitting to obtain the weights $\alpha_i(x)$. Rather, we have shown that, as
%claimed, our splitting rule is specifically sensitive to quantile shifts in a way that regression splits are
%not---and, moreover, deriving the above splitting rule was fully automatic
%given our generalized random forest formalism. Of course, there are presumably applications where changes in the conditional mean
%are good proxies for changes in the conditional quantiles; and, in these cases, we might expect the
%method of \citet{meinshausen2006quantile} to work very well. However, if we want to make sure our
%splitting rule is always sensitive to changes at the quantile of interest, our gradient-based approach may
%be more robust.

\section{Application: Estimating Conditional Average Partial Effects}
\label{sec:cape}

Next, we consider conditional average partial effect estimation under
exogeneity; procedurally, the statistical task is equivalent to solving linear regression problems
conditionally on features. Suppose that we observe samples
$(X_i, \, Y_i, \, W_i) \in \xx \times \RR \times \RR^q$, and posit a random effects model
%\begin{equation}
%\label{eq:random_effects}
\smash{$Y_i = W_i \cdot b_i + \varepsilon_i$}, \smash{$\beta(x) = \EE{b_i \cond X_i = x}$}.
%\end{equation}
Our goal is to estimate $\theta(x) = \xi \cdot \beta(x)$ for some contrast $\xi \in \RR^p$.
If $W_i \in \cb{0, \, 1}$ is a treatment assignment, then $\beta(x)$ corresponds to the
conditional average treatment effect.

In order for the average effect $\beta(x)$ to be identified, we need to make certain distributional
assumptions. Here, we assume that the $W_i$ are exogenous, i.e., independent
of the unobservables conditionally on $X_i$:
%\begin{equation}
%\label{eq:cape_unconf}
\smash{$\cb{b_i, \, \varepsilon_i} \indep W_i \cond X_i$}.
%\end{equation}
If $W_i$ is a binary treatment, this condition is equivalent to the unconfoundedness
assumption used to motivate propensity score methods \citep{rosenbaum1983central}. When
exogeneity does not hold, more sophisticated identification strategies are needed
(see following section).

\subsection{Growing a Forest}
\label{eq:cape_forest}

Our parameter of interest $\theta(x) = \xi \cdot \beta(x)$ is
identified by \eqref{eq:conditional_theta} with
\smash{$\psi_{\beta(x), \, c(x)}(Y_i, \, W_i) =  (Y_i - \beta(x) \cdot W_i - c(x)) (1 \ \ W_i^\top)^\top$}
where $c(x)$ is an intercept term;
this can also be written more explicitly as
%\begin{equation}
%\label{eq:cape_ident}
\smash{$\theta(x) = \xi^\top \Var{W_i \cond X_i = x}^{-1} \Cov{W_i, \, Y_i \cond X_i = x}$}.
%\end{equation}
Given forest weights $\alpha_i(x)$ as in \eqref{eq:gee_estimate}, the induced estimator
\smash{$\htheta(x)$} for $\theta(x)$ is
\begin{equation}
\label{eq:cape_estimate}
\htheta(x) \! = \! \xi^\top \! \p{\sum_{i = 1}^n \alpha_i(x) \p{W_i - \bW_\alpha}^{\otimes 2}}^{-1} 
\!\!\! \sum_{i = 1}^n \! \alpha_i(x) \! \p{W_i - \bW_\alpha} \! \p{Y_i - \bY_\alpha}, 
\end{equation}
where \smash{$\bW_\alpha = \sum \alpha_i(x) W_i$} and \smash{$\bY_\alpha = \sum \alpha_i(x) Y_i$},
and we write \smash{$v^{\otimes 2} = vv^\top$}.

Generalized random forests provide us with a quasi-automatic framework for
getting the weights \smash{$\alpha_i(x)$} needed in \eqref{eq:cape_estimate}; all that needs to be done
is to compute the pseudo-outcomes \smash{$\rho_i$} from \eqref{eq:relabel} used for recursive
partitioning. We use \eqref{eq:AP} and, for every parent $P$ and each observation
$i$ with $X_i \in P$ set
\begin{equation}
\label{eq:cape_rho}
\begin{split}
&\rho_i = \xi^\top A_P^{-1} \p{W_i - \bW_P}\p{Y_i - \bY_P - \p{W_i - \bW_{P}}\hbeta_P}, \\
&A_P = \frac{1}{\abs{\cb{i : X_i \in P}}} \sum_{\cb{i : X_i \in P}} \p{W - \bW_P}^{\otimes 2},
\end{split}
\end{equation}
where now \smash{$\bW_P$} and \smash{$\bY_P$} stand for averages taken over the parent $P$,
and \smash{$\hbeta_P$} is the least-squares regression solution of $Y_i$ on $W_i$ in the parent.
Note that the matrix inverse \smash{$A_P^{-1}$} only needs to be evaluated once per parent node.

Checking the conditions required in Section \ref{sec:theory}, note that
Assumption \ref{assu:lip} holds whenever the functions $\EE{Y_i \cond X_i = x}$, 
$\EE{W_i \cond X_i = x}$, $\Cov{Y_i, \, W_i \cond X_i = x}$ and $\Var{W_i \cond X_i = x}$
are all Lipschitz in $x$, Assumption \ref{assu:identification} holds provided that
$\Var{W_i \cond X_i = x}$ in invertible, while Assumptions \ref{assu:covariance}--\ref{assu:convexity}
hold by construction. Thus, Theorem \ref{theo:gauss} in fact applies in this setting.

\subsubsection{Local Centering}
\label{sec:precompute}

The above construction allows for asymptotically valid inference for $\theta(x)$,
but the performance of the forests can in practice
be improved by first regressing out the effect of the features $X_i$ on all the
outcomes separately. Writing
%\begin{equation}
%\label{eq:marginal}
\smash{$y(x) = \mathbb{E}[Y_i \cond X = x]$} and \smash{$w(x) = \mathbb{E}[W_i \cond X = x]$}
%\end{equation}
for the conditional marginal expectations of $Y_i$ and $W_i$ respectively,
define centered outcomes
%\begin{equation}
%\label{eq:residualization}
\smash{$\tY_i = Y_i - \hy^{(-i)}\p{X_i}$} and \smash{$\tW_i = W_i - \hw^{(-i)}\p{X_i}$},
%\end{equation}
where \smash{$\hy^{(-1)}\p{X_i}$}, etc., are leave-one-out estimates of the marginal expectations,
computed without using the $i$-th observation.
We then run a forest using centered outcomes
\smash{$\{\tY_i, \, \tW_i\}_{i = 1}^n$} instead of the original
\smash{$\{Y_i, \, W_i\}_{i = 1}^n$}.

In order to justify this transformation, we note if there is any set $\set \subseteq \xx$
over which $\beta(x)$ is constant (and so $\theta(x)$ is also constant), the following expression
also identifies $\theta(x)$ for any $x \in \set$:
\begin{equation}
\label{eq:orthog_motivation}
\begin{split}
\theta(x) = \xi^\top &\Var{\p{W_i - \EE{W_i \cond X_i}} \cond X_i \in \set}^{-1} \\
& \ \ \ \ \Cov{\p{W_i - \EE{W_i \cond X_i}}, \, \p{Y_i - \EE{Y_i \cond X_i}} \cond X_i \in \set}.
\end{split}
\end{equation}
Thus, if we locally center the $Y_i$ and the $W_i$ before running our forest, the estimator
\eqref{eq:cape_estimate} has the potential to be robust to confounding effects
even when the weights $\alpha_i(x)$ are not sharply concentrated around $x$.
Similar orthogonalization ideas have proven to be useful in many statistical contexts
\citep[e.g.,][]{chernozhukov2016double,newey1994asymptotic,neyman1979c};
in particular, \citet{robinson1988root} showed that if we have access to a neighborhood $\set$
over which $\beta(x) = \beta_{\set}$ is constant, then the moment condition \eqref{eq:orthog_motivation}
induces a semiparametrically efficient estimator for $\theta_{\set} = \xi \cdot \beta_{\set}$.

We note that if we ran a forest with any deterministic
centering scheme, i.e., we used \smash{$\tY_i = Y_i - \hy(X_i)$} for any Lipschitz function $\hy(X_i)$
that does not depend on the data, etc., then the theory developed in Section \ref{sec:theory}
would allow for valid inference about $\theta(x)$ (in particular, we do not need to
assume consistency of $\hy(X_i)$). Moreover, we could also emulate this result by using a form of $k$-fold
cross-fitting \citep{chernozhukov2016double,schick1986asymptotically}.
In the context of forests, it is much more practical to carry out residualization
via leave-one-out prediction than via $k$-fold cross-fitting, because
leave-one-out prediction in forests is computationally cheap \citep{breiman2001random};
however, a practitioner wanting to use results that are precisely covered by theory may prefer
to use cross-fitting for centering.

\subsection{Example: Causal Forests}
\label{sec:cf}

When $W_i \in \cb{0, \, 1}$ is a binary treatment assignment,
the present setup is equivalent to the standard problem of heterogeneous
treatment effect estimation under unconfoundedness.
%\footnote{To draw a tight connection
%between our setting and the classical potential outcomes framework
%\citep{neyman1923applications,rubin1974estimating}, note that if we write $Y_i(0)$ and
%$Y_i(1)$ for the untreated and treated potential outcomes for subject $i$, then we
%simply have $\varepsilon_i = Y_i(0)$ and $b_i = Y_i(1) - Y_i(0)$ in \eqref{eq:random_effects}.
%Moreover, the condition \eqref{eq:cape_unconf} can equivalently be written as
%$\cb{Y_i(0), \, Y_i(1)} \indep W_i \cond X_i$, which is in fact the standard unconfoundedness
%assumption \citep{rosenbaum1983central}.}
Heterogeneous treatment effect estimation via tree-based methods has received considerable
attention in the recent literature: \citet{athey2016recursive} and \citet{su2009subgroup}
develop tree-based methods for subgroup analysis, \citet{hill2011bayesian}
studies treatment effect estimation via Bayesian additive regression trees
\citep{chipman2010bart}, and \citet{wager2015estimation} propose a causal forest
procedure that is very nearly a special case of our generalized random forests.
The main interest of our method is in how it can handle situations for which no
comparable methods exist, such as instrumental variables regression as discussed below.
Here, however, we briefly discuss how some concepts developed as a part of our more
general approach directly improve the performance of causal forests.

The closest method to ours is Procedure 1
of \citet{wager2015estimation}, which is almost equivalent to a generalized random forest without centering,
the only substantive differences being that they split using the exact loss
criterion \eqref{eq:delta_criterion} rather than our gradient-based loss criterion \eqref{eq:gradient_criterion},
and let each tree compute its own treatment effect estimate rather than using the weighting scheme
from Section \ref{sec:ANN} (these methods are exactly equivalent for regression forests, but not
for causal forests).
\citet{wager2015estimation} also consider a second approach, Procedure 2, that obtains its
neighborhood function via a classification forest on the treatment assignments $W_i$.

A weakness of the methods in \citet{wager2015estimation}, as they note in their discussion, is
that these two procedures have different strengths---Procedure 1 is more sensitive to changes
in the treatment effect function, while Procedure 2 is more robust to confounding---but the
hard coded nature of these methods made it difficult to reconcile their relative advantages.
Conversely, given the framing of generalized random forests via
estimating equations, it is ``obvious'' that we can leverage best practices from the literature
on estimating equations and orthogonalize our moment conditions by regressing
out the main effect of $X_i$ on $W_i$ and $Y_i$ as in \citet{robinson1988root}.

To illustrate the value of orthogonalization, we revisit a simulation of
\citet{wager2015estimation} where $X_i \sim U([0, \, 1]^p)$,
$W_i \cond X_i \sim \text{Bernoulli}(e(X_i))$, and
$Y_i \cond X_i, \, W_i \sim \nn\p{m(X_i) + (W_i - 0.5) \tau(X_i), \, 1}$. The authors
consider two different simulation settings: One with no confounding,
$m(x) = 0$ and $e(x) = 0.5$, but with treatment heterogeneity
%\begin{equation}
%\label{eq:tau0_setup}
\smash{$\tau\p{x} = \varsigma\p{x_1} \, \varsigma\p{x_2}$}, \smash{$\varsigma\p{u} = 1 + {1}\,/\,({1 + e^{-20\p{u - 1/3}}})$},
%\end{equation}
and second with no treatment effect, $\tau(x) = 0$, but with confounding,
%\begin{equation}
%\label{eq:prop_setup}
\smash{$e(x) = \frac{1}{4} \p{1 + \beta_{2, \, 4}(x_3)}, \ \ m(x) = 2x_3 - 1$},
%\end{equation}
where $\beta_{a, \, b}$ is the $\beta$-density with shape parameters $a$ and $b$.
We also consider a third setting with both heterogeneity and confounding, that combines
$\tau(\cdot)$ from the first setting with $m(\cdot)$ and $e(\cdot)$ from the second.
For the first setting, \citet{wager2015estimation} used their Procedure 1, whereas for the
second they used Procedure 2, while noting that it is unfortunate that the practitioner
is forced to choose one procedure or the other.

Results presented in Table \ref{tab:cf_simu} are reassuring, suggesting that
generalized random forests with centering do well under both settings, and
can better handle the case with both confounding and treatment heterogeneity than
either of the other two procedures. In contrast, Procedure 1 of Wager and Athey
does poorly with pure confounding, whereas Procedure 2 of Wager and Athey
is good in the pure confounding setting, but does poorly with strong heterogeneity;
this is as expected, noting the design of both methods.

\begin{table}[t]
\centering
\begin{tabular}{||cc|cc||cc|cc||}
\hline
conf. & heterog. & $p$ & $n$ & WA-1 & WA-2 & GRF & C.~GRF \\
  \hline
\hline
no & yes & 10 & 800 & 1.37 & 6.48 & 0.85 & 0.87 \\
  no & yes & 10 & 1600 & 0.63 & 6.23 & 0.58 & 0.59 \\
  no & yes & 20 & 800 & 2.05 & 8.02 & 0.92 & 0.93 \\
  no & yes & 20 & 1600 & 0.71 & 7.61 & 0.52 & 0.52 \\
   \hline
yes & no & 10 & 800 & 0.81 & 0.16 & 1.12 & 0.27 \\
  yes & no & 10 & 1600 & 0.68 & 0.10 & 0.80 & 0.20 \\
  yes & no & 20 & 800 & 0.90 & 0.13 & 1.17 & 0.17 \\
  yes & no & 20 & 1600 & 0.77 & 0.09 & 0.95 & 0.11 \\
   \hline
yes & yes & 10 & 800 & 4.51 & 7.67 & 1.92 & 0.91 \\
  yes & yes & 10 & 1600 & 2.45 & 7.94 & 1.51 & 0.62 \\
  yes & yes & 20 & 800 & 5.93 & 8.68 & 1.92 & 0.93 \\
  yes & yes & 20 & 1600 & 3.54 & 8.61 & 1.55 & 0.57 \\
   \hline
\end{tabular}
\caption{Mean squared error of various ``causal forest'' methods, that
estimate heterogeneous treatment effects under unconfoundedness using forests.
We compare our generalized random forests with and without local centering (C.~GRF and GRF)
to Procedures 1 and 2 of \citet{wager2015estimation}, WA-1 and WA-2.
%WA-1 uses a direct extension of the CART splitting rule for regression targeted
%to treatment effect estimation, while WA-2 splits only on the treatment assignments.
%The simulation settings toggle the presence of confounding (conf.) and treatment heterogeneity (heterog.),
%as well as the number of features $p$ and the sample size $n$.
All forests have $B=2,000$ trees, and results are aggregated over 60 simulation replications
with 1,000 test points each. The mean-squared errors numbers are multiplied by 10 for readbility.}
\label{tab:cf_simu}
\vspace{-1.5\baselineskip}
\end{table}

\section{Application: Instrumental Variables Regression}
\label{sec:iv}

In many applications, we want to measure the causal effect of an intervention on
an outcome, all while recognizing that the intervention and the outcome may also be tied
together through non-causal pathways, thus ruling out the exogeneity assumption
made above. One approach in this situation is to rely on
instrumental variables (IV) regression, where we find an auxiliary source of randomness that can
be used to identify causal effects.

For example, suppose we want to measure the causal effect of child rearing on
a mother's labor-force participation. It is well known that, in the United States, mothers with more children are
less likely to work. But how much of this link is causal, i.e., some mothers work less because they
are busy raising children, and how much of it is merely due to confounding factors, e.g., some mothers have
preferences that both lead them to raise more children and be less likely to participate in the
labor force? Understanding these effects may be helpful in predicting the value of programs
like subsidized daycare that assist mothers' labor force participation while they have young children.

To study this question, \citet{angrist1998children} found a source of auxiliary randomness that can
be used to distinguish causal versus correlational effects: They found that, in the United States, parents who already
have two children of mixed sexes, i.e., one boy and one girl, will have fewer kids in the future than parents
whose first two children were of the same sex. Assuming that the sexes of the first two children in a family
are effectively random, this observed preference for having children of both sexes provides an exogenous source
of variation in family size that can be used to identify causal effects: If the mixed sex indicator is unrelated
to the mother's propensity to work for a fixed number of children, then the effect of the mixed sex 
indicator on the observed propensity to work can be attributed to its effect on family size.  The instrumental
variable estimator normalizes this effect by the effect of mixed sex on family size, so that the normalized estimate
is a consistent estimate of the treatment effect of family size on work.
Other classical uses of instrumental variables regression include measuring the impact of military
service on lifetime income by using the Vietnam draft lottery as an instrument \citep{angrist1990lifetime}, 
and measuring the extent to which 401(k) savings programs crowd out other savings, using
eligibility for 401(k) savings programs as an instrument \citep{abadie2003semiparametric,poterba1996retirement}.

\subsection{A Forest for Instrumental Variables Regression}
\label{sec:iv_methods}

Classical instrumental variables regression seeks a global understanding of the treatment
effect, e.g., on average over the whole U.S. population, does having more children reduce the labor force participation
of women? Here, we use forests to estimate heterogeneous
treatment effects: We might ask how the causal effect of child rearing varies with a mother's age and
socioeconomic status.

We observe $i = 1, \, ..., \, n$ independent and identically distributed subjects,
each of whom has features $X_i \in \xx$, an outcome $Y_i \in \RR$, a received treatment $W_i \in \cb{0, \, 1}$,
and an instrument $Z_i \in \cb{0, \, 1}$. We believe that the outcomes $Y_i$ and received treatment $W_i$ are
related via a structural model
%\begin{equation}
%\label{eq:treat_model}
\smash{$Y_i = \mu\p{X_i} +\tau\p{X_i}  W_i + \varepsilon_i$},
%\end{equation}
where $\tau(X_i)$ is understood to be the causal effect of $W_i$ on $Y_i$, and
$\varepsilon_i$ is a noise term that may be positively correlated with $W_i$. Because $\varepsilon_i$
is correlated with $W_i$, standard regression analyses will not in general be consistent for $\tau(X_i)$,
and we need to use the instrument $Z_i$. If $Z_i$ is independent of $\varepsilon_i$
conditionally on $X_i$ then, provided that $Z_i$ has an influence on the received treatment $W_i$, i.e., that
the covariance of $Z_i$ and $W_i$ conditionally on $X_i = x$ is non-zero, the treatment effect
$\tau(x)$ is identified via
%\begin{equation}
%\label{eq:2sls}
\smash{$\tau(x) = \Cov{Y_i, \, Z_i \cond X_i = x} \,\big/\, \Cov{W_i, \, Z_i \cond X_i = x}$}.
%\end{equation}
We can then estimate $\tau(x)$ by via moment functions
\smash{$ \EE{Z_i\p{Y_i - W_i \, \tau(x) - \mu(x)} \cond X_i = x} = 0$} and
\smash{$\EE{Y_i - W_i \, \tau(x) - \mu(x) \cond X_i = x} = 0$},
 where the intercept $\mu(x)$ is a nuisance parameter.
If we are not willing to assume that every individual $i$ with features
$X_i = x$ has the same treatment effect $\tau(x)$,
then heterogeneous instrumental variables regression
allows us to estimate a (conditional) local average treatment effect
\citep{abadie2003semiparametric,imbens1994identification}.
%We use the additive structure \eqref{eq:treat_model} for simplicity of exposition.
 
We then use our formalism to derive a forest that is targeted towards estimating
causal effects identified via conditional two-stage least squares. Gradient-based labeling
\eqref{eq:relabel} yields pseudo-outcomes for every parent node $P$ and each observation
$i$ with $X_i \in P$,
%\begin{equation}
%\label{eq:iv_pseudooutcome}
\smash{$\rho_i = \p{Z_i - \bZ_P} \p{\p{Y_i - \bY_P} - \p{W_i - \bW_P} \htau_P}$},
%\end{equation}
where $\bY_P, \, \bW_P, \, \bZ_P$ are moments in the parent node, and $\htau_P$ is a solution
to the estimating equation in the parent. Given these pseudo-outcomes,
the tree executes a CART regression split on the $\rho_i$ as usual. Finally, we obtain personalized
treatment effect estimates $\htau(x)$ by solving \eqref{eq:gee_estimate} with
forest weights \eqref{eq:forest_weights}.

To verify that Theorem \ref{theo:gauss} holds in this setting,
we note that Assumption \ref{assu:lip} holds whenever
the conditional moment functions $\EE{W_i \cond X_i = x}$, $\EE{Y_i \cond X_i = x}$, $\EE{Z_i \cond X_i = x}$,
$\EE{W_iZ_i \cond X_i = x}$ and $\EE{Y_iZ_i \cond X_i = x}$ are all Lipschitz continuous in $x$,
while Assumption \ref{assu:identification} holds whenever the instrument is correlated with
received treatment (i.e., the instrument is valid).
Assumptions \ref{assu:covariance}--\ref{assu:convexity} hold thanks to the definition of $\psi$.

As in Section \ref{sec:precompute}, we center our procedure 
using the transformation of \citet{robinson1988root}, and
regress out the marginal effects of $X_i$ first. Writing
%\begin{equation}
%\label{eq:marginal_iv}
\smash{$y(x) = \mathbb{E}[Y_i \cond X = x]$},
\smash{$w(x) = \mathbb{E}[W_i \cond X = x]$}, and
\smash{$z(x) = \mathbb{E}[Z_i \cond X = x]$},
%\end{equation}
we compute conditionally centered outcomes by leave-one-out estimation
%\begin{equation}
%\label{eq:residualization_iv}
\smash{$\tY_i = Y_i - \hy^{(-i)}\p{X_i}$},
\smash{$\tW_i = W_i - \hw^{(-i)}\p{X_i}$} and
\smash{$\tZ_i = Z_i - \hz^{(-i)}\p{X_i}$},
%\end{equation}
\sloppy{and then run the full instrumental variables forest using centered outcomes
\smash{$\{\tY_i, \, \tW_i, \, \tZ_i\}_{i = 1}^n$}.
We recommend working with centered outcomes by default, and 
we do so in our simulations. Our package
\texttt{grf} provides the option of making this transformation automatically,
where \smash{$\hy^{(-i)}\p{X_i}$},  \smash{$\hw^{(-i)}\p{X_i}$} and  \smash{$\hz^{(-i)}\p{X_i}$}
are first estimated using 3 separate regression forests.}

There is a rich literature on non-parametric instrumental variables regression. The above approach
generalizes classical approaches based on kernels or series estimation
\citep{abadie2003semiparametric,su2013local,wooldridge2010econometric}. In other threads,
\citet{darolles2011nonparametric} and \citet{newey2003instrumental} study instrumental variables
models that generalize the conditionally linear treatment model and allowing for non-linear
effects, and \citet{hartford2016counterfactual} develop deep learning tools.
\citet{belloni2012sparse} consider working with high-dimensional instruments $Z_i$.

Appendix \ref{sec:iv_simu} has a simulation study for IV forests, comparing
them to nearest-neighbor and series regression. We find our method to perform well relative to these
baselines, and centering to be helpful. We also evaluate
coverage of the bootstrap of little bag confidence intervals.

\subsection{The Effect of Child Rearing on Labor-Force Participation}
\label{sec:angrist_evans}

\begin{figure}
\centering
\begin{tabular}{ccc}
\includegraphics[width=\FIGW\textwidth]{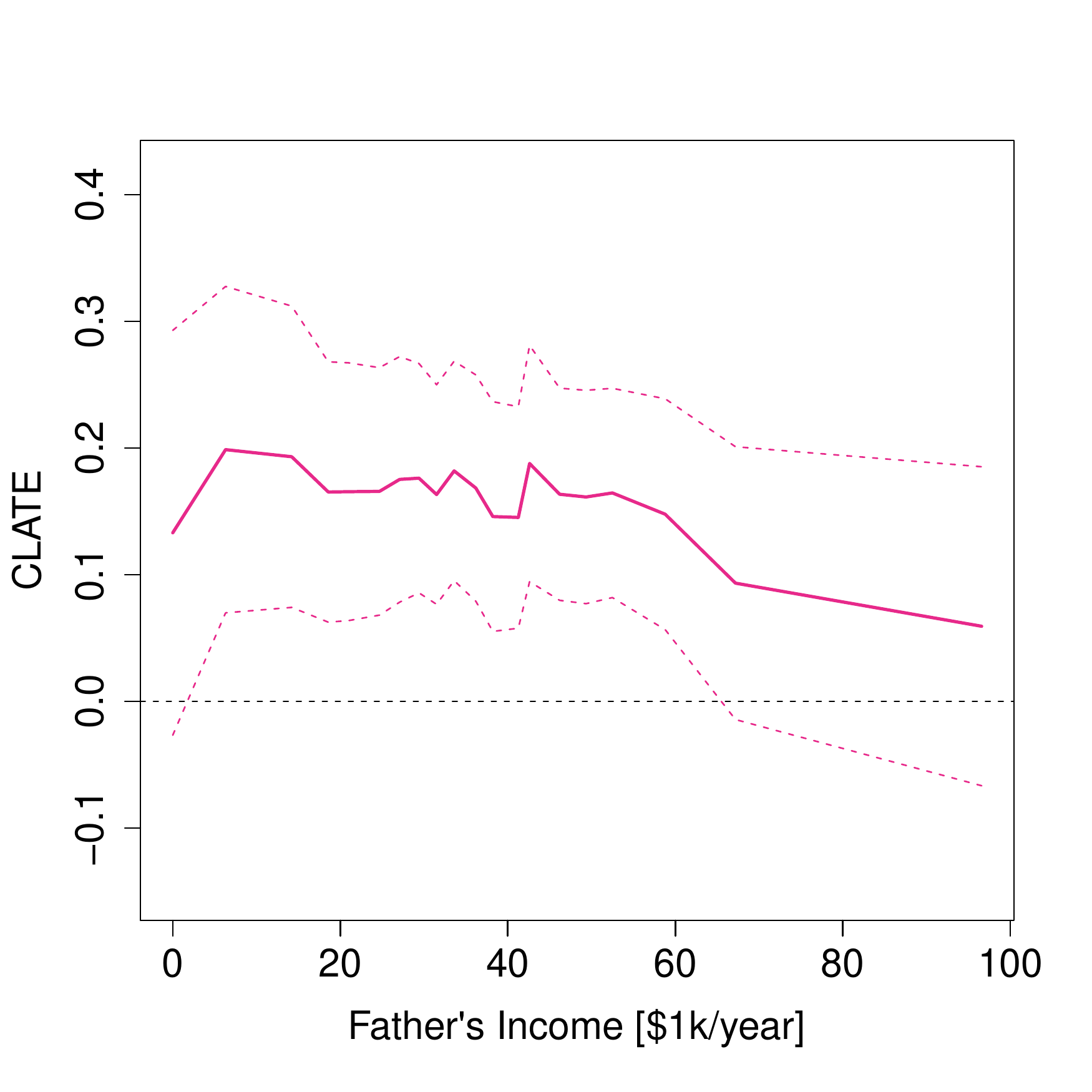} & &
\includegraphics[width=\FIGW\textwidth]{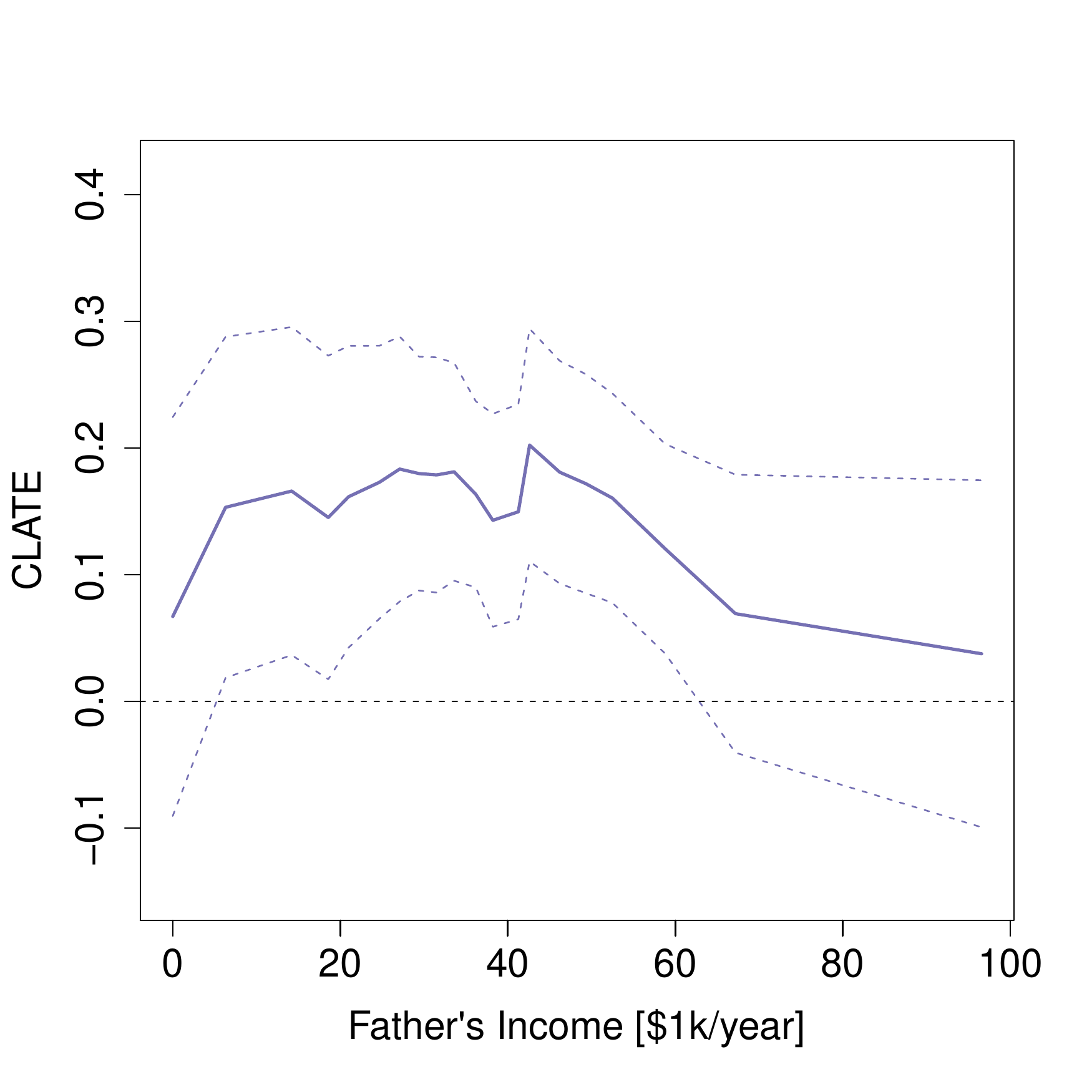} \\
Mother 18 years old at first birth & & Mother 22 years old at first birth
\end{tabular}
\caption{Generalized random forest estimates (along with pointwise 95\% confidence intervals)
for the causal effect of having a third child on the probability
that a mother works for pay.
as identified by the same sex instrument of \citet{angrist1998children};
a positive treatment effect means that the treatment \emph{reduces} the probability that the mother works.
We vary the mother's age at first birth
and the father's income; other covariates are set to their median values in the above plots.
The forest was grown with a sub-sample fraction $s/n = 0.05$, a minimum leaf size $k = 800$, and
consists of $B = 100,000$ trees.}
\label{fig:familysize}
\vspace{-1.5\baselineskip}
\end{figure}

We now revisit our motivating example discussed at the beginning of Section \ref{sec:iv}.
We follow \citet{angrist1998children} in constructing our dataset, and
study a sample of $n = 334,535$ married mothers with at least 2 children (1980 census data),
based on the following quantities: The outcome $Y_i$ is whether the mother did not work in the
year preceding the census, the received treatment $W_i$ is whether the mother had 3 or more
children at census time, and the instrument $Z_i$ measures whether or not
the mother's first two children were of different sexes.
Based on this data, \citet{angrist1998children} estimated the local average treatment effect
of having a third child among mothers with at least two children.
In our sample, \smash{$\hCov{W, \, Z} = 1.6 \cdot 10^{-2}$}, while
\smash{$\hCov{Y, \, Z} = 2.1 \cdot 10^{-3}$}, leading to a 95\% confidence interval for the
local average treatment effect $\tau \in (0.14 \pm 0.054)$ using the \texttt{R} function
\texttt{ivreg} \citep{kleiber2008applied}. Thus, it appears that having a third child reduces
women's labor force participation on average in the US.
\citet{angrist1998children} conduct extensive sensitivity analysis to corroborate the
plausibility of this identification strategy.

We seek to extend this analysis by fitting heterogeneity
on several covariates, including the mother's age at the birth of her first child, her age at census time,
her years of education and her race (black, hispanic, other), as well as the father's income. Formally, our analysis identifies
a conditional local average treatment effect $\tau(x)$ \citep{abadie2003semiparametric,imbens1994identification}.

Results from a generalized random forest analysis are presented in Figure \ref{fig:familysize}. These results suggest that
the observed treatment effect is driven by mothers whose husbands have a lower income. Such an effect would
be intuitively easy to justify: it seems plausible that mothers with wealthier husbands can afford to hire help in raising their children,
and so can choose whether or not to work based on other considerations. That being said, we caution
that the father's income was measured in the census, so there is potentially an endogeneity
problem: perhaps a mother's choice not to work after having a third child enables the husband to earn more.
Ideally, we would have wanted to measure the husband's income at the time of the second child's birth, but
we do not have access to this measurement in the present data. Moreover, the confidence intervals in Figure \ref{fig:familysize}
are rather wide, attesting to the importance of formal asymptotic theory when using forest-based methods for instrumental
variables regression.

\section{Discussion}

We introduced generalized random forests as a versatile method for adaptive,
local estimation in a wide variety of statistical models. We discussed our method
in the contexts of quantile regression and heterogeneous treatment effect estimation,
and our approach also applies to a wide variety of other settings, such as
demand estimation or panel data analysis. Our software, \texttt{grf},
is implemented in a modular way that should enable users to implement
splitting rules motivated by new statistical questions.

Many of the remaining challenges with generalized random forests are closely related to
those with standard nonparametric methods for local likelihood estimation. In particular, as discussed
above, our confidence interval construction relies on undersmoothing to get valid asymptotic
coverage (without undersmoothing, the confidence intervals account for sampling variability of
the forest, but do not capture bias). Developing a principled way to bias-correct our confidence
intervals, and thus avoid the need for undersmoothing, would be of considerable interest both
conceptually and in practice. Moreover, again like standard methods, forests can exhibit edge
effects whereby the slope of our estimates \smash{$\htheta(x)$} may taper off
as we approach the edge of $\xx$-space, even when the true function $\theta(x)$ keeps
changing. Finding an elegant way to deal with such edge effects could improve the
quality of the confidence intervals provided by generalized random forests.

\section*{Acknowledgment}

We are grateful to
Jerry Friedman for first recommending we take a closer look at splitting rules for quantile
regression forests, to Will Fithian for drawing our attention to connections between our
early ideas and gradient boosting, to Guido Imbens for suggesting the local centering scheme
in Section \ref{sec:precompute},
to the associate editor and three anonymous referees for helpful suggestions, and
to seminar participants at
the Atlantic Causal Inference Conference,
the BIRS Workshop on the Interface of Machine Learning and Statistical Inference,
the California Econometrics Conference,
Ca'Voscari University of Venice,
Columbia,
Cornell,
the Econometric Society Winter Meetings,
EPFL,
the European University Institute, 
INFORMS,
Kellogg,
the Microsoft Conference on Digital Economics,
the MIT Conference on Digital Experimentation,
Northwestern,
Toulouse,
Triangle Computer Science Distinguished Lecture Series,
University of Chicago,
University of Illinois Urbana--Champaign,
University of Lausanne, and
the USC Dornsife Conference on Big Data in Economics.

\bibliographystyle{imsart-nameyear}
\bibliography{references}

\newpage

\begin{appendix}

\section{Proof of Main Results}
\label{sec:main_proof}

\begin{figure}[t]
\begin{tikzpicture}[grow=right, sloped]
\node[bag] {$n = 7$}
    child {
        node[bag] {$n = 4$}        
            child {
                node[end, label=right:
                    {\minibox{ $n = 2$, \ $\bY = 1.1$ \\
                    data: $\{(0.1, 0.3), 0.9\}, \{(0.2, 0.5), 1.3\}$}}] {}
                edge from parent
                node[above] {}
                node[below]  {$X_2 \leq 0.7$}
            }
            child {
                node[end, label=right:
                    {\minibox{ $n = 2$, \ $\bY = -0.1$ \\
                    data: $\{(0.1, 0.9), 0.1\}, \{(0.2, 0.8), -0.3\}$}}] {}
                edge from parent
                node[above] {}
                node[below]  {$X_2 > 0.7$}
            }
            edge from parent 
            node[above] {}
            node[below]  {$X_1 \leq 0.3$}
    }
    child {
        node[bag] {$n = 3$}        
        child {
                node[end, label=right:
                    {\minibox{ $n = 1$, \ $\bY = -0.7$ \\
                    data: $\{(0.5, 0.4), -0.7\}$}}] {}
                edge from parent
                node[above] {}
                node[below]  {$X_1 \leq 0.8$}
            }
            child {
                node[end, label=right:
                    {\minibox{ $n = 2$, \ $\bY = 0.2$ \\
                    data: $\{(0.9, 0.4), -0.1\}, \{(1.0, 0.2), 0.3\}$}}] {}
                edge from parent
                node[above] {}
                node[below]  {$X_1 > 0.8$}
            }
        edge from parent         
            node[above] {}
            node[below]  {$X_1 > 0.3$}
    };
\end{tikzpicture}
\caption{Example of a small regression tree on a sample of size $n = 7$. The
examples used to build this tree are of the form $\{X_i, \, Y_i\} \in \RR^2 \times \RR$, and
axis-aligned splits based on the $X_i$ determine the leaf membership of each training example.
In ``standard'' regression trees as discussed in,
e.g., \citet{breiman1984classification} or \citet{hastie2009elements}, the tree predicts by averaging the outcomes $Y_i$ within the relevant leaf; thus, in the example of Figure 1,
any test point $x$ with $(x_1 \leq 0.3) \land (x_2 \leq 0.7)$ would be assigned a prediction
$\hmu(x) = 1.1$. 
In our method, we do not consider tree predictions directly, but instead use trees to construct a neighborhood weighting as in Figure \ref{fig:weighting}. Our approach also relies on a form of subsample splitting where different subsets of the data are used to grow the tree and make within-leaf predictions; see Section \ref{sec:implementation} for details.
}
\label{fig:tree}
\end{figure}
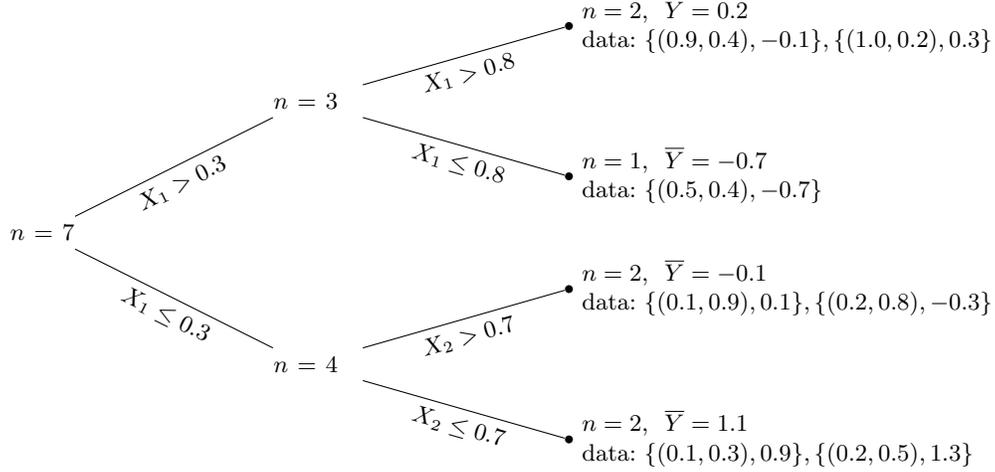

Here, we present arguments leading up to our main result, namely the central
limit theorem presented in Theorem \ref{theo:gauss}, starting with some technical
lemmas. The proofs of Propositions \ref{prop:motivation}
and \ref{prop:approximation}, as well as the technical results
stated below are given in Appendix \ref{sec:appendix_proofs}.
Throughout our theoretical analysis, we use the following notation:
Given our forest weights $\alpha_i(x)$ \eqref{eq:forest_weights}, let
\begin{equation}
\label{eq:Psi}
\Psi\p{\theta, \, \nu} := \sum_{i = 1}^n \alpha_i(x) \psi_{\theta, \, \nu}(O_i) \eqand
\bPsi\p{\theta, \, \nu} := \sum_{i = 1}^n \alpha_i(x) M_{\theta, \, \nu}(X_i).
\end{equation}
We will frequently use the following bounds on the moments of $\Psi$ at the true
parameter value $(\theta(x), \, \nu(x))$.

\begin{lemm}
\label{lemm:moments}
Let $\alpha_i(x)$ be weights from a forest obtained as in Specification \ref{spec:forest},
and suppose that the $M$-function is Lipschitz in $x$ as in Assumption \ref{assu:lip}.
Then, \smash{$\Psi\p{\theta(x), \, \nu(x)}$} satisfies the following moment bounds:
\begin{align}
\label{eq:PsiE}
&\Norm{\EE{\Psi\p{\theta(x), \, \nu(x)}}}_2 =  \oo\p{s^{-\frac{\pi}{2}\frac{\log\p{\p{1 - \omega}^{-1}}}{\log\p{\omega^{-1}}}}} \\
\label{eq:PsiV}
&\Norm{\Var{\Psi\p{\theta(x), \, \nu(x)}}}_F = \oo\p{s/n},
\end{align}
where $s$ is the subsampling rate used when building our forest.
\proof
To establish these bounds, we start by expanding $\Psi$ as
\begin{equation}
\Psi\p{\theta, \, \nu} = \frac{1}{B} \sum_{b = 1}^B \sum_{i = 1}^n \alpha_{bi}(x) \psi_{\theta, \, \nu}\p{O_i},
\end{equation}
where the $\alpha_{bi}$ are the individual tree weights used to build the forest weights in \eqref{eq:forest_weights}.
Now, $\Psi\p{\theta, \, \nu}$ is nothing but the output of a regression forest with response
$\psi_{\theta, \, \nu}\p{O_i}$. Thus, given our assumptions about the moments of $\psi_{\theta, \, \nu}(O_i)$
and the fact that our trees are built via honest subsampling, these bounds follow directly from arguments
made in \citet{wager2015estimation}. First, the proof of Theorem 3 of \citet{wager2015estimation}
shows that the weights $\alpha_i(x)$ are localized:
\begin{equation}
\label{eq:local_weights}
\EE{\sup\cb{\Norm{X_i - x}_2 : \alpha_i(x) > 0}} = \oo\p{s^{-\frac{\pi}{2}\frac{\log\p{\p{1 - \omega}^{-1}}}{\log\p{\omega^{-1}}}}},
\end{equation}
thus directly implying \eqref{eq:PsiE} thanks to Assumption \ref{assu:lip}.
Meanwhile, because individual trees are grown on subsamples,
we can verify that
\begin{equation}
\frac{n}{s} \Var{\Psi\p{\theta(x), \, \nu(x)}} \preceq \Var{\sum_{i = 1}^n \alpha_{bi}(x) \psi_{\theta, \, \nu}\p{O_i}} = \oo\p{1}.
\end{equation}
The first inequality results from classical results about $U$-statistics going back to
\citet{hoeffding1948class}, and simply states that the variance of the forest score is at most
$s/n$ times the variance of a tree score; see Appendix C3 of \citet{wager2015estimation} for
a discussion in the context of regression forests.
The second inequality follows from second-moment bounds on $\psi$
along with the fact that our trees are grown on honest subsamples.
\endproof
\end{lemm}

\subsection{Local Regularity of Forests}
\label{sec:regularity}

Before proving any of our main results, we need establish a result that gives us some control
over the ``sample paths'' of $\Psi$. To do so, define the local discrepancy measure
\begin{equation}
\label{eq:delta_alpha}
\delta\p{\p{\theta, \, \nu}, \, \p{\theta', \, \nu'}} = \Psi\p{\theta, \, \nu}  - \bPsi\p{\theta, \, \nu}  - \p{\Psi\p{\theta', \, \nu'}  - \bPsi\p{\theta', \, \nu'} },
\end{equation}
which describes how tightly the stochastic fluctuations of \smash{$\Psi - \bPsi$} are coupled for nearby parameter
values $(\theta, \, \nu)$ and $(\theta', \, \nu')$.
The following lemmas establish uniform local concentration of $\delta$:
First, in Lemma \ref{lemm:forest_covariance}, we control the variogram of the forest, and then
Lemma \ref{lemm:delta_conc} establishes concentration of $\delta$ over small balls.
Both proofs are given in Appendix \ref{sec:appendix_proofs}.

\begin{lemm}
\label{lemm:forest_covariance}
Let $(\theta, \, \nu)$ and $(\theta', \, \nu')$ be fixed pairs of parameters, and let $\alpha_i(x)$ be
forest weights generated according to Specification \ref{spec:forest}. Then, provided that
Assumptions \ref{assu:lip}--\ref{assu:covariance} hold,
\begin{equation}
\label{eq:delta_cov}
\begin{split}
&\EE{\delta\p{\p{\theta, \, \nu}, \, \p{\theta', \, \nu'}}} = 0, \\
&\EE{\Norm{\delta\p{\p{\theta, \, \nu}, \, \p{\theta', \, \nu'}}}_2^2} \leq 
L \frac{s}{n} \Norm{\begin{pmatrix} \theta \\ \nu \end{pmatrix} -  \begin{pmatrix} \theta' \\ \nu' \end{pmatrix}}_2, 
\end{split}
\end{equation}
where $L$ is the Lipschitz parameter from \eqref{eq:variogram}.
\end{lemm}

Next, to generalize this concentration bound from a single point into a uniform bound,
we will need some standard formalism from empirical process theory
as presented in, e.g., \citet{van1996weak}. To do so, we start by defining a
bracketing, as follows. For any pair of parameters $(\theta_{-}, \, \nu_{-})$,
$(\theta_{+}, \, \nu_{+})$, define the bracket
$$ \beta\p{\begin{pmatrix} \theta_{-} \\ \nu_{-} \end{pmatrix},
\begin{pmatrix} \theta_{+} \\ \nu_{+} \end{pmatrix}}
: = \cb{\begin{pmatrix} \theta \\ \nu \end{pmatrix} \in \bb : 
\Psi\p{\theta_{-}, \, \nu_{-}}
\leq \Psi\p{\theta, \, \nu}
\leq \Psi\p{\theta_{+}, \, \nu_{+}}} $$
for all realizations of $\Psi$, where the inequality is
understood coordinate-wise; and define the radius $r$
of the bracket in terms of the $L_2$-distance of the individual ``$\psi$-trees''
that comprise $\Psi$:
\begin{equation}
\label{eq:bracketing_radius}
\begin{split}
&r^2\p{\beta\p{\begin{pmatrix} \theta_{-} \\ \nu_{-} \end{pmatrix},
\begin{pmatrix} \theta_{+} \\ \nu_{+} \end{pmatrix}}} \\
&\ \ \ \ \ \ \ \
:= \EE{\Norm{\sum_{\cb{i : i \in \jj_1}} \alpha_i(x; \, \jj_2) \p{\psi_{\theta_+, \, \nu_+}\p{O_i} - \psi_{\theta_-, \, \nu_-}\p{O_i}}}_2^2},
\end{split}
\end{equation}
where $\jj_1$ and $\jj_2$ are two disjoint half-subsamples as in Algorithm \ref{alg:forest}.
For any $\varepsilon > 0$, the $\varepsilon$-bracketing number
$N_{[]}(\varepsilon, \, \Psi, \, L_2)$ is the minimum number of brackets of
radius at most $\varepsilon$ required to cover $\bb$.

Given this notation, our concentration bound for $\delta$ will depend
on controlling this covering number. Specifically, we assume that there is a
constant $\kappa$ for which the bracketing entropy $\log N_{[]}$ is bounded by
\begin{equation}
\label{eq:bracketing_entropy}
\log\p{N_{[]}(\varepsilon, \, \Psi, \, L_2)} \leq \frac{\kappa}{\varepsilon} \ \text{ for all } \ 0 < \varepsilon \leq 1.
\end{equation}
We use Assumption \ref{assu:donsker} to give us bounds of this type; and, in fact,
this is the only place we use Assumption \ref{assu:donsker}. Replacing Assumption
\ref{assu:donsker} with \eqref{eq:bracketing_entropy} would be enough to prove our results,
which will only depend on this assumption through Lemma \ref{lemm:delta_conc} below.

To see how Assumption \ref{assu:donsker} leads to \eqref{eq:bracketing_entropy}, we first write
$$ \Psi\p{\theta, \, \nu} = \Psi^{\lambda}\p{\theta, \, \nu} + \Psi^{\zeta}\p{\theta, \, \nu}, $$
where $\Psi^\lambda$ is Lipschitz and $\Psi^\zeta$ is a monotone function of
a univariate representation of $O_i$. Writing analogously $N_{[]}(\varepsilon, \, \Psi^\lambda, \, L_2)$
and $N_{[]}(\varepsilon, \, \Psi^\zeta, \, L_2)$ for the bracketing numbers of these two additive
components on their own, we can verify that
$$ \log\p{N_{[]}(\varepsilon, \, \Psi, \, L_2)} \leq
\log\p{N_{[]}(\varepsilon/2, \, \Psi^\lambda, \, L_2)} +\log\p{N_{[]}(\varepsilon/2, \, \Psi^\zeta, \, L_2)}. $$
Because $\Psi^\zeta$ is a bounded,
monotone, univariate family, Theorem 2.7.5 of \citet{van1996weak} implies that
$\log N_{[]}(\varepsilon, \, \Psi^\lambda, \, L_2) = \oo\p{1/\varepsilon}$.
Meanwhile, because $\Psi^\lambda$ is Lipschitz and our parameter space $\bb$ is compact,
Lemma 2.7.11 of \citet{van1996weak} implies that
$\log N_{[]}(\varepsilon, \, \Psi^\lambda, \, L_2) = \oo\p{\log \varepsilon^{-1}}$. Thus, both
terms are controlled at the desired order, and so \eqref{eq:bracketing_entropy} holds.

\begin{lemm}
\label{lemm:delta_conc}
Under the conditions of Lemma \ref{lemm:forest_covariance},
suppose moreover that \eqref{eq:bracketing_entropy} holds. Then,
\begin{equation}
\label{eq:discr_concentration}
\begin{split}
&\EE{\sup_{\p{\theta', \, \nu'}}\cb{\Norm{\delta\p{\p{\theta, \, \nu}, \, \p{\theta', \, \nu'}}}_2  : \Norm{\begin{pmatrix} \theta - \theta' \\ \nu - \nu' \end{pmatrix}}_2 \leq \eta}} \\
&\ \ \ \ \ \ \ = \oo\p{\sqrt{\frac{\kappa L \eta}{ n/s }} + \frac{8\kappa G}{ (n/s)  {L\eta}}},
\end{split}
\end{equation}
 for any $\eta > 0$ and $1 \leq s \leq n$, where
 $G$ is an upper bound for $\Norm{\psi_{\theta, \, \nu}(O_i) - \psi_{\theta', \, \nu'}(O_i)}_\infty \leq G$;
 note that Assumption \ref{assu:donsker} guarantees that a finite bound $G$ exists.
\end{lemm}

\subsection*{Proof of Theorem \ref{theo:consistency}}

First, thanks to Lemma \ref{lemm:moments}, we know that
\begin{equation}
\label{eq:gradient_conv}
\Norm{\Psi\p{\theta(x), \, \nu(x)}}_2 \rightarrow_p 0.
\end{equation}
Thus, thanks to Assumption \ref{assu:existence}, we know there must exist a sequence
$\varepsilon_n > 0$ with $\limn \varepsilon_n = 0$ such that
$$ \Norm{\Psi\p{\theta(x), \, \nu(x)}}_2, \, \Norm{\Psi\p{\htheta(x), \, \hnu(x)}}_2 < \varepsilon_n $$
with probability tending to 1;
and so Lemma \ref{lemm:unique} below implies the desired result.

\begin{lemm}
\label{lemm:unique}
Suppose that Assumptions \ref{assu:lip}--\ref{assu:convexity} hold,
and that the forest is trained according to Specification \ref{spec:forest}.
Then, all approximate solutions to \eqref{eq:gee_estimate}
are close to each other, in the following sense: for any sequence $\varepsilon_n > 0$ with $\limn \varepsilon_n = 0$,
\begin{equation}
\label{eq:close_minima}
\sup\cb{\Norm{\begin{pmatrix} \theta - \theta' \\ \nu - \nu' \end{pmatrix}}_2  :
\Norm{\Psi\p{\theta, \, \nu}}_2, \ \Norm{\Psi\p{\theta', \, \nu'}}_2
< \varepsilon_n} \rightarrow_p 0.
\end{equation}
\begin{proof}
Starting with some notation, let
$$ \Psi(\theta, \, \nu) \in -\partial F(\theta, \, \nu), \ \ \bPsi(\theta, \, \nu) = -\nabla \bF(\theta, \, \nu), $$
where $F$ and $\bF$ are the respectively convex and $\sigma^2$-strongly convex functions
implicitly defined in the hypothesis statement. Recall that \smash{$(\htheta, \, \hnu)$} is assumed to satisfy
Assumption \ref{assu:existence}, and let $\eta_n > 0$ be any sequence with
$\limn \eta_n = 0$, $\eta_n > \max\{4\varepsilon_n/\sigma^2, \, \sqrt[4]{s/n}\}$ for all $n = 1, \, 2, \, ...$,
and \smash{$\eta_n^{-1} \lVert\Psi(\htheta, \, \hnu)\rVert_2 \rightarrow_p 0$}.

Now, thanks to Assumptions \ref{assu:lip}--\ref{assu:donsker}, we can
apply Lemma \ref{lemm:delta_conc}. Because $\eta_n \geq \sqrt[4]{s/n}$, we can
pair \eqref{eq:discr_concentration} with the fundamental theorem of calculus for line integrals to check that
\begin{align*}
&F(\theta, \, \nu) - F(\htheta, \, \hnu) + \Psi(\htheta, \, \hnu) \cdot \begin{pmatrix} \theta - \htheta \\ \nu - \hnu \end{pmatrix} \\
&\ \ \ \ = \bF(\theta, \, \nu) - \bF(\htheta, \, \hnu) + \bPsi(\htheta, \, \hnu) \cdot \begin{pmatrix} \theta - \htheta \\ \nu - \hnu \end{pmatrix}
+ o_P\p{\eta_n^2},
\end{align*}
for points $(\theta, \, \nu)$ within $L_2$-distance $\eta_n$ of \smash{$(\htheta, \, \hnu)$}.
By strong convexity of $\bF$, this implies that
$$ F(\theta, \, \nu) \geq F(\htheta, \, \hnu) - \Psi(\htheta, \, \hnu) \cdot \begin{pmatrix} \theta - \htheta \\ \nu - \hnu \end{pmatrix} + \frac{\sigma^2}{2} \Norm{\begin{pmatrix} \theta - \htheta \\ \nu - \hnu \end{pmatrix}}_2^2 + o_P\p{\eta_n^2}, $$
again for $(\theta, \, \nu)$ within $\eta_n$ of \smash{$(\htheta, \, \hnu)$}.
Thus, with probability tending to 1,
$$ \inf\cb{ F(\theta, \, \nu) -  F(\htheta, \, \hnu) :\Norm{\begin{pmatrix} \theta - \htheta \\ \nu - \hnu \end{pmatrix}}_2 = \eta_n}
\geq \frac{\sigma^2}{4} \, \eta_n^2; $$
note that, here, we also used the fact that \smash{$\eta_n^{-1} \lVert\Psi(\htheta, \, \hnu)\rVert_2 \rightarrow_p 0$}. 
Finally, by convexity of $F$, this last fact implies that, with probability tending to 1,
$$ \Norm{\Psi\p{\theta, \, \nu}}_2 \geq \frac{\sigma^2}{4} \, \eta_n \ \text { for all } \ \Norm{\begin{pmatrix} \theta - \htheta \\ \nu - \hnu \end{pmatrix}}_2 \geq \eta_n. $$
Recall that, by construction, $\varepsilon_n < \sigma^2 \eta_n/4$, and so
\eqref{eq:close_minima} must hold.
\end{proof}
\end{lemm}

\subsection*{Proof of Lemma \ref{lemm:coupling}}
If $\psi_{\theta, \, \nu}(O_i)$ were twice differentiable in
$(\theta, \, \nu)$, then we could verify \eqref{eq:coupling} fairly directly via
Taylor expansion of $\psi$. Now, of course, $\psi$ is not twice differentiable,
and so we cannot apply this argument directly. Rather, we need to first apply
a Taylor expansion on the expected $\psi$ function, $M_{\theta, \, \nu}(X_i)$,
which \emph{is} twice differentiable; we then use the regularity properties established
in Section \ref{sec:regularity} to extend this result to $\psi$.

Given consistency of \smash{$(\htheta(x), \, \hnu(x))$}, there
is a sequence $\varepsilon_n \rightarrow 0$ such that
\begin{equation}
\label{eq:consistency_first}
\Norm{\begin{pmatrix}
\htheta(x) - \theta(x) \\ \hnu(x) - \nu(x)
\end{pmatrix}}_2 = \oo_P\p{\varepsilon_n}. 
\end{equation}
Using notation established in \eqref{eq:Psi} and \eqref{eq:delta_alpha}, we then write
\begin{equation}
\label{eq:decomposition}
\begin{split}
&\bPsi\p{\htheta(x), \, \hnu(x)} -  \bPsi\p{\theta(x), \, \nu(x)}
= \Psi\p{\htheta(x), \, \hnu(x)} -  \Psi\p{\theta(x), \, \nu(x)} \\
& \ \ \ \ \ \ \ \ \ + \delta\p{\p{\theta(x), \, \nu(x)} , \, \p{\htheta(x), \, \hnu(x)}}.
\end{split}
\end{equation}
By the assumed smoothness of the moment functions, we know that
$\bPsi$ is twice differentiable in $(\theta, \, \nu)$ with a bound on the second derivative
that holds uniformly over all realizations of $\alpha_i(x)$ and $X_i$,
and so we can take a Taylor expansion:
\begin{align*}
&\bPsi\p{\htheta(x), \, \hnu(x)} -  \bPsi\p{\theta(x), \, \nu(x)} \\
&\ \ \ \ \ \ \ \ \ = \p{\sum_{i = 1}^n \alpha_i(x) \nabla M_{\theta(x), \, \nu(x)}(X_i)}
\begin{pmatrix}
\htheta(x) - \theta(x) \\ \hnu(x) - \nu(x)
\end{pmatrix}
+ H
\end{align*}
with $\Norm{H} \leq c\,\varepsilon_n^2/2$, where $c$ is the uniform bound on the curvature
of $M$ required in Assumption \ref{assu:identification}.
Moreover, because the weights $\alpha_i(x)$ are localized as in \eqref{eq:local_weights},
\begin{equation}
\label{eq:first_bound}
\Norm{\sum_{i = 1}^n \alpha_i(x) \nabla M_{\theta(x), \, \nu(x)}(X_i) - V(x)}_F = \oo_P\p{s^{-\frac{\pi}{2}\frac{\log\p{\p{1 - \omega}^{-1}}}{\log\p{\omega^{-1}}}}},
\end{equation}
where $s \rightarrow \infty$ is the sub-sample size used to grow trees in the forest.
This expansion suggests that \eqref{eq:decomposition} should be helpful in relating
our quantities of interest.

It now remains to bound the extraneous terms.
By Assumption \ref{assu:existence}, we know that
$$ \Psi\p{\htheta(x), \, \hnu(x)} \leq C \max_{1 \leq i \leq n} \cb{\alpha_i} \leq C \frac{s}{n}. $$
Next, by the consistency of \smash{$(\htheta(x), \, \hnu(x))$},
we can apply Lemma \ref{lemm:delta_conc} with ``$\eta$''
set to $\varepsilon_n^{2/3}$ to verify that
$$\Norm{\delta\p{\p{\theta(x), \, \nu(x)} , \, \p{\htheta(x), \, \hnu(x)}}}_2 =
\oo_P\p{\max\cb{\varepsilon_n^{1/3} \sqrt{\frac{s}{n}}, \, \frac{s}{n \, \varepsilon_n^{2/3}}}}. $$
Thus, thanks to Assumption \ref{assu:identification} which lets us invert $V(x)$,
we conclude that 
\begin{equation}
\label{eq:exact_coupling}
\begin{split}
&\Norm{\begin{pmatrix}
\htheta(x) - \theta(x) \\ \hnu(x) - \nu(x)
\end{pmatrix}
+ V(x)^{-1} \Psi\p{\theta(x), \, \nu(x)}}_2 \\
&\ \ \ \ \ \ \ \ = \oo_P\p{\max\cb{s^{-\frac{\pi}{2}\frac{\log\p{\p{1 - \omega}^{-1}}}{\log\p{\omega^{-1}}}} \, \varepsilon_n, \, \varepsilon_n^2, \,  \varepsilon_n^{1/3} \, \sqrt{\frac{s}{n}}, \, \frac{s}{n \, \varepsilon_n^{2/3}}}}. 
\end{split}
\end{equation}
\sloppy{Finally, recall that $\Norm{\Psi\p{\theta(x), \, \nu(x)}}_2^2 = \oo_P\p{s/n}$ by
Lemma \ref{lemm:moments} and \eqref{eq:scaling}. Thus, we can use the bound \eqref{eq:exact_coupling} to
get stronger consistency guarantees, and in fact verify that \smash{$(\htheta(x), \, \hnu(x))$}
must have been $\sqrt{s/n}$-consistent; and so, in particular, we can take \eqref{eq:consistency_first}
to hold with $\varepsilon_n  = \sqrt{s/n}$.
The desired result then follows directly from \eqref{eq:exact_coupling}, noting that
\smash{$\ttheta^*(x) = \theta(x) + \xi^\top V(x)^{-1} \Psi\p{\theta(x), \, \nu(x)}$}.}

\subsection*{Proof of Theorem \ref{theo:gauss}}

As argued in Section \ref{sec:gauss}, \smash{$\ttheta^*(x)$} is formally equivalent
to the output of a regression forest, and so we can directly apply
Theorem 1 of \citet{wager2015estimation}.
Given the assumptions made here, their result shows that
\begin{equation}
\label{eq:pseudo_CLT}
 \p{\ttheta^*(x) - \theta(x)} \,\big/\, \sigma_n(x) \Rightarrow \nn\p{0, \, 1}, \ \ \sigma^2_n(x) \rightarrow_p 0.
\end{equation}
Moreover, from Theorem 5 and Lemma 7 of \citet{wager2015estimation}, we see that
$\sigma_n^2$ scales as discussed in the hypothesis statement.
Given this central limit theorem, it only remains to show that the discrepancy between
\smash{$\htheta(x)$} and \smash{$\ttheta^*(x)$} established in Lemma \ref{lemm:coupling},
decays faster than $\sigma_n(x)$.
But, thanks to the consistency result from Theorem \ref{theo:consistency}, the coupling
result in Lemma \ref{lemm:coupling} implies that
$$ \frac{n}{s} \, \p{\ttheta^*(x) - \htheta(x)}^2 = \oo_P\p{ \max\cb{s^{-\pi\frac{\log\p{\p{1 - \omega}^{-1}}}{\log\p{\omega^{-1}}}}, \, \sqrt[3]{\frac{s}{n}}} }, $$
and so \smash{$(\ttheta^*(x) - \htheta(x))/\sigma_n \rightarrow_p 0$}.

\subsection*{Proof of Theorem \ref{theo:blb}}

Following our discussion in Section \ref{sec:blb}, we here only consider the ideal ``$B \rightarrow \infty$''
half-sampling estimator. We start by considering its expectation,
$$ \EE{\hH_n^{HS}(x)} = \EE{\p{\Psi_{\hh}\p{\htheta(x), \, \hnu(x)} - \Psi\p{\htheta(x), \, \hnu(x)}}^{\otimes 2}}, $$
for $\hh = \cb{1, \, ..., \, \lfloor n/2 \rfloor}$. By the proof of Theorem \ref{theo:gauss},
we know that \smash{$\lVert (\htheta(x), \, \hnu(x)) - (\theta(x), \, \nu(x)) \rVert_2^2 = \oo_P(s/n)$},
and so we can use Lemma \ref{lemm:delta_conc} with $\eta = (s/n)^{1/3}$ to verify that
\begin{align*}
&\Psi_{\hh}\p{\htheta(x), \, \hnu(x)} - \Psi\p{\htheta(x), \, \hnu(x)}
=  Q_{\hh} + R_{\hh} + \oo_P\p{\p{\frac{s}{n}}^{2/3}}, \\
&Q_{\hh} := \Psi_{\hh}\p{\theta(x), \, \nu(x)} - \Psi\p{\theta(x), \, \nu(x)}, \\
&R_{\hh} := \bPsi_{\hh}\p{\htheta(x), \, \hnu(x)} - \bPsi_{\hh}\p{\theta(x), \, \nu(x)} \\
&\ \ \ \ \ \ \ \  - \p{\bPsi\p{\htheta(x), \, \hnu(x)} - \bPsi\p{\theta(x), \, \nu(x)}} ,
\end{align*}
where $\smash{\bPsi_{\hh}}$ is defined analogously to $\bPsi$ in \eqref{eq:Psi}.

The first term above, $Q_{\hh}$, is the type of term used by an oracle half-sampling estimator that gets to use the
true parameter values $(\theta(x), \, \nu(x))$ rather than their plug-in analogues.
Given our assumptions and because $(\theta(x), \, \nu(x))$ is non-random, we can use results from
\citet{wager2015estimation} to directly verify that (see their Lemma 7 and Theorem 8)
\begin{equation}
\label{eq:hajek}
\begin{split}
&\p{1 + o_P(1)}\p{\Psi\p{\theta(x), \, \nu(x)} - \EE{\Psi\p{\theta(x), \, \nu(x)}}} \\
&\ \ \ \ \  = \sum_{i = 1}^n \p{\EE{\Psi\p{\theta(x), \, \nu(x)} \cond \p{X_i, \, O_i}} - \EE{\Psi\p{\theta(x), \, \nu(x)}}}, \\
&\p{1 + o_P(1)} \p{\Psi_{\hh}\p{\theta(x), \, \nu(x)} - \EE{\Psi\p{\theta(x), \, \nu(x)}}} \\
&\ \ \ \ \ = \frac{n}{\abs{\hh}} \sum_{i \in \hh} \p{\EE{\Psi\p{\theta(x), \, \nu(x)} \cond \p{X_i, \, O_i}} - \EE{\Psi\p{\theta(x), \, \nu(x)}}}.
\end{split}
\end{equation}
This holds because, as discussed in \citet{wager2015estimation}, forests
have regularity properties by which the scaled first-order effects
\smash{$n(\EE{\Psi\p{\theta(x), \, \nu(x)} \cond \p{X_i, \, O_i}} - \EE{\Psi\p{\theta(x), \, \nu(x)}})$}
depend only on the type of tree being grown; and here of course $\Psi$ and $\Psi_{\hh}$
are built using exactly the same type of trees ($\Psi_{\hh}$ just averages over fewer of them).
Given tail bounds to control moments, it follows immediately that
\begin{align*}
&\EE{Q_{\hh}^{\otimes 2}} \\
&\ \ = n\p{1 + o(1)} \EE{ \p{\EE{\Psi\p{\theta(x), \, \nu(x)} \cond \p{X_1, \, O_1}} - \EE{\Psi\p{\theta(x), \, \nu(x)}}}^{\otimes 2}} \\
&\ \ \ = \p{1 + o(1)} H_n(x; \, \theta(x), \, \nu(x)),
\end{align*}
where the latter is again immediate by the proof of Theorem 8 in \citet{wager2015estimation}.
Thus, taking second moments term $Q_{\hh}$ gives us the limiting expectation we want.

It remains to show that the residual term $R_{\hh}$, used to account for the plug-in effects,
is negligible. Recall that $\bPsi$ is twice differentiable with a uniform second derivative, so
we can take a Taylor expansion as in the proof of Lemma \ref{lemm:coupling}:
$$ R_{\hh} = \p{\nabla \Psi_{\hh}\p{\theta(x), \, \nu(x)} - \nabla \Psi\p{\theta(x), \, \nu(x)}}
\p{\begin{pmatrix}
\htheta(x) \\ \hnu(x)
\end{pmatrix} - 
\begin{pmatrix}
\theta(x) \\ \nu(x)
\end{pmatrix}}
+ \oo_P\p{\frac{s}{n}}, $$
where the $s/n$ error term is a bound on the squared error of \smash{$(\htheta(x), \, \hnu(x))$}.
Now, by the same argument as in \eqref{eq:first_bound}, we see that
\smash{$\lVert \nabla \Psi_{\hh}\p{\theta(x), \, \nu(x)} - \nabla \Psi\p{\theta(x), \, \nu(x)} \rVert \rightarrow_P 0$},
whereas the squared distance between \smash{$(\htheta(x), \, \hnu(x))$} and
\smash{$(\theta(x), \, \nu(x))$} is of the same order as \smash{$H_n(x; \, \theta(x), \, \nu(x))$};
and so in fact
$$\Norm{\EE{R_{\hh}^{\otimes 2}}} = o_P\p{\Norm{ H_n(x; \, \theta(x), \, \nu(x))}}, $$
implying that
$$ \Norm{\EE{\hH_n^{HS}(x)} - H_n(x; \, \theta(x), \, \nu(x))} =  o_P\p{\Norm{ H_n(x; \, \theta(x), \, \nu(x))}}. $$
To establish consistency, it remains to verify concentration of
\smash{$\hH_n^{HS}(x)$}; which, given that the contribution of $R_{\hh}$ is
negligible, also follows immediately from \eqref{eq:hajek}.
Finally, given consistency of \smash{$\hH_n^{HS}(x)$} and Theorem \ref{theo:gauss},
the validity of the delta method confidence intervals is immediate by
Slutsky's theorem whenever \smash{$\lVert\hV(x) - V(x)\rVert \rightarrow_p 0$};
in particular, recall that $V(x)$ is invertible by Assumption \ref{assu:identification}.

\section{Technical Results}
\label{sec:appendix_proofs}

The proofs presented here depend on arguments and notation established
in Appendix \ref{sec:main_proof}.
From now on, we also use shorthand
\begin{equation}
\oo\p{a, \, b, \, c} = \oo\p{\max\cb{a, \, b, \, c}},
\end{equation}
etc. The proof of Proposition \ref{prop:motivation} builds on
that of Proposition \ref{prop:approximation}, so we present the latter first.

\subsection*{Proof of Proposition \ref{prop:approximation}}

Our goal is to couple the actual solution \smash{$\htheta_{C_j}$} of the estimating equation
over the leaf $C_j$ with the gradient-based approximation \smash{$\ttheta_{C_j}$} obtained
by taking a single gradient step from the parent. Here, instead of directly establishing
a relationship between these two quantities, we couple the both to the average of the
influence functions $\rho_i^*(x_P)$ averaged over $C_j$, namely
\begin{equation}
\label{eq:influence_child}
\ttheta_{C_j}^*(x_P) = \theta(x_P) + \frac{1}{\abs{C_j}} \sum_{i \in C_j} \rho_i^*(x_P),
\end{equation}
where $x_P$ is the center of mass of the parent node $P$.

Because the leaf $C_j$ is considered fixed, we can use second-moment bounds on $\psi$
to verify that $\operatorname{Var}[\ttheta_{C_j}^*(x_P)] = \oo\p{1/n_{C_j}}$; meanwhile,
by Lipschitz-continuity of the $M$-function \eqref{eq:M}, we see that
$\mathbb{E}[\ttheta_{C_j}^*(x_P) - \theta(x_P)] = \oo\p{r}$, where $r$ is the radius
of the leaf. Finally, given assumptions made so far about the estimating equation,
it is straight-forward to show that $\htheta_{C_j}$ is consistent for $\theta(x_P)$ in a
limit where $r \rightarrow 0$ and $n_{C_j} \rightarrow \infty$. Thus, a direct analogue to
our result, Lemma \ref{lemm:coupling}, implies that
\begin{equation}
\ttheta_{C_j}^*(x_P) - \htheta_{C_j} = o_P\p{r, \, 1/\sqrt{n_{C_j}}}.
\end{equation}
Next, in order to couple $\ttheta_{C_j}$ and $\ttheta_{C_j}^*(x_P)$, we note that
\begin{equation}
\begin{split}
&\ttheta_{C_j} - \ttheta_{C_j}^*(x_P) = \htheta_P - \theta(x_P) \\
&\ \ \ \ \ \ \ \ 
- \frac{1}{n_{C_j}} \, \xi^\top V(x_P)^{-1}\sum_{i \in C_j}  \p{\psi_{\htheta_P, \, \hnu_P}\p{O_i} -  \psi_{\theta(x_P), \, \nu(x_P)}\p{O_i}} \\
& \ \ \ \ \ \ \ \ 
 - \frac{1}{n_{C_j}} \, \xi^\top \p{A_P^{-1} - V(x_P)^{-1}} \sum_{i \in C_j} \psi_{\htheta_P, \, \hnu_P}\p{O_i};
\end{split}
\end{equation}
our goal is then to bound the terms on the first and second lines at the desired rate.
The first line term is bounded by $o_P(r)$ by smoothness of the $M$-function as
we change $\theta$ and $\nu$, as well as an analogue to Lemma \ref{lemm:delta_conc};
while the second line term can be bounded by recalling that $\|A_P^{-1} - V(x_P)^{-1}\| = o_P(1)$,
and verifying that \smash{$\sum_{i \in C_j} \psi_{\htheta_P, \, \hnu_P}\p{O_i} = \oo_P\p{1/\sqrt{n_{C_j}}, \, r}$}.
Everything we have showed so far implies that
\begin{equation}
\ttheta_{C_j} - \htheta_{C_j} = o_P\p{r, \, 1/\sqrt{n_{C_j}}}, \eqfor j = 1, \, 2.
\end{equation}
Finally, it is straight-forward to check that
\begin{equation}
\ttheta_{C_2} - \ttheta_{C_1} = O_P\p{r, \, 1/\sqrt{n_{C_1}}, \, 1/\sqrt{n_{C_2}}},
\end{equation}
which implies the desired for the coupling of $\Delta(C_1, \, C_2)$ and $\tDelta(C_1, \, C_2)$.

\subsection*{Proof of Proposition \ref{prop:motivation}}

First, we show that we can replace \smash{$\htheta_{C_j}(\mathcal{J})$} with the influence-based
approximation \smash{$\ttheta^*_{C_j}(x_P;\,\mathcal{J})$} (where we make explicit the dependence
of $\ttheta^*_{C_j}$ on the sample $\mathcal{J}$ for clarity) when computing the error function $\err(C_j)$.  To simplify notation without changing the essence of the argument,
we restrict attention to samples $\mathcal{J}$ where the number of observations in $C_1$ and $C_2$ are held fixed at $n_{C_1}$ and $n_{C_2}$, respectively (and recall from the main text that $P$, $C_1$, and $C_2$, subsets of $\mathcal{X}$, are also held fixed).  To start, let $x_P \in P$ be the center of mass of the parent leaf, and observe that
\begin{equation*}
\begin{split}
\err(C_j) &= \EE[X \in C_j]{\p{\htheta_{C_j}(\mathcal{J}) - \theta(X)}^2} \\
&= \EE[X \in C_j]{\p{\ttheta_{C_j}^*(x_P;\,\mathcal{J}) - \theta(X)}^2} 
 + \underbrace{\EE{\p{\htheta_{C_j}(\mathcal{J}) - \ttheta^*_{C_j}(x_P; \mathcal{J})}^2}}_{o\p{r^2, \, 1/n_{C_j}}} \\
& + 2 \underbrace{\EE{\htheta_{C_j}(\mathcal{J}) - \ttheta^*_{C_j}(x_P;\,\mathcal{J})}}_{o\p{r, \, 1/\sqrt{n_{C_j}}}} 
 \underbrace{\p{\EE{\ttheta_{C_j}^*(x_P;\,\mathcal{J})} - \EE[X \in C_j]{\theta(X)}}}_{\oo\p{r^2}},
\end{split}
\end{equation*}
where the first two bounds given in underbraces follow from the proof of Proposition \ref{prop:approximation},
while the last one is a direct consequence of Assumption \ref{assu:identification}, by noting that
\begin{align*}
&\EE[X \in C_j]{\theta(X)} - \EE{\ttheta_{C_j}^*(x_P;\,\mathcal{J})} \\
& \ \ = \EE[X \in C_j]{\theta(X) - \theta(x_P) - \xi^\top \p{\nabla M_{\theta(x_P), \, \nu(x_P)}(x_P)}^{-1} M_{\theta(x_P), \, \nu(x_P)}(X)},
 \end{align*}
and so this term is just the average error from Taylor expanding $M_{\theta(x_P), \, \nu(x_P)}(\cdot)$
over $C_j$. 
%Recall that all expectations are taken over \smash{$\htheta_{C_j}$}, \smash{$\ttheta^*_{C_j}(x_P)$}, and $X$; while the leaves $C_j$, the observation counts $n_{C_j}$, and the anchor point $x_P \in P$ are taken as fixed.
Now, using the above expansion, we find that
\begin{equation*}
\err\p{C_1, \, C_2}  = \sum_{j = 1}^2 \frac{n_{C_j}}{n_P}  \, \EE[X \in C_j]{\p{\ttheta_{C_j}^*(x_P;\,\mathcal{J}) - \theta(X)}^2} +o\p{r^2, \, \frac{1}{n_{C_1}}, \, \frac{1}{n_{C_2}}} 
\end{equation*}
Following arguments of \citet{athey2016recursive}, we see that
\begin{equation*}
\begin{split}
&\EE[X \in C_j]{\p{\ttheta_{C_j}^*(x_P;\,\mathcal{J}) - \theta(X)}^2} = \Var[X \in C_j]{\theta(X)} + \Var{\ttheta_{C_j}^*(x_P;\,\mathcal{J})} \\
&\ \ \ \ \ \ \ \ \ + \p{\EE{\ttheta_{C_j}^*(x_P;\,\mathcal{J})} - \EE[X \in C_j]{\theta(X)}}^2,
\end{split}
\end{equation*}
and the last term is bounded by $\oo\p{r^4}$ as argued above. Thus,
\begin{align*}
\err&\p{C_1, \, C_2} \\
&= \sum_{j = 1}^2 \frac{n_{C_j}}{n_P} \p{ \Var[X \in C_j]{\theta(X)} + \Var{\ttheta_{C_j}^*(x_P;\,\mathcal{J})} } + o\p{r^2, \, \frac{1}{n_{C_1}}, \, \frac{1}{n_{C_2}}} \\
&= \Var[X \in P]{\theta(X)} - \frac{n_{C_1}n_{C_2}}{n_P^2} \p{\EE[X \in C_2]{\theta(X)} - \EE[X \in C_1]{\theta(X)}}^2  \\
& \ \ \ \  + \sum_{j = 1}^2 \frac{n_{C_j}}{n_P}  \Var{\ttheta_{C_j}^*(x_P;\,\mathcal{J})} + o\p{r^2, \, \frac{1}{n_{C_1}}, \, \frac{1}{n_{C_2}}} \\
&= \Var[X \in P]{\theta(X)} - \frac{n_{C_1}n_{C_2}}{n_P^2}\EE{\p{\ttheta_{C_2}^*(x_P;\,\mathcal{J}) - \ttheta_{C_1}^*(x_P;\,\mathcal{J})}^2} \\
&\ \ \ \ + \frac{n_{C_1}n_{C_2}}{n_P^2} \,\Bigg(\EE{\p{\ttheta_{C_2}^*(x_P;\,\mathcal{J}) - \ttheta_{C_1}^*(x_P;\,\mathcal{J})}^2} \\
&\ \ \ \ \ \ \ \ \ \ \ \ - \EE{\ttheta_{C_2}^*(x_P;\,\mathcal{J}) - \ttheta_{C_1}^*(x_P;\,\mathcal{J})}^2\Bigg) \\
& \ \ \ \ + \sum_{j = 1}^2 \frac{n_{C_j}}{n_P}  \Var{\ttheta_{C_j}^*(x_P;\,\mathcal{J})} + o\p{r^2, \, \frac{1}{n_{C_1}}, \, \frac{1}{n_{C_2}}}.
\end{align*}
Now, to parse this expression, note that, by the proof of Proposition \ref{prop:approximation},
\begin{align*}
&\EE{\Delta\p{C_1, \, C_2}} = \frac{n_{C_1}n_{C_2}}{n_P^2} \, \EE{\p{\ttheta_{C_2}^*(x_P;\,\mathcal{J}) - \ttheta_{C_1}^*(x_P;\,\mathcal{J})}^2} \\
&\ \ \ \ \ \ \ \ + o\p{r^2, \, \frac{1}{n_{C_1}}, \, \frac{1}{n_{C_2}}}.
\end{align*}
Thus, writing
\smash{$K(P) := \Var[X \in P]{\theta(X)}$}
as the split-independent error term, all that remains is the
sampling variance of $\Delta\p{C_1, \, C_2}$ due to noise in the training
sample $\mathcal{J}$ (which becomes negligible as $n$ gets large),
and a term
\begin{equation*}
\mathcal{E} := \frac{1}{n_P} \sum_{j = 1}^2 n_{C_j} \p{2 - \frac{n_{C_j}}{n_P}}  \Var{\ttheta_{C_j}^*(x_P;\,\mathcal{J})}
\end{equation*}
that captures the effect of overfitting to random noise when estimating \smash{$\ttheta_{C_j}^*(x_P)$}.
This last term scales as \smash{$\mathcal{E}  = \oo_P(1/{n_{C_1}}, \, 1/{n_{C_2}})$}, 
and so can be ignored since we assume that $n_P \gg r^{-2}$.
Note that if we attempt to correct for a plug-in version of $\mathcal{E}$, we recover
exactly the variance correction of \citet{athey2016recursive}, up to an additive term that
is the same for all splits and so doesn't affect split selection.

\subsection*{Proof of Lemma \ref{lemm:forest_covariance}}
We first note that, because we grew our forest honestly (Specification \ref{spec:forest})
and so $\alpha_i$ is independent of $O_i$ conditionally on $X_i$, we
can use the chain rule to verify that
\begin{align*}
&\EE{\Psi\p{\theta, \, \nu}  - \bPsi\p{\theta, \, \nu}} \\
&\ \ \ \ \ \ \ \ = \sum_{i = 1}^n \EE{\EE{\alpha_i(x) \cond X_i} \p{\EE{ \psi_{\theta, \, \nu}(O_i) \cond X_i} - M_{\theta, \, \nu}(X_i)}} = 0,
\end{align*}
and so $\delta$ must be mean-zero.

Next, to establish bounds on the second moments, we start by considering individual
trees. To do so, define
$$ \ecal_{\theta, \, \nu}(O_i,\, X_i) = \psi_{\theta, \, \nu}(O_i) - M_{\theta, \, \nu}(X_i). $$
Because $\EE{\ecal_{\theta, \, \nu}(O_i,\, X_i) \cond M_{\theta, \, \nu}(X_i)} = 0$ and $M_{\theta, \, \nu}(X_i)$
is locally $(\theta, \, \nu)$-Lipschitz continuous by Assumption \ref{assu:identification}, we can verify
that the worst-case variogram of the $\ecal_{\theta, \, \nu}(O_i,\, X_i)$ must also satisfy \eqref{eq:variogram}.
Now, as in our Algorithm \ref{alg:forest}  let $\jj_1, \, \jj_2$ be any
non-overlapping subset of points of size $\lfloor s/2 \rfloor$ and $\lceil s/2 \rceil$ respectively.
Let $\alpha_i \geq 0$ be weights summing to 1 such that
$\cb{\alpha_i : i \in \jj}$ depends only on $\jj_2$ and on $\cb{X_i : i \in \jj_1}$, and write
$$ T_{\theta, \, \nu}(\jj_1, \, \jj_2) = \sum_{\cb{i \in \jj_1}} \alpha_i \ecal_{\theta, \, \nu}(O_i,\, X_i). $$
By the previous argument, we already know that $\EE{T_{\theta, \, \nu}(\jj_1, \, \jj_2)} = 0$;
meanwhile, thanks to the variogram bound, for any pair of points
$(\theta, \, \nu)$ and $(\theta', \, \nu')$,
\begin{equation}
\label{eq:tree_variogram}
\begin{split}
&\EE{\Norm{T_{\theta, \, \nu}(\jj_1, \, \jj_2) - T_{\theta', \, \nu'}(\jj_1, \, \jj_2)}_2^2} \\
&\ \ \ \ \ \ \ \ \ \leq \EE{\sum_{\cb{i \in \jj_1}} \alpha_i^2 \EE{\Norm{\ecal_{\theta, \, \nu}(O_i,\, X_i) - \ecal_{\theta', \, \nu'}(O_i,\, X_i)}_2^2 \cond X_i}}  \\
& \ \ \ \ \ \ \ \ \leq L \Norm{\begin{pmatrix} \theta \\ \nu \end{pmatrix} -  \begin{pmatrix} \theta' \\ \nu' \end{pmatrix}}_2.
\end{split}
\end{equation}
As in arguments used by \citet{wager2015estimation}, we see that
our quantity of interest $U$-statistic over $T$, and in particular
\begin{align*}
&\delta\p{\p{\theta, \, \nu}, \, \p{\theta', \, \nu'}} \\
&\ \ \ \ \ \ \ \ = \binom{n}{\lfloor s/2 \rfloor, \lceil s/2 \rceil}^{-1} \sum_{\cb{\set_1, \, \set_2 \in \cb{1, \, ..., \, n}}} T_{\theta, \, \nu}(\jj_1, \, \jj_2) - T_{\theta', \, \nu'}(\jj_1, \, \jj_2). 
\end{align*}
Thus, combing our above variogram bound for $T$ with results on $U$-statistics
going back to \citet{hoeffding1948class}, we see that \eqref{eq:delta_cov} holds.

\subsection*{Proof of Lemma \ref{lemm:delta_conc}}

We start by establishing a concentration bound for
$\delta$ at a single point. Given Assumption \ref{assu:donsker}, we
know that $\Norm{\delta}_\infty$ is bounded by $2G$, where $G$ is as defined in the problem statement. Thus,
recalling that $\delta$ is a $U$-statistic and using \eqref{eq:tree_variogram}
to bound the variance of a single tree, we can use the Bernstein bound for $U$-statistics
established by \citet{hoeffding1963probability} to verify that, for any $\eta > 0$,
\begin{equation}
\label{eq:hoeff_bernstein}
\begin{split}
&\PP{\Norm{\delta\p{\p{\theta, \, \nu}, \, \p{\theta', \, \nu'}}}_\infty > \eta} \\
& \ \ \ \ \ \ \ \ 
\leq 2k \exp\p{-\lfloor n/s \rfloor \eta^2 \, \big/ \, \p{2 L \Norm{\begin{pmatrix} \theta -\theta' \\ \nu - \nu' \end{pmatrix}}_2 + \frac{4G}{3}\, \eta}}.
\end{split}
\end{equation}
In other words, as expected, the forest concentrates at a rate $\sqrt{s/n}$.

Now, the kernel of $\delta$, i.e., the function evaluated on
subsamples, is a sum of 4 components that can all be bracketed into a number of brackets bounded
as in \eqref{eq:bracketing_entropy}, using the radius \eqref{eq:bracketing_radius}. Thus, the kernel of $\delta$ can
be bracketed with respect to $L_2$-measure with a bracketing entropy of at most $16\kappa/\eta$.
Given these preliminaries, we proceed by replicating the argument
from Lemma 3.4.2 of \citet{van1996weak} and, in particular, replacing all applications of Bernstein's
inequality with Bernstein's inequality for $U$-statistics as in \eqref{eq:hoeff_bernstein},
we find that for any set $\set$ with
$\Norm{T_{\theta, \, \nu}(\jj_1, \, \jj_2) - T_{\theta', \, \nu'}(\jj_1, \, \jj_2)}_2^2 \leq r^2$
for all $((\theta, \, \nu), \, (\theta', \, \nu')) \in \set$, we have
\begin{align*}
&\EE{\sup\cb{\delta\p{\p{\theta, \, \nu}, \, \p{\theta', \, \nu'}} : ((\theta, \, \nu), \, (\theta', \, \nu')) \in \set}} \\
&\ \ \ \ \ \ \ \ = \oo\p{\frac{J_{[]}(r, \, \delta, \, L_2)}{\sqrt{ n/s }} + \frac{J^2_{[]}(r, \, \delta, \, L_2)}{r^2 \,  n/s } 2G},
\end{align*}
where $J_{[]}$ is the bracketing entropy integral
$$ J_{[]}(r, \, \delta, \, L_2) := \int_0^r \sqrt{1 + \log \p{N_{[]}(\eta, \, \delta, \, L_2)}} \ d\eta, $$
and we omitted the $\lfloor \cdot \rfloor$ notation since $(n/s) / \lfloor n/s \rfloor \leq 2$.
From our bounds on the bracketing number we get $J_{[]}(r, \, \delta, \, L_2) \leq 4\sqrt{\kappa r} + o(\sqrt{r})$.
Thus, thanks to Lemma \ref{lemm:forest_covariance}, we conclude by applying the above result with
$r = L \eta$.

\section{Simulating Instrumental Variables Forests}
\label{sec:iv_simu}

\subsection{Evaluating the Instrumental Variables Splitting Rule}
\label{sec:iv_split}

\begin{figure}
\centering
\begin{tabular}{cc}
\includegraphics[width=\FIGW\textwidth]{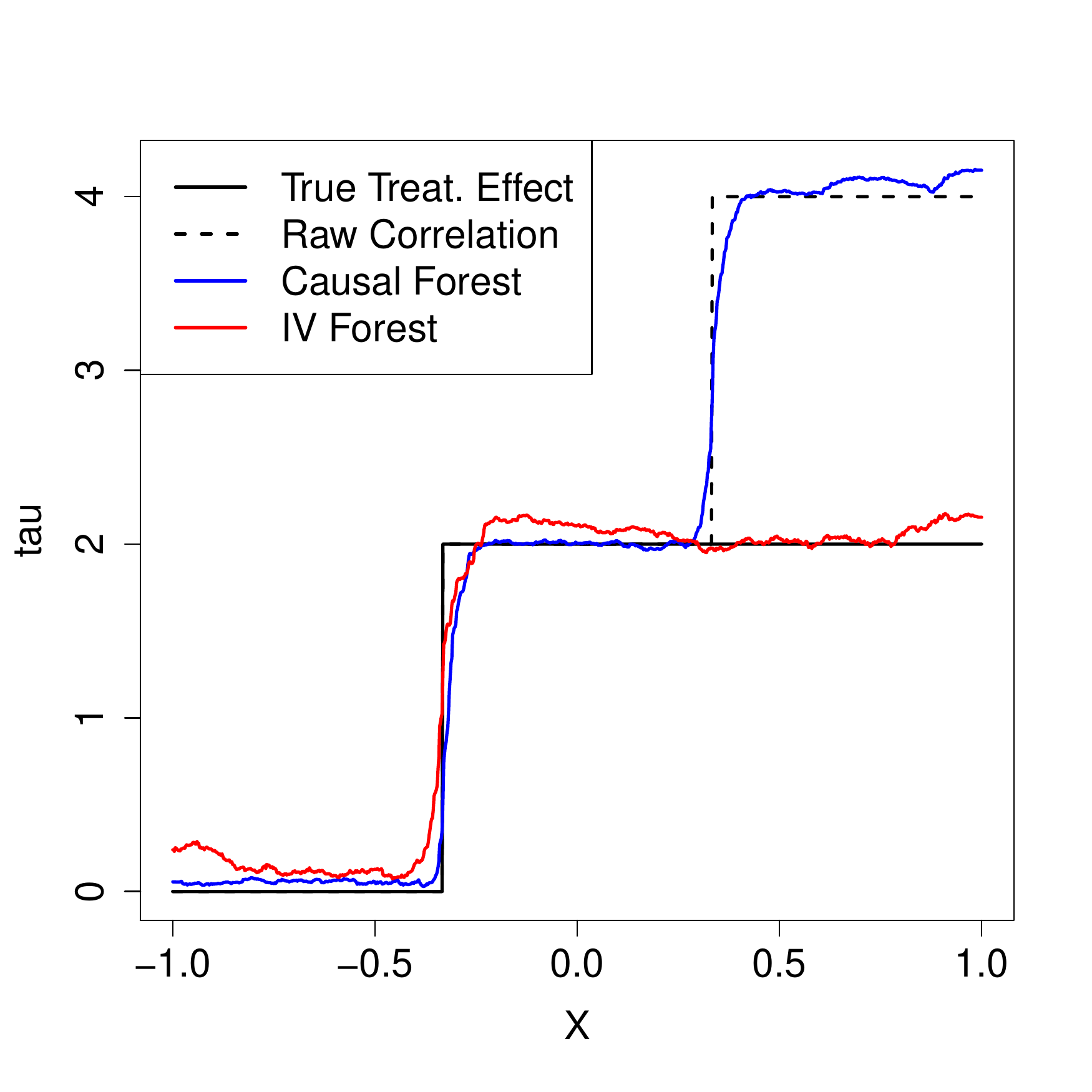} &
\includegraphics[width=\FIGW\textwidth]{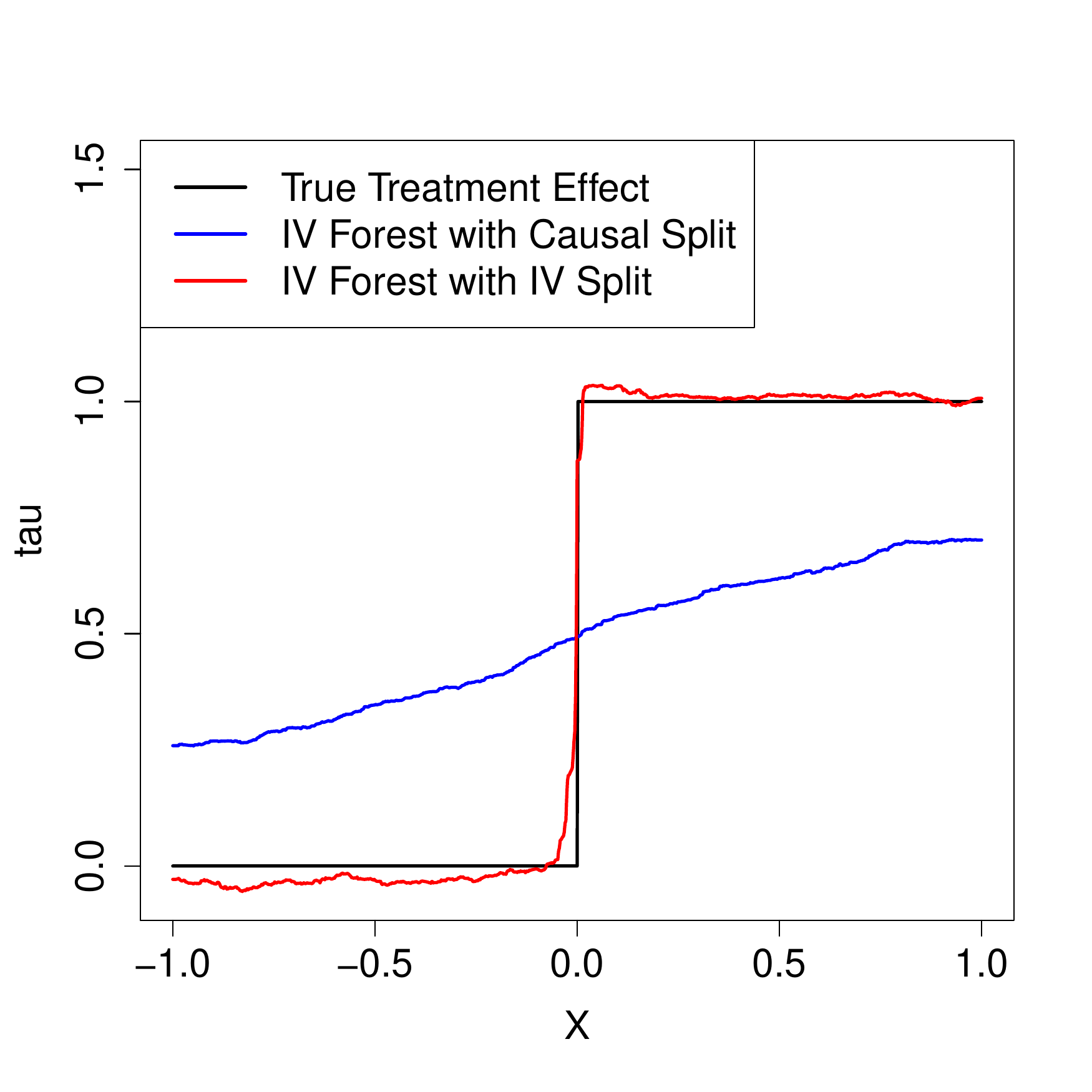}
\end{tabular}
\caption{In both panels, we generate data as $X_i \sim [-1, \, 1]^p$, with $n = 10,000$ and $p = 20$.}
\label{fig:iv_simu}
\end{figure}

We start our simulation analysis with a simple diagnostic of IV splitting rules, and
illustrate the behavior of IV forests in Figure \ref{fig:iv_simu} using two simple simulation designs.
In both examples, $X$ is uniformly spread over a cube, $X_i \sim [-1, \, 1]^p$, but the causal effect
$\tau(X_i)$ only depends on the first coordinate $(X_i)_1$. In both panels of Figure \ref{fig:iv_simu},
we show estimates of $\tau(x)$ produced by different methods, where we vary $x_1$ and set all
other coordinates to 0.

In the first panel, we illustrate the importance of using an IV forests when the received treatment
may be endogenous. We consider a case where the true causal effect of has a single jump,
$\tau(X_i) = 2 \times \1\p{\cb{(X_i)_1 > -1/3}}$. Meanwhile, at $(X_i)_1 = +1/3$, there is a change
in the correlation structure between $W_i$ and $\varepsilon_i$ that leads to a spurious
(i.e., non-causal) jump in the correlation between $W_i$ and $Y_i$. As expected, our IV forest
correctly picks out the first jump while ignoring the second one. Conversely, a plain causal forest
as in Section \ref{sec:cf} that assumes that the received treatment $W_i$ is
exogenous will mistakenly also pick out the second spurious jump in the correlation structure
of $W_i$ and $Y_i$.

The second panel tests our splitting rule. We have a simulation design where there
is a jump in the true causal effect, $\tau(X_i) = \1\p{\cb{(X_i)_1 > 0}}$. However this causal effect is
masked by a change in the correlation of $W_i$ and $\varepsilon_i$, such that the joint distribution
of $W_i$ and $Y_i$ does not depend on $X_i$. The IV forest described above
again performs well; however, the simpler causal tree splitting from
Section \ref{sec:cape} that was not designed for IV regression
fails to accurately detect the jump.

\subsection{Numerical Comparisons}
\label{sec:bigsimu}

We now examine the value of adaptivity in local instrumental variables regression using generalized random forests
across several simulation designs. We compare the following four methods:
{\bf nearest neighbors} instrumental variables regression, which sets $\alpha_i(x) = 1/k$ in
\eqref{eq:gee_estimate} for the $k$ nearest neighbors of $x$ in Euclidean distance,
{\bf series} instrumental variables regression, 
plain {\bf generalized random forests} as described above, and finally
{\bf centered generalized random forests}, using residualization via marginal regressions as in \citet{robinson1988root}.

Due to computational constraints, we used a fairly limited amount of tuning for each method.
For the nearest neighbors method, we tried $k = 10, \, 30, \, 100, \, 300$,
and report results for the best resulting choice of $k$ in each setting.
For series estimation, we expanded out each feature into a natural spline basis with
3 degrees of freedom, using the \texttt{R} function \texttt{ns}. We also considered adding
interactions of these spline terms to the series basis; however, this led to poor estimates
in all of our experiments and so we do not detail these results. Thus, our series method
effectively amounts to additive modeling.
We made no effort to tune generalized random forests, and simply ran them with the default
tuning parameters in our \texttt{grf} software, including a subsample size $s = n/2$.
We implemented the nearest neighbors method with the \texttt{R} package \texttt{FNN}
\citep{beygelzimer2013fnn}, and used the function \texttt{ivreg} from the \texttt{R}
package \texttt{AER} \citep{kleiber2008applied} for series regression.

In all of our simulations, we drew our data from the following generative model, motivated
by an intention to treat design:\footnote{Intuitively, we could think of $Z_i$ as a random intention
to treat and of $Q_i$ as a compliance variable; then, if $\omega > 0$, subjects with better
outcomes $\varepsilon_i$ are more likely to comply, and we need to use the instrument $Z_i$
to deal with this non-compliance effect.}
\begin{equation}
\label{eq:simu_design}
\begin{split}
&X_i \sim \nn\p{0, \, I_{p \times p}}, \ \varepsilon_i \sim \nn\p{0, \, 1}, \
Z_i \sim \operatorname{Binom}\p{1/3}, \\
&Q_i \sim \operatorname{Binom}\p{1/\p{1 + e^{- \omega \varepsilon_i}}}, \
W_i = Z_i \land Q_i, \\
&Y_i = \mu\p{X_i} + \p{W_i - 1/2} \tau\p{X_i} + \varepsilon_i.
\end{split}
\end{equation}
In other words, we exogenously draw features $X_i$, a noise term $\varepsilon_i$ and a
binary instrument $Z_i$. Then, the treatment $W_i$ itself depends on both $Z_i$ and $Q_i$,
where $Q_i$ is a random noise term that is correlated with the noise $\varepsilon_i$ when
$\omega > 0$.
We varied the following problem parameters.
{\bf Confounding:} We toggled the confounding parameter $\omega$
in \eqref{eq:simu_design} between $\omega = 0$ (no confounding) and $\omega = 1$
(confounding). {\bf Sparsity of signal:} The signal $\tau(x)$ depended on $\kappa_\tau$
features; we used $\kappa_\tau \in \cb{2, \, 4}$. {\bf Additivity of signal:} When true, we set
$\tau(x) = \sum_{j = 1}^{\kappa_\tau} \max\{0, \, x_j\}$; when false, we set
$\tau(x) = \max\{0, \, \sum_{j = 1}^{\kappa_\tau} x_j\}$.
{\bf Presence of nuisance terms:} When true, we set
$\mu(x) = 3\max\{0, \, x_5\} + 3\max\{0, \, x_6\}$ or
$\mu(x) = 3\max\{0, \, x_5 + x_6\}$ depending on
the additive signal condition; when false we set $\mu(x) = 0$.
We also varied the {\bf ambient dimension} $p$ and {\bf sample size}
$n$.\footnote{Note that these simulation setups do not perfectly match the assumptions
made in Section \ref{sec:theory}, because if we map the features into the unit cube
via a monotone transformation, then the signals are no longer Lipschitz. Reassuringly, this
does not appear to hurt performance of our method.}

\begin{table}[p]
\newgeometry{left=1.3in,right=1.3in}
\centering
% latex table generated in R 3.3.1 by xtable 1.8-2 package
% Tue Jan 17 10:56:59 2017
\begin{tabular}{||cc||ccc||cccc||cccc||}
  \hline
\hline
 & & & & & \multicolumn{4}{c||}{No nuisance from $\mu(\cdot)$} & \multicolumn{4}{c||}{Presence of main effect $\mu(\cdot)$} \\
add. & conf. & $\kappa_\tau$ & $p$ & $n$ & kNN & series & GRF & C.~GRF & kNN & series & GRF & C.~GRF  \\ 
  \hline
\hline
yes & no & 2 & 10 & 1000 & 0.50 & 0.87 & \bf 0.33 & 0.33 & 0.77 & 1.08 & 0.74 & \bf 0.40 \\
  yes & no & 2 & 10 & 2000 & 0.42 & 0.36 & \bf 0.23 & 0.23 & 0.64 & 0.43 & 0.56 & \bf 0.27 \\
  yes & no & 2 & 20 & 1000 & 0.56 & 2.18 & 0.41 & \bf 0.40 & 0.82 & 2.67 & 0.76 & \bf 0.48 \\
  yes & no & 2 & 20 & 2000 & 0.51 & 0.75 & 0.31 & \bf 0.31 & 0.78 & 0.89 & 0.64 & \bf 0.34 \\
   \hline
yes & no & 4 & 10 & 1000 & 0.87 & 0.86 & 0.65 & \bf 0.64 & 1.23 & 1.01 & 1.11 & \bf 0.71 \\
  yes & no & 4 & 10 & 2000 & 0.79 & \bf 0.37 & 0.49 & 0.48 & 1.03 & \bf 0.43 & 0.86 & 0.51 \\
  yes & no & 4 & 20 & 1000 & 1.09 & 2.06 & 0.85 & \bf 0.83 & 1.35 & 2.52 & 1.33 & \bf 0.94 \\
  yes & no & 4 & 20 & 2000 & 0.96 & 0.80 & 0.64 & \bf 0.62 & 1.23 & 0.94 & 1.07 & \bf 0.70 \\
   \hline
yes & yes & 2 & 10 & 1000 & 0.51 & 0.89 & \bf 0.35 & 0.36 & 0.72 & 1.01 & 0.69 & \bf 0.38 \\
  yes & yes & 2 & 10 & 2000 & 0.43 & 0.37 & \bf 0.23 & 0.24 & 0.66 & 0.42 & 0.57 & \bf 0.26 \\
  yes & yes & 2 & 20 & 1000 & 0.57 & 2.25 & 0.40 & \bf 0.39 & 0.86 & 2.47 & 0.79 & \bf 0.47 \\
  yes & yes & 2 & 20 & 2000 & 0.51 & 0.79 & 0.28 & \bf 0.28 & 0.79 & 0.94 & 0.65 & \bf 0.34 \\
   \hline
yes & yes & 4 & 10 & 1000 & 0.87 & 0.88 & 0.63 & \bf 0.62 & 1.21 & 0.99 & 1.12 & \bf 0.69 \\
  yes & yes & 4 & 10 & 2000 & 0.78 & \bf 0.37 & 0.47 & 0.46 & 1.02 & \bf 0.44 & 0.87 & 0.51 \\
  yes & yes & 4 & 20 & 1000 & 1.05 & 2.41 & 0.80 & \bf 0.78 & 1.33 & 2.52 & 1.28 & \bf 0.91 \\
  yes & yes & 4 & 20 & 2000 & 0.97 & 0.78 & 0.64 & \bf 0.62 & 1.22 & 0.93 & 1.07 & \bf 0.67 \\
   \hline
\hline
no & no & 2 & 10 & 1000 & 0.49 & 0.94 & \bf 0.38 & 0.39 & 0.76 & 1.86 & 0.85 & \bf 0.47 \\
  no & no & 2 & 10 & 2000 & 0.41 & 0.44 & 0.29 & \bf 0.29 & 0.61 & 0.77 & 0.63 & \bf 0.32 \\
  no & no & 2 & 20 & 1000 & 0.57 & 2.34 & 0.47 & \bf 0.47 & 0.88 & 4.47 & 0.89 & \bf 0.57 \\
  no & no & 2 & 20 & 2000 & 0.50 & 0.89 & 0.35 & \bf 0.35 & 0.80 & 1.59 & 0.71 & \bf 0.43 \\
   \hline
no & no & 4 & 10 & 1000 & 0.83 & 1.17 & 0.77 & \bf 0.74 & 1.18 & 2.09 & 1.31 & \bf 0.87 \\
  no & no & 4 & 10 & 2000 & 0.74 & 0.66 & 0.64 & \bf 0.61 & 1.00 & 1.02 & 1.05 & \bf 0.66 \\
  no & no & 4 & 20 & 1000 & 1.04 & 2.43 & 0.98 & \bf 0.95 & 1.32 & 4.57 & 1.35 & \bf 1.04 \\
  no & no & 4 & 20 & 2000 & 0.93 & 1.10 & 0.80 & \bf 0.77 & 1.18 & 1.85 & 1.18 & \bf 0.87 \\
   \hline
no & yes & 2 & 10 & 1000 & 0.49 & 0.96 & \bf 0.37 & 0.37 & 0.73 & 1.86 & 0.88 & \bf 0.48 \\
  no & yes & 2 & 10 & 2000 & 0.41 & 0.44 & \bf 0.28 & 0.28 & 0.62 & 0.85 & 0.65 & \bf 0.34 \\
  no & yes & 2 & 20 & 1000 & 0.55 & 2.42 & 0.44 & \bf 0.43 & 0.85 & 4.16 & 0.89 & \bf 0.57 \\
  no & yes & 2 & 20 & 2000 & 0.49 & 0.88 & 0.34 & \bf 0.33 & 0.75 & 1.59 & 0.70 & \bf 0.41 \\
   \hline
no & yes & 4 & 10 & 1000 & 0.83 & 1.15 & 0.77 & \bf 0.74 & 1.19 & 2.00 & 1.23 & \bf 0.84 \\
  no & yes & 4 & 10 & 2000 & 0.73 & 0.64 & 0.62 & \bf 0.60 & 1.01 & 1.04 & 1.05 & \bf 0.66 \\
  no & yes & 4 & 20 & 1000 & 1.04 & 2.70 & 0.96 & \bf 0.94 & 1.36 & 4.67 & 1.37 & \bf 1.05 \\
  no & yes & 4 & 20 & 2000 & 0.94 & 1.08 & 0.81 & \bf 0.78 & 1.22 & 1.86 & 1.17 & \bf 0.84 \\
   \hline
\hline
\end{tabular}
\caption{Results from simulation study described in Appendix \ref{sec:bigsimu}, in terms of
mean-squared error for the treatment effect on a test set, i.e., $\mathbb{E}[(\htau(X) - \tau(X))^2]$,
where $X$ is a test example. The methods under
comparison are centered generalized random forests (C.~GRF), plain generalized random forests (GRF), series instrumental
variables regression, and the nearest neighbors method (kNN). The simulation specification varies by whether or not
the function $\mu(\cdot)$ in \eqref{eq:simu_design} is 0, whether all signals are additive (add.),
whether the received treatment $W$ is confounded (conf.), the signal dimension ($\kappa_\tau$),
the ambient dimension ($p$), and the sample size ($n$). All errors are aggregated over 100
runs of the simulation with 1,000 test points each, and all forests have $B = 2,000$ trees.}
\label{tab:simu_results}
\restoregeometry
\end{table}

Results from the simulation study are presented in Table \ref{tab:simu_results}.
We see that the forest-based methods achieve consistently good performance across a
wide variety of simulation designs, and do not appear to be too sensitive to non-additive signals
or the presence of fairly strong confounding in the received treatment. Moreover, we see that
the centering behaves as we might have hoped. When there is no nuisance from $\mu(\cdot)$,
the centered and uncentered forests perform comparably, while when we add in the nuisance
term, the centering substantially improves performance.

It is also interesting to examine the few situations where the series method
substantially improves over generalized random forests. This only happens in situations
where the true signal is additive (as expected),
and, moreover, the ambient dimension is small ($p = 10$) while the signal dimension is
relatively high ($\kappa_\tau = 4$). In other words, these are the simulation designs where the
potential upside from adaptively learning a sparse neighborhood function are the smallest.
These results corroborate the intuition that forests provide a form of variable selection for nearest-neighbor
estimation.\footnote{In this simulation design, sparse variants of the series regression based on the lasso
might be expected to perform well. Here, however, we only examine
the ability of generalized random forests to improve over non-adaptive baselines; a thorough comparison of when lasso- versus
forest-based methods perform better is a question that falls beyond the scope of this paper, and
hinges on the experience of practitioners in different application areas. In the traditional regression context,
both lasso- and forest-based methods have been found to work best in different application areas, and
can be considered complementary methods in an applied statistician's toolbox.}

\subsection{Evaluating Confidence Intervals}

\begin{table}[t]
\centering
\begin{tabular}{cc}
\begin{tabular}{|ccc|cc|}
\hline
\multicolumn{3}{|r|}{additive} & yes & no \\
$\kappa_\tau$ & $p$ & $n$ & \multicolumn{2}{c|}{coverage} \\ 
  \hline
\hline
2 & 6 & 2000 & 0.87 & 0.82 \\ 
  2 & 6 & 4000 & 0.91 & 0.87 \\ 
  2 & 6 & 8000 & 0.91 & 0.89 \\ 
  2 & 6 & 16000 & 0.93 & 0.92 \\ 
   \hline
2 & 12 & 2000 & 0.75 & 0.77 \\ 
  2 & 12 & 4000 & 0.82 & 0.77 \\ 
  2 & 12 & 8000 & 0.89 & 0.86 \\ 
  2 & 12 & 16000 & 0.93 & 0.90 \\ 
   \hline
2 & 18 & 2000 & 0.73 & 0.73 \\ 
  2 & 18 & 4000 & 0.82 & 0.79 \\ 
  2 & 18 & 8000 & 0.89 & 0.80 \\ 
  2 & 18 & 16000 & 0.91 & 0.87 \\ 
   \hline
\hline
4 & 6 & 2000 & 0.78 & 0.75 \\ 
  4 & 6 & 4000 & 0.82 & 0.76 \\ 
  4 & 6 & 8000 & 0.87 & 0.77 \\ 
  4 & 6 & 16000 & 0.88 & 0.81 \\ 
   \hline
4 & 12 & 2000 & 0.64 & 0.59 \\ 
  4 & 12 & 4000 & 0.73 & 0.63 \\ 
  4 & 12 & 8000 & 0.78 & 0.66 \\ 
  4 & 12 & 16000 & 0.82 & 0.70 \\ 
   \hline
4 & 18 & 2000 & 0.59 & 0.53 \\ 
  4 & 18 & 4000 & 0.63 & 0.54 \\ 
  4 & 18 & 8000 & 0.70 & 0.60 \\ 
  4 & 18 & 16000 & 0.78 & 0.60 \\  
   \hline
\end{tabular}
&
\begin{tabular}{|ccc|cc|}
\hline
\multicolumn{3}{|r|}{additive} & yes & no \\
$\kappa_\tau$ & $p$ & $n$ & \multicolumn{2}{c|}{coverage} \\ 
  \hline
\hline
2 & 6 & 2000 & 0.96 & 0.95 \\ 
  2 & 6 & 4000 & 0.96 & 0.96 \\ 
  2 & 6 & 8000 & 0.94 & 0.95 \\ 
  2 & 6 & 16000 & 0.95 & 0.95 \\ 
   \hline
2 & 12 & 2000 & 0.96 & 0.97 \\ 
  2 & 12 & 4000 & 0.97 & 0.95 \\ 
  2 & 12 & 8000 & 0.96 & 0.97 \\ 
  2 & 12 & 16000 & 0.97 & 0.98 \\ 
   \hline
2 & 18 & 2000 & 0.98 & 0.96 \\ 
  2 & 18 & 4000 & 0.97 & 0.97 \\ 
  2 & 18 & 8000 & 0.97 & 0.96 \\ 
  2 & 18 & 16000 & 0.97 & 0.98 \\ 
   \hline
\hline
4 & 6 & 2000 & 0.96 & 0.96 \\ 
  4 & 6 & 4000 & 0.95 & 0.94 \\ 
  4 & 6 & 8000 & 0.95 & 0.95 \\ 
  4 & 6 & 16000 & 0.95 & 0.95 \\ 
   \hline
4 & 12 & 2000 & 0.96 & 0.95 \\ 
  4 & 12 & 4000 & 0.97 & 0.96 \\ 
  4 & 12 & 8000 & 0.97 & 0.97 \\ 
  4 & 12 & 16000 & 0.96 & 0.97 \\ 
   \hline
4 & 18 & 2000 & 0.96 & 0.97 \\ 
  4 & 18 & 4000 & 0.97 & 0.97 \\ 
  4 & 18 & 8000 & 0.96 & 0.96 \\ 
  4 & 18 & 16000 & 0.96 & 0.97 \\ 
   \hline
\end{tabular} \\
target: population $\tau(x)$ & target: expected $\htau(x)$
\end{tabular}
\caption{Empirical coverage of 95\% confidence intervals for instrumental variables forests,
averaged over 20 replications with 1,000 test points each.
The left panel reports coverage of the true effects $\tau(X_i)$ on the test set, while the right panel
measures the fraction of times the expected forest prediction $\EE{\htau(X_i) \cond X_i}$ falls
within the confidence intervals.}
\label{tab:coverage}
\vspace{-1.5\baselineskip}
\end{table}

\begin{figure}
\includegraphics[width=\FIGW\textwidth]{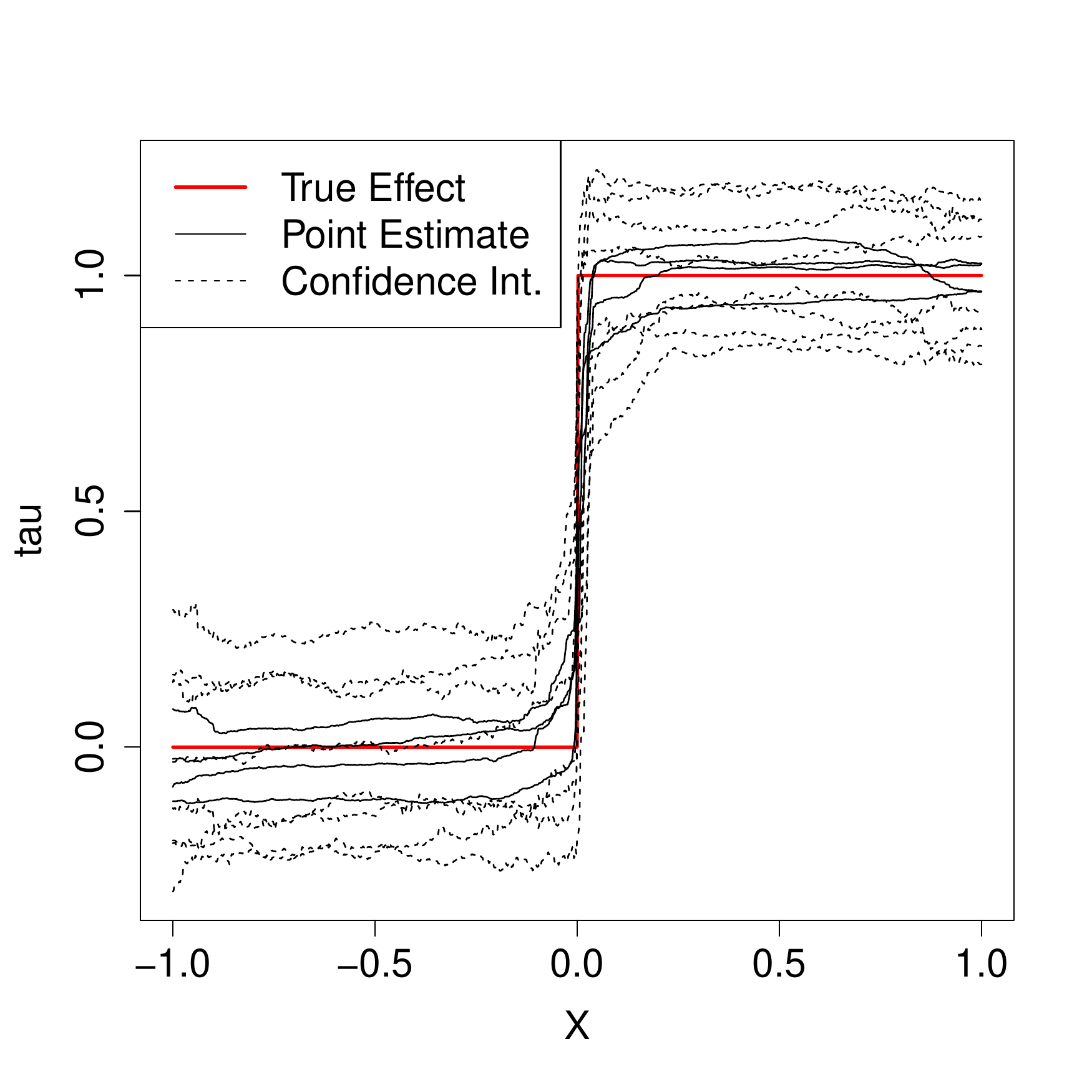}
\caption{Illustration of 95\% confidence intervals for instrumental variables forests across 4 simulation replications.
We use the same simulation setting as in the right panel of Figure \ref{fig:iv_simu},
except now with $n = 4,000$, $p = 20$, and $B = 10,000$ trees.}
\label{fig:ci_picture}
\vspace{-1.5\baselineskip}
\end{figure}

We also examine the quality of the delta method confidence intervals discussed in
Section \ref{sec:delta_method}, built using the bootstrap of little bags \citep{sexton2009standard}.
In Table \ref{tab:coverage}, we report coverage results in a subset of the simulation settings from the previous
section. We always have confounding ($\omega = 1$) and nuisance terms
($\mu(x) = \max\{0, \, x_5\} + \max\{0, \, x_6\}$ or $\mu(x) = \max\{0, \, x_5 + x_6\}$);
we also only consider centered forests. As discussed in \citet{wager2014confidence}, forests
typically require more trees to provide accurate confidence intervals; thus, we use $B = 4,000$ trees
per forest, rather than the default $B = 2,000$ used in Table \ref{tab:simu_results}.
Figure \ref{fig:ci_picture} gives an illustration of our confidence intervals by superimposing
the output from 4 different simulation runs from a single data-generating distribution.

As expected, coverage results are better when $n$ is larger, the ambient dimension $p$ is smaller,
the true signal is sparser, and the true signal is additive. Of these effects, the most important one
in Table \ref{tab:coverage} is the sparsity of $\tau$. When $\kappa_\tau = 2$, i.e., the true signal
can be expressed as a bivariate function, our confidence intervals achieve closer to nominal coverage;
however, when $\kappa_\tau = 4$, performance declines considerably at the sample sizes $n$ under
investigation.

To gain more intuition about this result, the right panel of Table \ref{tab:coverage} reports the fraction
of confidence intervals that cover the expected prediction made by the forest; in other words, it measures
the accuracy with which our confidence intervals quantify the sampling uncertainty of the forest.
If instrumental forests were unbiased, the left and right panels would be the same.
These results suggest that low coverage numbers in the left panel are mostly due to our forests having non-negligible bias,
rather than to failures of Gaussianity or of the variance estimates underlying our confidence intervals.
It would be of considerable interest to develop confidence intervals for random forests that allow
for asymptotically non-vanishing bias.

\end{appendix}

\end{document}